\documentclass[10pt]{article}

\newlength{\minitwocolumn}
\font\teneufm=eufm10
\font\seveneufm=eufm7
\font\fiveeufm=eufm5
\newfam\eufmfam
\textfont\eufmfam=\teneufm
\scriptfont\eufmfam=\seveneufm
\scriptscriptfont\eufmfam=\fiveeufm

\makeatletter
\@addtoreset{equation}{section}
\makeatother

\newtheorem{thm}{Theorem}[section]
\newtheorem{prop}[thm]{Proposition}

\newtheorem{cor}[thm]{Corollary}

\newtheorem{dfn}[thm]{Definition}
\title{\bf
\large{\bf 
DIAGONALIZATION OF TRANSFER MATRIX OF\\
SUPERSYMMETRY $U_{q}(\widehat{sl}(M+1|N+1))$} CHAIN WITH A BOUNDARY}
\begin{document}

\maketitle

\begin{center}
{TAKEO KOJIMA}
\\~\\
{\it
Faculty of Engineering,
Yamagata University, Jonan 4-3-16, Yonezawa 992-8510,
Japan\\
kojima@yz.yamagata-u.ac.jp}
\end{center}

~\\

\begin{abstract}
We study the supersymmetry $U_q(\widehat{sl}(M+1|N+1))$ analogue
of the supersymmetric $t$-$J$ model with a boundary,
in the framework of the algebraic analysis method.
We diagonalize the commuting transfer matrix by
using the bosonization of the vertex operator
associated with the quantum affine supersymmetry.
\end{abstract}

~\\

\section{Introduction}

There have been many developments in the exactly solvable models.
Various methods were invented to solve models.
The algebraic analysis method \cite{JM} provides a powerful method to study
exactly solvable models.
This paper is devoted to
the algebraic analysis method to open boundary problem of 
exactly solvable model.
In this paper we study the quantum supersymmetry 
$U_q(\widehat{sl}(M+1|N+1))$ analogue
of the supersymmetric $t$-$J$ model with a boundary,
where $M,N= 0,1,2,\cdots$ such that $M \neq N$.
The supersymmetric $t$-$J$ model was proposed in an attempt to understand 
high-temperature superconductivity.
In the framework of
the quantum inverse scattering method \cite{Gaudin, Korepin},
the investigations of the supersymmetric $t$-$J$ model and 
its $U_q(\widehat{sl}(M+1|N+1))$ analogue
have been carried out in several papers
\cite{BS, PS, LYH, BR, EK, ZYZ, FWW}.
In the framework of
the algebraic analysis method \cite{JM, Baxter},
the $U_q(\widehat{sl}(M+1|N+1))$ chain
"without boundaries" has been studied in few papers
\cite{KSU, YZ1, YZ2}.
In this paper we focus our attention on the boundary condition
of the exactly solvable model.
We study the supersymmetry $U_q(\widehat{sl}(M+1|N+1))$ chain "with a boundary" 
in the framework of the algebraic analysis method \cite{JM, JKKKM}.
We diagonalize the commuting transfer matrix 
of the supersymmetry $U_q(\widehat{sl}(M+1|N+1))$ chain with a boundary.
Several solvable models with a boundary
have been studied by means of algebraic analysis method
\cite{JKKKM, FK, MW, Kojima1, YZ3, Kojima2, HYZZ}.
Here we would like to draw reader's attention to
new technical aspect 
in our problem.
Generally speaking, in the algebraic analysis method,
the transfer matrix $T_B(z)$ of the solvable model with a boundary is
written by the product of the vertex operators
$\Phi_j^*(z)$ and $\Phi_j(z)$ associated with the quantum affine symmetry
\begin{eqnarray}
T_B^{(i)}(z)=g\sum_{j=1}^{M+N+2}
\Phi_j^*(z^{-1})K^{(i)}(z)_j^j 
\Phi_j(z)(-1)^{[v_j]}.\nonumber
\end{eqnarray}
Here $K^{(i)}(z)_j^j$ is the matrix element of the boundary $K$-matrix.
The key of solving the problem is the bosonization of the boundary state
${_{(i)}}\langle B|$ that satisfies the following condition
\begin{eqnarray}
{_{(i)}}\langle B| T_B^{(i)}(z)={_{(i)}}\langle B|.\nonumber
\end{eqnarray}
By using the bosonizations of the vertex operators,
we construct the boundary state ${_{(i)}}\langle B|$.
Our calculations depend heavily on 
the bosonization formulae of the vertex operators.
For solvable models that are governed by the quantum symmetry
$U_q(\widehat{sl}(N))$, $U_q(A_2^{(2)})$ or the elliptic symmetry $U_{q,p}(\widehat{sl}(N))$
\cite{JKKKM, FK, MW, Kojima1, YZ3, Kojima2},
the bosonizations of the vertex operators
are realized by "monomial". 
However the bosonizations of the vertex operators
for the quantum supersymmetry $U_q(\widehat{sl}(M+1|N+1))$
are realized by "sum".
For instance the bosonizations of the vertex operators 
$\Phi^*_{M+1+j}(z)$ $(j=1,2,\cdots, N+1)$
are written by the sum 
$\sum_{\epsilon_1,\epsilon_2,\cdots,\epsilon_j}$ as followings (see (\ref{boson:VO6})).
\begin{eqnarray}
\Phi_{M+1+j}^*(z)
&=&
\sum_{\epsilon_1,\epsilon_2,\cdots,\epsilon_j=\pm}
e^{\frac{M \pi \sqrt{-1}}{M-N}}q^{j-1}(q-q^{-1})^M(qz)^{-1}
\prod_{k=1}^j \epsilon_k \prod_{k=1}^{M+j}
\int_{C_{M+1+j}} \frac{dw_k}{2\pi \sqrt{-1}w_k}
\nonumber\\
&\times&
\frac{1}{
\displaystyle
\prod_{k=0}^M (1-qw_k/w_{k+1})(1-qw_{k+1}/w_k)
\prod_{k=1}^{j-1}(1-q^{\epsilon_k}w_{M+k}/w_{M+k+1})}\nonumber\\
&\times&
\eta_0 \xi_0 :\phi_1^*(z)
X_1^{-}(qw_1)\cdots X_M^{-}(qw_M)
X_{M+1,\epsilon_1}^{-}(qw_{M+1})\cdots X_{M+j,\epsilon_j}^{-}(qw_{M+j}):
\eta_0 \xi_0.\nonumber
\end{eqnarray}
Technically this is cool part of our problem.
Surprisingly we shall conclude that
the bosonization of the boundary state ${_{(i)}}\langle B|$ is realized by
"monomial", though those of the vertex operator is realized by "sum".
The bosonization of the boundary state ${_{(i)}}\langle B|$
is constructed by acting 
a monomial of exponential $e^{G^{(i)}}$ 
on the highest weight vector
$\langle \Lambda_{M+1}| \in V^*(\Lambda_{M+1})$ of 
the quantum supersymmetry $U_q(\widehat{sl}(M+1|N+1))$.
\begin{eqnarray}
{_{(i)}}\langle B|=\langle \Lambda_{M+1}|e^{G^{(i)}}.\nonumber
\end{eqnarray}
Here $G^{(i)}$ is quadratic in the bosonic operators (see (\ref{def:G})).
We would like to give a comment on the earlier study
in the framework of the algebraic analysis \cite{HYZZ}.
The supersymmetric $t$-$J$ model with a boundary 
(the supersymmetry $U_q(\widehat{sl}(2|1))$ chain with a boundary)
was studied and the bosonization conjecture of the boundary state was given
in \cite{HYZZ}.
However, their conjecture of the boundary state
is different from our bosonization
upon the special case of $M=1, N=0$.
In this paper we give not only
the bosonization formulae of the boundary state,
but also give complete  proof that the vector ${_{(i)}}\langle B|$ 
becomes the eigenvector
of the transfer matrix $T_B^{(i)}(z)$.
In this paper we classify the boundary $K$-matrix and find a new solution
that has three different diagonal elements (see (\ref{def:tildeK3})).
Of-course we construct the boundary state associated with this new $K$-matrix. 

The text is organized as follows.
In section 2 we introduce 
the supersymmetry $U_q(\widehat{sl}(M+1|N+1))$ analogue
of finite supersymmetric finite $t$-$J$ model with double boundaries.
We introduce
the $U_q(\widehat{sl}(M+1|N+1))$ analogue of semi-infinite
$t$-$J$ model as the limit of the finite chain.
In section 3 we give mathematical formulation of
the supersymmetry $U_q(\widehat{sl}(M+1|N+1))$ chain with a boundary.
This formulation is based on the representation theory of
the quantum supersymmetry
$U_q(\widehat{sl}(M+1|N+1))$.
This formulation is free from the difficulty of divergence.
In section 4 we review the bosonizations of the vertex operators
and give the integral representations of the vertex operators.
In section 5 we give the bosonizations of the boundary state,
that is the main result of this paper.
We give complete proof of the bosonization of the boundary state.
In appendix \ref{appendix0}
we summarize the figures that we use in section \ref{section2}.
In appendix \ref{appendix1}
we classify the diagonal solutions of the boundary Yang-Baxter equation 
associated with $U_q(\widehat{sl}(M+1|N+1))$.
New solution is presented even for the small rank case $U_q(\widehat{sl}(2|1))$.
In appendix \ref{appendix2} we summarize the normal orderings,
that we use in section \ref{section4} and \ref{section6}.

\section{$U_q(\widehat{sl}(M+1|N+1))$ chain with a boundary}
\label{section2}

In this section 
we introduce
the $U_q(\widehat{sl}(M+1|N+1))$ chain with a boundary.
We fix a complex number $0<|q|<1$
and two natural numbers $M,N=0,1,2,\cdots$ such that $M \neq N$.

\subsection{Finite $U_q(\widehat{sl}(M+1|N+1))$ chain}

In this section we introduce
the finite $U_q(\widehat{sl}(M+1|N+1))$ chain with double boundaries.
We follow the general scheme given by \cite{Cherdnik, Sklyanin, MN, FWW}.
In what follows we use the standard notation of the $q$-integer
\begin{eqnarray}
[a]_q=\frac{q^a-q^{-a}}{q-q^{-1}}.
\end{eqnarray}
Let us introduce the signatures $\nu_i$ $(i=1,2,\cdots,M+N+2)$ by 
\begin{eqnarray}
\nu_1=\nu_2=\cdots=\nu_{M+1}=+,~~~\nu_{M+2}=\nu_{M+3}=\cdots=\nu_{M+N+2}=-.
\label{def:signature}
\end{eqnarray}
Let us set the vector space 
$V=\oplus_{j=1}^{M+N+2}{\bf C}v_j$.
The ${\bf Z}_2$-grading of the basis $\{v_j\}$ of $V$ is chosen to be 
$\left[v_j\right]=\frac{\nu_j+1}{2}$ $(j=1, 2, \cdots, M+N+2)$.

\begin{dfn}~~~
We set the $R$-matrix 
$R(z) \in {\rm End}(V \otimes V)$ associated with 
the quantum supersymmetry $U_q(\widehat{sl}(M+1|N+1))$
as followings \cite{BS, PS}.
\begin{eqnarray}
R(z)=r(z)\bar{R}(z),~~~
\bar{R}(z)v_{j_1}\otimes v_{j_2}=\sum_{k_1,k_2=1}^{M+N+2}
v_{k_1}\otimes v_{k_2}
\bar{R}(z)_{k_1,k_2}^{j_1,j_2}.
\label{def:R-matrix1}
\end{eqnarray}
Here we have set
\begin{eqnarray}
\bar{R}(z)_{j,j}^{j,j}
&=&
\left\{
\begin{array}{cc}
-1& (1\leq j \leq M+1),\\
-\frac{\displaystyle
(q^2-z)}{\displaystyle
(1-q^2 z)}& (M+2\leq j \leq M+N+2),
\end{array}
\right.
\label{def:R-matrix2}
\\
\bar{R}(z)_{i,j}^{i,j}&=&\frac{(1-z)q}{(1-q^2 z)}
~~~(1\leq i \neq j \leq M+N+2),
\label{def:R-matrix3}
\\
\bar{R}(z)_{i,j}^{j,i}&=&
\left\{
\begin{array}{cc}
(-1)^{ [v_i] [v_j] }\frac{\displaystyle
(1-q^2)}{
\displaystyle
(1-q^2 z)}&
(1\leq i < j \leq M+N+2),
\label{def:R-matrix4}
\\
(-1)^{ [v_i] [v_j] }
\frac{
\displaystyle
(1-q^2)z}{
\displaystyle
(1-q^2 z)}&
(1\leq j < i \leq M+N+2),
\end{array}
\right.
\label{def:R-matrix5}\\
\bar{R}(z)_{i,j}^{i,j}&=&0~~~~~{\rm otherwise}.
\end{eqnarray}
Here we have set
\begin{eqnarray}
r(z)=z^{\frac{1-M+N}{M-N}}\exp\left(-\sum_{m=1}^\infty
\frac{[(M-N-1)m]_q}{m[(M-N)m]_q}q^m(z^m-z^{-m})\right).
\label{def:r}
\end{eqnarray}
\end{dfn}
For instance the function $r(z)$ is written 
as following in the case $M>N$.
\begin{eqnarray}
r(z)=z^{\frac{1-M+N}{M-N}}
\frac{(q^2z;q^{2(M-N)})_\infty 
(q^{2(M-N)}z^{-1};q^{2(M-N)})_\infty}{
(q^{2(M-N)}z;q^{2(M-N)})_\infty 
(q^2z^{-1};q^{2(M-N)})_\infty}.
\end{eqnarray}
Here we have used the infinite product
\begin{eqnarray}
(z;p)_\infty=\prod_{n=0}^\infty (1-p^{n}z)~~~~~(|p|<1).\nonumber
\end{eqnarray}
Graphically, the $R$-matrix $R(z)$ is represented in Fig.1 
in appendix \ref{appendix0}.
The $R$-matrix
${R}(z)$
satisfies the graded Yang-Baxter equation.
\begin{eqnarray}
{R}_{1 2}(z_1/z_2)
{R}_{1 3}(z_1/z_3)
{R}_{2 3}(z_2/z_3)=
{R}_{2 3}(z_2/z_3)
{R}_{1 3}(z_1/z_3)
{R}_{1 2}(z_1/z_2),
\label{def:gYBE}
\end{eqnarray}
where ${R}_{1 2}(z)$, 
${R}_{1 3}(z)$ and 
${R}_{2 3}(z)$ act in $V \otimes V \otimes V$
with ${R}_{1 2}(z)={R}(z)\otimes 1$, 
${R}_{2 3}(z)=1 \otimes {R}(z)$, etc.\\
The relation (\ref{def:gYBE}), expressed in terms of matrix elements,
can be rewritten in the following form.
\begin{eqnarray}
&&\sum_{j_1,j_2,j_3=1}^{M+N+2}
{R}(z_1/z_2)_{i_1,i_2}^{j_1,j_2}
{R}(z_1/z_3)_{j_1,i_3}^{k_1,j_3}
{R}(z_2/z_3)_{j_2,j_3}^{k_2,k_3}
(-1)^{([v_{j_1}]+[v_{k_1}])[v_{j_2}]}\nonumber\\
&=&
\sum_{j_1,j_2,j_3=1}^{M+N+2}
{R}(z_2/z_3)_{i_2,i_3}^{j_2,j_3}
{R}(z_1/z_3)_{i_1,j_3}^{j_1,k_3}
{R}(z_1/z_2)_{j_1,j_2}^{k_1,k_2}
(-1)^{([v_{i_1}]+[v_{j_1}])[v_{j_2}]}.
\end{eqnarray}
Multiplying the signature 
$(-1)^{[v_{k_1}][v_{k_2}]}$
to our $R$-matrix ${R}(z)_{k_1,k_2}^{j_1,j_2}$, we have
the $R$-matrix $R^{PS}(z)$ of the Perk-Schultz model \cite{PS} 
: $R^{PS}(z)_{k_1,k_2}^{j_1,j_2}
=(-1)^{[v_{k_1}][v_{k_2}]}{R}(z)_{k_1,k_2}^{j_1,j_2}$.
The $R$-matrix $R^{PS}(z)$ of the Perk-Schultz model \cite{PS}
satisfies the ungraded Yang-Baxter equation.
The $R$-matrix
${R}(z)$
satisfies the initial condition
${R}(1)=P$ where $P$ is the graded permutation operator :
$P_{k_1,k_2}^{j_1,j_2}=\delta_{j_1,k_2}\delta_{j_2,k_1}
(-1)^{[v_{k_1}][v_{k_2}]}$.
The $R$-matrix
${R}(z)$
satisfies the unitary condition
\begin{eqnarray}
{R}_{1 2}(z){R}_{2 1}(1/z)=1,
\end{eqnarray}
where ${R}_{21}(z)=PR_{12}(z)P$.
The $R$-matrix
${R}(z)$ satisfies 
and the crossing symmetry
\begin{eqnarray}
{R}^{-1, st_1}_{1 2}(z)M_1
{R}^{st_1}_{1 2}(q^{2(N-M)}z)M_1^{-1}=1.
\end{eqnarray}
Here we have set the matrix $M \in {\rm End}(V)$ defined by
\begin{eqnarray}
M_{i,j}=\delta_{i,j}M_j,~~~
M_j=\left\{\begin{array}{cc}
q^{-2(j-1)}&  (1\leq j \leq M+1),\\
q^{-2(2M+2-j)}& (M+2 \leq j \leq M+N+2).
\end{array}\right.
\end{eqnarray}
We have the commutativity $[M \otimes M, R(z)]=0$.
For instance the various supertranspositions of the $R$-matrix are given by
\begin{eqnarray}
&&R^{st_1}(z)_{i,j}^{k,l}
=R(z)_{k,j}^{i,l}(-1)^{[v_i]([v_i]+[v_k])},~
R^{st_2}(z)_{i,j}^{k,l}
=R(z)_{i,l}^{k,j}(-1)^{[v_j]([v_j]+[v_l])},\nonumber
\\
&&R^{st_{12}}(z)_{i,j}^{k,l}
=(-1)^{([v_i]+[v_j])([v_i]+[v_j]+[v_k]+[v_l])}R(z)_{k,l}^{i,j}
=R(z)_{k,l}^{i,j}.
\end{eqnarray}
In appendix \ref{appendix1} we classify the boundary $K$-matrix
that satisfies the graded boundary Yang-Baxter equation (\ref{def:gBYBE}),
and find new diagonal solution (see (\ref{def:tildeK3})).
\begin{dfn}~~~
We set the $K$-matrix $K^{(i)}(z) \in {\rm End}(V)$ $(i=1,2,3)$ 
associated with
the quantum supersymmetry $U_q(\widehat{sl}(M+1|N+1))$ as followings. 
\begin{eqnarray}
&&K^{(i)}(z)=z^{-\frac{2M}{M-N}}
\frac{\varphi^{(i)}(z)}{
\varphi^{(i)}(z^{-1})}
\bar{K}^{(i)}(z)~~~(i=1,2,3),
\\
&&\bar{K}^{(i)}(z)v_j=\sum_{k=1}^{M+N+2}v_k \bar{K}^{(i)}(z)_k^j,\\
&&
\bar{K}^{(i)}(z)_k^j=\delta_{j,k}
\bar{K}^{(i)}(z)_j^j.
\end{eqnarray}
The $\bar{K}^{(i)}(z)_j^j$ and $\varphi^{(i)}(z)$ are given by
following {\rm CONDITION 1, 2, 3}.\\
{\rm CONDITION 1~:~}
$\bar{K}^{(1)}(z)_j^j$ are $\varphi^{(1)}(z)$ are defined by followings.
\begin{eqnarray}
\bar{K}^{(1)}(z)_j^j=1~~~(j=1,2, \cdots, M+N+2),
\end{eqnarray}
and
\begin{eqnarray}
\varphi^{(1)}(z)&=&
\exp\left(
\sum_{m=1}^\infty
\frac{[2(N+1)m]_q}{m [2(M-N)m]_q}z^{2m}+
\sum_{j=1}^M
\sum_{m=1}^\infty
\frac{[2(M-N-j)m]_q}{2m [2(M-N)m]_q}(1-q^{2m})z^{2m}
\right.\nonumber\\
&+&\left.\sum_{j=M+2}^{M+N+1}\sum_{m=1}^\infty
\frac{[2(-M-N-2-j)m]_q}{2m [2(M-N)m]_q}(1+q^{2m})z^{2m}
-\sum_{m=1}^\infty
\frac{[(M-N-1)m]_q}{2m[(M-N)m]_q}q^mz^{2m}
\right).\nonumber\\
\label{def:varphi1}
\end{eqnarray}
{\rm CONDITION 2~:~}
$\bar{K}^{(2)}(z)_j^j$ and $\varphi^{(2)}(z)$ are defined by followings.
We fix a natural number $L=1,2,\cdots,M+N+1$ and 
a complex number $r \in {\bf C}$.
\begin{eqnarray}
\bar{K}^{(2)}(z)_j^j=\left\{\begin{array}{cc}
1& (1\leq j=k \leq L),\\
\frac{\displaystyle 1-r/z}{\displaystyle 1-r z}& (L+1\leq j=k \leq M+N+2).
\end{array}
\right.
\end{eqnarray}
{\rm Condition 2.1~:~}For $L \leq M+1$ we set
\begin{eqnarray}
\varphi^{(2)}(z)= 
\varphi^{(1)}(z)
\times \exp\left(-\sum_{m=1}^\infty \frac{[(M-N-L)m]_q}{m [(M-N)m]_q}
(rq^{-L}z)^m\right).
\label{def:varphi2-1}
\end{eqnarray}
{\rm Condition 2.2~:~}For $M+2 \leq L \leq M+N+2$ we set
\begin{eqnarray}
\varphi^{(2)}(z)= 
\varphi^{(1)}(z)
\times
\exp\left(-\sum_{m=1}^\infty \frac{[(-M-N-2+L)m]_q}{
m [(M-N)m]_q}
(r q^{L-2M-2}z)^m \right).
\label{def:varphi2-2}
\end{eqnarray}
{\rm CONDITION 3~:~}$\bar{K}^{(3)}(z)_j^j$ and $\varphi^{(3)}(z)$ 
are defined by followings.
We fix two natural numbers
$L,K=1,2,\cdots,M+N$ such that $L+K \leq M+N+1$.
We fix a complex number $r \in {\bf C}$.
\begin{eqnarray}
\bar{K}^{(3)}(z)_j^j=
\left\{\begin{array}{cc}
1& (1\leq j=k \leq L),\\
\frac{\displaystyle 1-r/z}{\displaystyle 1-r z}&
(L+1 \leq j=k \leq L+K),\\
z^{-2}& (L+K+1 \leq j=k \leq M+N+2),
\end{array}
\right.\label{def:tildeK3}
\end{eqnarray}
{\rm Condition 3.1~:~}For $L+K\leq M+1$ we set
\begin{eqnarray}
&&
\varphi^{(3)}(z)=
\varphi^{(1)}(z)\nonumber\\
&\times&
\exp\left(\sum_{m=1}^\infty
\left\{\frac{[(-M+N+L)m]_q}{m [(N-M)m]_q} 
(rq^{-L}z)^m
+\frac{[(-M+N+L+K)m]_q}{m [(M-N)m]_q}(q^{L-K}z/r)^m \right\}
\right).
\label{def:varphi3-1}
\end{eqnarray}
{\rm Condition 3.2~:~}For $L\leq M+1 \leq L+K+1$ we set
\begin{eqnarray}
&&\varphi^{(3)}(z)=\varphi^{(1)}(z)
\label{def:varphi3-2}
\\
&\times&
\exp
\left(
\sum_{m=1}^\infty
\left\{
\frac{[(-M+N+L)m]_q}{m[(M-N)m]_q} (rq^{-L}z)^m
+
\frac{[(M+N+L+2-L-K)m]_q}{m [(M-N)m]_q}(q^{3L+K-2M-2}z/r)^m 
\right\}
\right).
\nonumber
\end{eqnarray}
{\rm Condition 3.3~:~}For $M+1\leq L-1$ we set
\begin{eqnarray}
&&
\varphi^{(3)}(z)=\varphi^{(1)}(z)
\label{def:varphi3-3}
\\
&\times&
\exp
\left(\sum_{m=1}^\infty
\left\{
\frac{[(M+N+2-L)m]_q}{m[(M-N)m]_q} (rq^{L-2M-2}z)^m
+
\frac{[(M+N+L+2-L-K)m]_q} 
{m [(M-N)m]_q}(q^{L-2M-2}z/r)^m 
\right\}
\right).
\nonumber
\end{eqnarray}
\end{dfn}
In what follows we sometimes just write $K(z)$ by dropping 
the suffix "$(i)$" from $K^{(i)}(z)$.
For classification of $K$-matrix, see appendix \ref{appendix1}
and the references \cite{VR, LYH, BR}.
Graphically, the $K$-matrix $K(z)$
is represented in Fig.2 in appendix \ref{appendix0}.
The $K$-matrix $K(z) \in {\rm End}(V)$
satisfies the graded boundary Yang-Baxter equation
\begin{eqnarray}
{K}_2(z_2)
{R}_{21}(z_1z_2)
{K}_1(z_1)
{R}_{12}(z_1/z_2)=
{R}_{21}(z_1/z_2)
{K}_1(z_1)
{R}_{12}(z_1z_2)
{K}_2(z_2).
\label{def:gBYBE}
\end{eqnarray}
The relation (\ref{def:gBYBE}), expressed in terms of matrix elements,
can be rewritten in the form 
\begin{eqnarray}
&&\sum_{j_1,j_2,k_1,k_2=1}^{M+N+2}
{K}(z_2)_{i_2}^{j_2}
{R}(z_1z_2)_{i_1,j_2}^{j_1,k_2}
{K}(z_1)_{j_1}^{k_1}
{R}(z_1/z_2)_{k_2,k_1}^{l_2,l_1}
(-1)^{([v_{j_1}]+[v_{k_1}])[v_{j_2}]}\nonumber\\
&=&
\sum_{j_1,j_2,k_1,k_2=1}^{M+N+2}
{R}(z_1/z_2)_{i_1,i_2}^{j_1,j_2}
{K}(z_1)_{j_1}^{k_1}
{R}(z_1z_2)_{j_2,k_1}^{k_2,l_1}
{K}(z_2)_{k_2}^{l_2}(-1)^{([v_{j_1}]+[v_{k_1}])[v_{k_2}]}.
\end{eqnarray}
The $K$-matrix $K(z)$ satisfies $K(1)=1$.
The $K$-matrix $K(z)$ satisfies the boundary unitary condition
\begin{eqnarray} 
K(z)K(1/z)=1.
\end{eqnarray}
We set the dual $K$-matrix $K^+(z) \in {\rm End}(V)$ by
\begin{eqnarray}
{K}^{+}(z)={K}(1/q^{N-M}z)^{st} M.
\end{eqnarray}
Graphically, the dual $K$-matrix $K^+(z)$
is represented in Fig.2 in appendix \ref{appendix0}.
The dual $K$-matrix $K^+(z)$ 
satisfies the dual graded boundary Yang-Baxter equation
\begin{eqnarray}
&&K_2^{+}(z_2)^{st_2} M_2^{-1} R_{2,1}(1/q^{2(M-N)}z_1z_2)
M_2 K_1^+(z_1)^{st_1} R_{12}(z_2/z_1)\nonumber\\
&=&
R_{21}(z_2/z_1)K_1^+(z_1)^{st_1} M_1^{-1}R_{12}(1/q^{2(M-N)}z_1z_2)
M_1 K_2^+(z_2)^{st_2}.
\label{def:dgBYBE}
\end{eqnarray}
We set the monodromy matrix ${\cal T}(z)$ by
\begin{eqnarray}
{\cal T}(z)=R_{01}(z)R_{02}(z)\cdots R_{0,P}(z) \in {\rm End}(V_P \otimes 
\cdots \otimes V_1 \otimes V_0),
\end{eqnarray}
where $V_j$ are copies of $V$.
\begin{dfn}~~~
We introduce the transfer matrix $T_B^{fin}(z)$ by
\begin{eqnarray}
T_B^{fin}(z)={\rm str}_{V_0}(K^+(z){\cal T}(z^{-1})^{-1}K(z){\cal T}(z)),
\end{eqnarray}
where the supertrace is defined by
${\rm str}(A)=\sum_j (-1)^{[v_j]}A_{j,j}$.
\end{dfn}
Graphically, the transfer matrix $T_B^{fin}(z)$ is represented in 
Fig.3 in appendix \ref{appendix0}.

\begin{prop}~~~
The transfer matrix $T_B^{fin}(z)$ form a commutative family.
\begin{eqnarray}
[T_B^{fin}(z_1),T_B^{fin}(z_2)]=0~~~{\rm for~any}~z_1,z_2.
\end{eqnarray}
\end{prop}
The commutativity can be proved by using unitarity and cross-symmetry, 
boundary Yang-Baxter equation and
dual boundary Yang-Baxter equation
\cite{Sklyanin, MN, FWW}.
We set the Hamiltonian $H_B^{fin}$ by
\begin{eqnarray}
H_B^{fin}=\frac{d}{dz}
T_B^{fin}(z)|_{z=1}=\sum_{j=1}^{P-1}h_{j,j+1}
+\frac{1}{2}\frac{d}{dz}K_0(z)|_{z=1}+
\frac{{\rm str}_{V_0}(K_0^+(1)h_{0,P})}{{\rm str}_{V_0}(K_0^+(1))},
\label{def:finiteHamiltonian}
\end{eqnarray}
where $h_{j,j+1}=P_{j,j+1}\frac{d}{dz}R_{j,j+1}(z)|_{z=1}$.

\subsection{Semi-infinite $U_q(\widehat{sl}(M+1|N+1))$ chain}

In this section we introduce
the semi-infinite $U_q(\widehat{sl}(M+1|N+1))$ chain with a boundary.
We consider the Hamiltonian (\ref{def:finiteHamiltonian})
in the semi-infinite limit.
\begin{eqnarray}
H_B^{(i)}={\lim_{P \to \infty}}
H_B^{fin}=
{\lim_{P \to \infty}}\frac{d}{dz}T_B^{fin}(z)|_{z=1}
=
\sum_{j=1}^\infty h_{j,j+1}+
\frac{1}{2}\frac{d}{dz}K_0^{(i)}(z)|_{z=1},
\end{eqnarray}
which acts formally on the left-infinite tensor product space.
\begin{eqnarray}
\cdots \otimes V \otimes V \otimes V.
\end{eqnarray}
We would like to diagonalize the Hamiltonian $H_B^{(i)}$ in the semi-infinite limit.
It is convenient to study the transfer matrix 
$\widetilde{T}_B^{(i)}(z)=\lim_{P\to \infty}T_B^{fin}(z)$,
including the spectral parameter $z$, 
instead of the Hamiltonian $H_B^{(i)}$.
The transfer matrix $\widetilde{T}_B^{(i)}(z)$ corresponding to the semi-infinite limit 
is depicted in Fig.4 in appendix \ref{appendix0}.
By convention, the lattice sites in Fig.4 are numbered $k=1,2,3,\cdots$
from right to left.
Fig.4 in appendix \ref{appendix0}
describes a semi-infinite 
two dimensional lattice, with alternating spectral parameters.
The transfer matrix 
$\widetilde{T}_B^{(i)}(z)$ is rewritten as follows.
\begin{eqnarray}
\widetilde{T}_B^{(i)}(z)
=\sum_{j=1}^{M+N+2}
\widetilde{\Phi}_j^*(z^{-1})K^{(i)}(z)_j^j \widetilde{\Phi}_j(z)(-1)^{[v_j]}.
\end{eqnarray}
Here the vertex operator $\widetilde{\Phi}_j(z)$
and the dual vertex operator $\widetilde{\Phi}_j^*(z)$
are depicted in Fig.5 and Fig.6 in appendix \ref{appendix0}, 
respectively.
The vertex operators $\widetilde{\Phi}_j(z)$ and $\widetilde{\Phi}_j^*(z)$ satisfy
the following commutation relations.
\begin{eqnarray}
\widetilde{\Phi}_{j_2}(z_2)\widetilde{\Phi}_{j_1}(z_1)&=&
\sum_{k_1,k_2=1}^{M+N+2}
R(z_1/z_2)_{j_1,j_2}^{k_1,k_2}
\widetilde{\Phi}_{k_1}(z_1)\widetilde{\Phi}_{k_2}(z_2)(-1)^{[v_{j_1}][v_{j_2}]},
\\
\widetilde{\Phi}_{j_2}^*(z_2)\widetilde{\Phi}_{j_1}^*(z_1)&=&
\sum_{k_1,k_2=1}^{M+N+2}
R(z_1/z_2)_{k_1,k_2}^{j_1,j_2}
\widetilde{\Phi}_{k_1}^*(z_1)\widetilde{\Phi}_{k_2}^*(z_2)(-1)^{[v_{j_1}][v_{j_2}]},
\\
\widetilde{\Phi}_{j_2}(z_2)\widetilde{\Phi}_{j_1}^*(z_1)&=&
\sum_{k_1,k_2=1}^{M+N+2}
R^{-1, st_1}(z_1/z_2)_{j_1,j_2}^{k_1,k_2}
\widetilde{\Phi}_{k_1}^*(z_1)\widetilde{\Phi}_{k_2}(z_2)(-1)^{[v_{k_1}][v_{k_2}]}.
\end{eqnarray}
From the graded boundary Yang-Baxter relation 
(\ref{def:gBYBE}) and the commutation relations of the vertex operators,
we have the commutativity of the transfer matrix $\widetilde{T}_B^{(i)}(z)$.
\begin{eqnarray}
~[\widetilde{T}_B^{(i)}(z_1),\widetilde{T}_B^{(i)}(z_2)]=0~~~{\rm for~any}~z_1,z_2.
\end{eqnarray}
In order to diagonalize the transfer matrix $\widetilde{T}_B^{(i)}(z)$,
we follow the strategy that we call 
the algebraic analysis method.

\section{Mathematical formulation}

In this section we give mathematical formulation of 
the supersymmetry $U_q(\widehat{sl}(M+1|N+1))$
chain with a boundary,
that is free from the difficulty of divergence
\cite{JM, Baxter, JKKKM}.

\subsection{
Quantum supersymmetry 
$U_q(\widehat{sl}(M+1|N+1))$}

In this section we review the definition
of the quantum supersymmetry
$U_q(\widehat{sl}(M+1|N+1))$ \cite{Kac, Yamane}.
The Cartan matrix of the affine
superalgebra $\widehat{sl}(M+1|N+1)$ is given by
\begin{eqnarray}
(A_{i,j})_{0\leq i,j \leq M+N+1}=\left(\begin{array}{ccccccccccc}
0&-1&0&\cdots& & & &\cdots &0&1\\
-1&2&-1&\cdots & & & &\cdots &0&0\\
0&-1&2& \cdots & & & & & &\cdots \\
 &\cdots &\cdots &\cdots &-1& & & &\\
 & & &-1&2&-1& & & & &\\
 & & & &-1&0&1& & & &\\
 & & & & & 1&-2&1& \\
 & & & & & &1&\cdots &\cdots &\cdots \\
0&0&\cdots & & & & &\cdots &-2&1\\
1&0&\cdots & & & & &\cdots &1&-2
\end{array}\right).
\label{def:Cartan}
\end{eqnarray}
Here the diagonal part is 
$(A_{i,i})_{0\leq i \leq M+N+1}=(0,\overbrace{2,\cdots,2}^{M},0,
\overbrace{-2,\cdots,-2}^{N})$.
Let us introduce orthonormal basis $\{\epsilon_j|j=1,2,\cdots,M+N+2\}$@
~with the bilinear form $(\epsilon_i|\epsilon_j)=\nu_i \delta_{i,j}$,
where the signature $\nu_i=\pm$ is given in (\ref{def:signature}).
Define $\bar{\epsilon}_i=\epsilon_i-\nu_i \frac{1}{M-N}\sum_{j=1}^{M+N+2}\epsilon_j$.
The classical simple roots $\bar{\alpha}_i$ $(i=1,2,\cdots, M+N+1)$ and 
the classical fundamental weights $\bar{\Lambda}_i$ $(i=1,2,\cdots,M+N+1)$
are defined by
\begin{eqnarray}
\bar{\alpha}_i=\nu_i \epsilon_i-\nu_{i+1} \epsilon_{i+1},~~~
\bar{\Lambda}_i=\sum_{j=1}^i \bar{\epsilon}_j~~~(i=1,2,\cdots,M+N+1).
\end{eqnarray}
Introduce the affine weight $\Lambda_0$ and the null root $\delta$
having $(\Lambda_0|\epsilon_i)=(\delta|\epsilon_i)=0$
for $i=1,2,\cdots,M+N+2$ and $(\Lambda_0|\lambda_0)=(\delta|\delta)=0$, $(\Lambda_0|\delta)=1$.
The affine roots $\alpha_i$ $(i=0,1,2,\cdots,M+N+1)$ and
the affine fundamental weights $\Lambda_i$ $(i=0,1,2,\cdots,M+N+1)$ are given by
\begin{eqnarray}
&&
\alpha_0=\delta-\sum_{j=1}^{M+N+1}\alpha_j,~~~\alpha_i=\bar{\alpha}_i~~~(i=1,2,\cdots,M+N+1),\\
&&
\Lambda_0=\Lambda_0,~~~\Lambda_i=\Lambda_0+\bar{\Lambda}_i~~~(i=1,2,\cdots,M+N+1).
\end{eqnarray}

\begin{dfn}
The quantum supersymmetry 
$U_q(\widehat{sl}(M+1|N+1))$ $(M,N=0,1,2,\cdots, {\rm and}~M \neq N)$
is a $q$-analogue of the universal enveloping algebra $\widehat{sl}(M+1|N+1)$
generated by the Chevalley generators $\{e_i,f_i,h_i|i=0,1,2,\cdots,M+N+1\}$.
The ${\bf Z}_2$-grading of the generators are 
$[e_0]=
[f_0]=[e_{M+1}]=[f_{M+1}]=1$ and zero otherwise.\\
{\rm The Cartan-Kac relations} :
For $i,j=0,1,\cdots, M+N+1$, 
the generators subject to the following relations.
\begin{eqnarray}
~[h_i,h_j]=0,
~~~[h_i,e_j]=A_{i,j}e_j,
~~~[h_i,f_j]=-A_{i,j}f_j,
~~~[e_i,f_j]=\delta_{i,j}\frac{q^{h_i}-q^{-h_i}}{q-q^{-1}}.
\end{eqnarray}
For $i,j=0,1,\cdots, M+N+1$ such that $|A_{i,j}|=0$,
the generators subject to the following relations. 
\begin{eqnarray}
~[e_i,e_j]=0,~~~[f_i,f_j]=0.
\end{eqnarray}
{\rm The Serre relations} :
For $i,j=0,1, \cdots, M+N+1$ 
such that $|A_{i,j}|=1$ and $i \neq 0, M+1$,
the generators subject to the following relations.
\begin{eqnarray}
~[e_i,[e_i,e_j]_{q^{-1}}]_q=0,
~~~[f_i,[f_i,f_j]_{q^{-1}}]_q=0.
\end{eqnarray}
For $M+N \geq 2$, 
the Serre relations of the fourth degree hold.
\begin{eqnarray}
&&[[[e_i,e_j]_{q}, e_k]_{q^{-1}}, e_j]=0,~~~
[[[f_i,f_j]_{q}, f_k]_{q^{-1}}, f_j]=0,\nonumber\\
&&(i,j,k)=(M+N-1,0,1), (M-1,M,M+1).
\end{eqnarray}
For $(M,N)=(1,0)$ the extra Serre relations of the fifth degree hold.
\begin{eqnarray}
&&[e_0,[e_2,[e_0,[e_2,e_1]_{q}]]]_{q^{-1}}=
[e_2,[e_0,[e_2,[e_0,e_1]_{q}]]]_{q^{-1}},\\
&&
[f_0,[f_2,[f_0,[f_2,f_1]_{q}]]]_{q^{-1}}=
[f_2,[f_0,[f_2,[f_0,f_1]_{q}]]]_{q^{-1}}.
\end{eqnarray}
For $(M,N)=(0,1)$ the extra Serre relations of the fifth degree hold.
\begin{eqnarray}
&&[e_0,[e_1,[e_0,[e_1,e_2]_{q}]]]_{q^{-1}}=
[e_1,[e_0,[e_1,[e_0,e_2]_{q}]]]_{q^{-1}},\\
&&
[f_0,[f_1,[f_0,[f_1,f_2]_{q}]]]_{q^{-1}}=
[f_1,[f_0,[f_1,[f_0,f_2]_{q}]]]_{q^{-1}}.
\end{eqnarray}
Here and throughout this paper, we use the notations
\begin{eqnarray}
[X,Y]_\xi=XY-(-1)^{[X][Y]}\xi YX.
\end{eqnarray}
We write $[X,Y]_1$ as $[X,Y]$ for simplicity.
The quantum supersymmetry  $U_q(\widehat{sl}(M+1|N+1))$
has the ${\bf Z}_2$-graded Hopf-algebra structure. 
We take the following coproduct.
\begin{eqnarray}
\Delta(e_i)=e_i\otimes 1+q^{h_i}\otimes e_i,~~
\Delta(f_i)=f_i\otimes q^{-h_i}+1 \otimes f_i,~~
\Delta(h_i)=h_i \otimes 1+1 \otimes h_i, 
\end{eqnarray}
and the antipode
\begin{eqnarray}
S(e_i)=-q^{-h_i}e_i,~~
S(f_i)=-f_i q^{h_i},~~
S(h_i)=-h_i.
\end{eqnarray}
The coproduct 
$\Delta$ satisfies an algebra automorphism
$\Delta(XY)=\Delta(X)\Delta(Y)$
and the antipode $S$ satisfies
a ${\bf Z}_2$-graded algebra anti-automorphism
$S(XY)=(-1)^{[X][Y]}S(Y)S(X)$.The multiplication rule 
for the tensor product is ${\bf Z}_2$-graded 
and is defined for homogeneous elements
$X,Y,X',Y' \in U_q(\widehat{sl}(N|1))$ and
$v \in V, w \in W$ by
$X \otimes Y \cdot X' \otimes Y'=(-1)^{[Y][X']}
X X' \otimes Y Y'$ and
$X \otimes Y \cdot v \otimes w=(-1)^{[Y][v]}
X v \otimes Y w$,
which extends to inhomogeneous elements through linearity.
\end{dfn}

\subsection{Mathematical formulation}

In this section we give mathematical formulation of our problem.
We introduce the evaluation representation $V_z$
of the $(M+N+2)$ dimensional basic representation 
$V=\oplus_{j=1}^{M+N+2} {\bf C}v_j$.
Let $E_{i,j}$ be the $(M+N+2) \times (M+N+2)$ matrix whose 
$(i,j)$-elements is unity and zero elsewhere :
$E_{i,j}v_k=\delta_{j,k}v_i$.
For $i=1,2,\cdots,M+N+2$, we set the evaluation representation 
$V_z$ with the vectors 
$\{v_i \otimes z^n | i=1,2,\cdots, M+N+2; n \in {\bf Z}\}$.
\begin{eqnarray}
&&e_i=E_{i,i+1},~~~f_i=\nu_i E_{i+1,i},
~~~h_i=\nu_i E_{i,i}-\nu_{i+1}E_{i+1,i+1},
\nonumber
\\
&&e_0=-zE_{M+N+2,1},~~~f_0=z^{-1}E_{1,M+N+2},~~~
h_0=-E_{1,1}-E_{M+N+2,M+N+2}.
\end{eqnarray}
Let $V_z^*$ the dual space of $V_z$
with vectors
$\{v_i^* \otimes z^n | i=1,2,\cdots, M+N+2; n \in {\bf Z}\}$
such that
$(v_i^*\otimes z^m|v_j\otimes z^n)=\delta_{i,j}\delta_{m+n,0}$.
The $U_q(\widehat{sl}(M+1|N+1))$-module structure
is given by
$(xv|w)=(v|(-1)^{|x||v|}S(x)w)$ for $v\in V_z^*, w \in V_z$
and we call the module as $V_z^{* S}$.
For $i=1,2,\cdots, M+N+2$, we have
the explicit action on $V_z^{* S}$ as follows.
\begin{eqnarray}
&&e_i=-\nu_i\nu_{i+1}q^{-\nu_i}E_{i+1,i},~~~
f_i=-\nu_i q^{\nu_i}E_{i,i+1},~~~
h_i=-\nu_iE_{i,i}+\nu_{i+1}E_{i+1,i+1},
\nonumber
\\
&&e_0=qz E_{1,M+N+2},~~~
f_0=q^{-1}z^{-1}E_{M+N+2,1},~~~
h_0=E_{1,1}+E_{M+N+2,M+N+2}.
\end{eqnarray}

\begin{dfn}~~~
Let $V(\lambda)$ be the highest weight $U_q(\widehat{sl}(M+1|N+1))$-module
with the highest weight $\lambda$.
We define the type-I vertex operators 
$\Phi(z)$ and $\Phi^*(z)$
as the intertwiners of $U_q(\widehat{sl}(M+1|N+1))$-module if they exist.
\begin{eqnarray}
&&\Phi(z) : V(\lambda)\rightarrow V(\mu)\otimes V_z,~~~
\Phi^*(z) : V(\mu)\rightarrow V(\lambda)\otimes V_z^{* S},
\\
&&\Phi(z)\cdot x=\Delta(x)\cdot \Phi(z),~~~
\Phi^*(z)\cdot x=\Delta(x)\cdot \Phi^*(z),
\end{eqnarray}
for $x \in U_q(\widehat{sl}(M+1|N+1))$.
\end{dfn}
We expand the vertex operators
$\Phi(z)=\sum_{j=1}^{M+N+2}
\Phi_j(z) \otimes v_j$ and
$\Phi^*(z)=\sum_{j=1}^{M+N+2}
\Phi_j^*(z) \otimes v_j^*$.
The vertex operators 
$\Phi_j(z)$ and 
$\Phi_j^*(z)$ satisfy
the following commutation relations.
\begin{eqnarray}
\Phi_{j_2}(z_2)
\Phi_{j_1}(z_1)&=&
\sum_{k_1,k_2=1}^{M+N+2}
R(z_1/z_2)_{j_1,j_2}^{k_1,k_2}
\Phi_{k_1}(z_1)
\Phi_{k_2}(z_2)(-1)^{[v_{j_1}][v_{j_2}]},
\\
\Phi_{j_2}^*(z_2)
\Phi_{j_1}^*(z_1)&=&
\sum_{k_1,k_2=1}^{M+N+2}
R(z_1/z_2)_{k_1,k_2}^{j_1,j_2}
\Phi_{k_1}^*(z_1)
\Phi_{k_2}^*(z_2)(-1)^{[v_{j_1}][v_{j_2}]},
\\
\Phi_{j_2}(z_2)
\Phi_{j_1}^*(z_1)&=&
\sum_{k_1,k_2=1}^{M+N+2}
R^{-1, st_1}(z_1/z_2)_{j_1,j_2}^{k_1,k_2}
\Phi_{k_1}^*(z_1)
\Phi_{k_2}(z_2)(-1)^{[v_{k_1}][v_{k_2}]}.
\end{eqnarray}
The vertex operators satisfy the inversion relations
\begin{eqnarray}
\Phi_i(z)\Phi_j^*(z)
=g^{-1}(-1)^{[v_i]}\delta_{i,j},~~~
\sum_{k=1}^{M+N+2}(-1)^{[v_k]}
\Phi_k^*(z)
\Phi_k(z)=g^{-1}.
\label{def:VO-INV}
\end{eqnarray}
Here we have used 
\begin{eqnarray}
g=e^{\frac{\pi \sqrt{-1} M}{2(M-N)}}
\exp\left(
-\sum_{m=1}^\infty
\frac{[(M-N-1)m]_q}{m[(M-N)m]_q}q^m
\right).
\end{eqnarray}

\begin{dfn}~~~We set the transfer matrix 
$T_B^{(i)}(z)$ by
\begin{eqnarray}
T_B^{(i)}(z)
=\sum_{j=1}^{M+N+2}\Phi_j^*(z^{-1})
K^{(i)}(z)_j^j \Phi_j(z)(-1)^{[v_j]}.
\end{eqnarray}
\end{dfn}
From the graded boundary Yang-Baxter relation 
(\ref{def:gBYBE}) and the commutation relations of the vertex operators,
we have the commutativity of the transfer matrix 
$T_B^{(i)}(z)$.

\begin{prop}~~~
The transfer matrix $T_B^{(i)}(z)$
forms a commutative family.
\begin{eqnarray}
~[T_B^{(i)}(z_1),T_B^{(i)}(z_2)]=0~~~{\rm for~any}~z_1,z_2.
\end{eqnarray}
\end{prop}
Following the strategy proposed in
\cite{JM, JKKKM}, we consider our problem upon the following identification.
\begin{eqnarray}
T_B^{(i)}(z)=\widetilde{T}_B^{(i)}(z),~~~
\Phi_j(z)=\widetilde{\Phi}_j(z),~~~
\Phi_j^*(z)=\widetilde{\Phi}_j^*(z).
\end{eqnarray}
The point of using the vertex operators 
$\Phi_j(z), \Phi_j^*(z)$
is that they are well-defined objects, 
free from the difficulty of divergence. 
Let us set $V(\lambda)^*$ the restricted dual module of the highest weight $V(\lambda)$.

\begin{dfn}~~~
We call the eigenvector 
${_{(i)}}\langle B| \in V^*(\lambda)$ 
with eigenvalue $1$ the boundary state.
\begin{eqnarray}
{_{(i)}}\langle B|T_B^{(i)}(z)={_{(i)}}\langle B|.
\end{eqnarray}
\end{dfn}
We would like to construct the boundary state 
${_{(i)}}\langle B| \in V^*(\lambda)$.
Multiplying the vertex operator $\Phi_j^*(z)$ from the right
and using the inversion relation (\ref{def:VO-INV}), we have
the following.

\begin{prop}~~~
The boundary state
${_{(i)}}\langle B|$ is characterized by
\begin{eqnarray}
{_{(i)}}\langle B|\Phi_j^*(z^{-1}) K^{(i)}(z)_j^j=
{_{(i)}}\langle B|\Phi_j^*(z)~~~(j=1,2,\cdots,M+N+2).
\label{eqn:main'}
\end{eqnarray}
\end{prop}
In order to construct 
the boundary state ${_{(i)}}\langle B|$,
it is convenient 
to introduce the bosonizations of the vertex operators
$\Phi_j^*(z)$.

\section{Vertex operator}
\label{section4}

In this section we review the bosonization of the vertex operators.
We give the integral representation of the vertex operators, 
which are convenient for the construction
of the boundary state ${_{(i)}}\langle B|$.

\subsection{Drinfeld realization}

In order to give the bosonizations, it is convenient to introduce the
Drinfeld realization of 
the quantum supersymmetry $U_q(\widehat{sl}(M+1|N+1))$
\cite{Drinfeld, Yamane}.

\begin{dfn}~~\cite{Yamane}~
The generators of
the quantum supersymmetry $U_q(\widehat{sl}(M+1|N+1))$,
which we call the Drinfeld generators,
are given by
\begin{eqnarray}
X_{i,m}^\pm,~h_{i,n},~h_i,~c,~~(i=1,2, \cdots, M+N+1,
m \in {\bf Z}, n \in {\bf Z}_{\neq 0}).
\end{eqnarray}
The ${\bf Z}_2$-grading of the Drinfeld generators are $|X_m^{\pm, M+1}|=1$
for $m \in {\bf Z}$ and zero otherwise.
For $i,j=1,2, \cdots, M+N+1$, the Drinfeld generators are subject to
the following relations.
\begin{eqnarray}
&&~c : {\rm central},~[h_i,h_{j,m}]=0,\\
&&~[h_{i,m},h_{j,n}]=\frac{[A_{i,j}m]_q[cm]_q}{m}
\delta_{m+n,0},\\
&&~[h_i,X_j^\pm(z)]=\pm A_{i,j}X_j^\pm(z),\\
&&~[h_{i,m}, X_j^+(z)]=\frac{[A_{i,j}m]_q}{m}
q^{-c|m|/2} z^m X_j^+(z),\\
&&~[h_{i,m}, X_j^-(z)]=-\frac{[A_{i,j}m]_q}{m}
q^{c|m|/2} z^m X_j^-(z),\\
&&(z_1-q^{\pm A_{i,j}}z_2)
X_i^\pm(z_1) X_j^\pm(z_2)
=
(q^{\pm A_{j,i}}z_1-z_2)
X_j^\pm(z_2) X_i^\pm(z_1),~~{\rm for}~|A_{i,j}|\neq 0,
\\
&&
[X_i^\pm(z_1), X_j^\pm(z_2)]
=0,~~~{\rm for}~|A_{i,j}|=0,
\\
&&~[X_i^+(z_1), X_j^-(z_2)]
=\frac{\delta_{i,j}}{(q-q^{-1})z_1z_2}
\left(
\delta(q^{-c}z_1/z_2)\psi_i^+(q^{\frac{c}{2}}z_2)-
\delta(q^{c}z_1/z_2)\psi_i^-(q^{-\frac{c}{2}}z_2)
\right),\\
&& 
\left(
X_i^\pm(z_{1})
X_i^\pm(z_{2})
X_j^\pm(z)-(q+q^{-1})
X_i^\pm(z_{1})
X_j^\pm(z)
X_i^\pm(z_{2})
+
X_j^\pm(z)
X_i^\pm(z_{1})
X_i^\pm(z_{2})\right)\nonumber\\
&&+\left(z_1 \leftrightarrow z_2\right)=0,
~~~{\rm for}~|A_{i,j}|=1,~i\neq M+1,
\\
&&\left(
X_{M+1}^\pm(z_1)X_{M+2}^\pm(w_1)X_{M+1}^\pm(z_2)X_{M}^\pm(w_2)
-q^{-1}
X_{M+1}^\pm(z_1)X_{M+2}^\pm(w_1)X_{M}^\pm(w_2)X_{M+1}^\pm(z_2)
\right.\nonumber\\
&&-q
X_{M+1}^\pm(z_1)
X_{M+1}^\pm(z_2)
X_{M}^\pm(w_2)
X_{M+2}^\pm(w_1)
+X_{M+1}^\pm(z_1)
X_{M}^\pm(w_2)
X_{M+1}^\pm(z_2)
X_{M+2}^\pm(w_1)\nonumber\\
&&+X_{M+2}^\pm(w_1)
X_{M+1}^\pm(z_2)
X_{M}^\pm(w_2)
X_{M+1}^\pm(z_1)
-q^{-1}
X_{M+2}^\pm(w_1)
X_{M}^\pm(w_2)
X_{M+1}^\pm(z_2)
X_{M+1}^\pm(z_1)
\nonumber\\
&&
\left.
-q
X_{M+1}^\pm(z_2)
X_{M}^\pm(w_2)
X_{M+2}^\pm(w_1)
X_{M+1}^\pm(z_1)
+
X_{M}^\pm(w_2)
X_{M+1}^\pm(z_2)
X_{M+2}^\pm(w_1)
X_{M+1}^\pm(z_1)
\right)\nonumber\\
&&+(z_1 \leftrightarrow z_2)=0,
\end{eqnarray}
where we have used
$\delta(z)=\sum_{m \in {\bf Z}}z^m$.
Here we have set the generating functions
\begin{eqnarray}
X_j^\pm(z)&=&
\sum_{m \in {\bf Z}}X_{j,m}^\pm z^{-m-1},\\
\psi_i^+(z)&=&q^{h_i}
\exp\left(
(q-q^{-1})\sum_{m=1}^\infty h_{i,m}z^{-m}
\right),\\
\psi_i^-(z)&=&q^{-h_i}
\exp\left(-(q-q^{-1})\sum_{m=1}^\infty
h_{i,-m}z^m\right).
\end{eqnarray}
\end{dfn}
The relation between
the Chevalley generators and 
the Drinfeld generators are obtained by the followings.
\begin{eqnarray}
e_i&=&X_{i,0}^{+},~~~f_i=X_{i,0}^{-}~~~~~(i=1,2,\cdots,M+N+1),\\
h_0&=&c-h_1-h_2-\cdots-h_{M+N+1},\\
e_0&=&
(-1)^{N+1}[X_{M+N+1,0}^{-} \cdots, [X_{M+2,0}^{-}, 
[X_{M+1,0}^{-}\cdots,
[X_{2,0}^{-},X_{1,0}^{-}]_{q^{-1}}\cdots ]_{q^{-1}}]_q \cdots]_q
\nonumber\\
&\times& q^{-h_1-h_2-\cdots-h_{M+N+1}}
,\\
f_0&=&q^{h_1+h_2+\cdots+h_{M+N+1}}\nonumber\\
&\times&
[\cdots[[\cdots[X_{1,-1}^{+},X_{2,0}^{+}]_q,\cdots,X_{M+1,0}^{+}]_q,
X_{M+2,0}^{+}]_{q^{-1}},
\cdots X_{M+N+1,0}^{+}]_{q^{-1}}.
\end{eqnarray}

\subsection{Bosonization}

In this section we review the bosonizations of 
the Drinfeld realizations of 
the quantum supersymmetry $U_q(\widehat{sl}(M+1|N+1))$ \cite{KSU}.
In what follows we assume the level $c=1$,
where we have the simplest realization.
Let us introduce the bosons
\begin{eqnarray}
a_n^i, b_n^j, c_n^j, Q_{a^i}, Q_{b^j}, Q_{c^j},~~~~(n \in {\bf Z}, i=1,2,\cdots,M+1, j=1,2,\cdots, N+1),
\end{eqnarray}
satisfying the following commutation relations.
\begin{eqnarray}
&&[a_m^i, a_n^j]=\delta_{i,j}\delta_{m+n,0}\frac{[m]_q^2}{m},
~~~[a_0^i, Q_{a^j}]=\delta_{i,j},\\
&&[b_m^i,b_n^j]=-\delta_{i,j}\delta_{m+n,0}\frac{[m]_q^2}{m},
~~~[b_0^i,Q_{b^j}]=-\delta_{i,j},\\
&&[c_m^i,c_n^j]=\delta_{i,j}\delta_{m+n,0}\frac{[m]_q^2}{m},
~~~[c_0^i,Q_{c^j}]=\delta_{i,j}.
\end{eqnarray}
Other commutation relations vanish.
We set $h_i=h_{i,0}$.
For $i=1,2,\cdots,M$ and $j=1,2,\cdots,N$, we set 
\begin{eqnarray}
Q_{h_i}=Q_{a^i}-Q_{a^{i+1}},~~~
Q_{h_{M+1}}=Q_{a^{M+1}}+Q_{b^1},~~~
Q_{h_{M+1+j}}=-Q_{b^j}+Q_{b^{j+1}}.
\end{eqnarray}
It is convenient to introduce the generating function $h^i(z;\beta)$ by
\begin{eqnarray}
h^i(z; \beta)=-\sum_{n \neq 0}\frac{h_{i,n}}{[n]_q}q^{-\beta |n|}z^{-n}
+Q_{h_i}+h_{i,0} {\rm log} z~~~~(\beta \in {\bf R}).
\end{eqnarray}
We introduce the $q$-difference operator defined by
\begin{eqnarray}
\partial_z f(z)=\frac{f(qz)-f(q^{-1}z)}{(q-q^{-1})z}.
\end{eqnarray}
In what follows we use the standard normal ordering $: :$.
For instance we set
\begin{eqnarray}
:a_m^i a_n^i:=\left\{\begin{array}{cc}
a_m^i a_n^j& (m<0)\\
a_n^i a_m^i& (m>0),
\end{array}
\right.~~~:a_0^i Q_{a^i}:=:Q_{a^i} a_0^i:=Q_{a^i} a_0^i. 
\end{eqnarray}

\begin{thm}~\cite{KSU}~~~The Drinfeld generators for the level $c=1$
are realized as follows.
\begin{eqnarray}
c&=&1,\\
h_{i,m}&=&a_m^i q^{-|m|/2}-a_{m}^{i+1} q^{|m|/2},\\
h_{M+1,m}&=&a_m^{M+1}q^{-|m|/2}+b_m^1q^{-|m|/2},\\
h_{M+1+j,m}&=&-b_m^j q^{|m|/2}+b_m^{j+1}q^{-|m|/2},\\
X_i^{+}(z)&=&:e^{h^i(z;1/2)}:e^{\pi \sqrt{-1} a_0^i},\\
X_{M+1}^{+}(z)&=&:e^{h^{M+1}(z;1/2)} e^{c^1(z;0)}:
\prod_{i=1}^M e^{-\pi \sqrt{-1} a_0^i},\\
X_{M+1+j}^{+}(z)&=&
:e^{h^{M+1+j}(z;1/2)} [\partial_z e^{-c^j(z;0)}]
e^{c^{j+1}(z;0)}:,\\
X_i^{-}(z)&=&
-:e^{-h^i(z;-1/2)}:e^{-\pi \sqrt{-1} a_0^i},\\
X_{M+1}^{-}(z)&=&:e^{-h^{M+1}(z;-1/2)}[\partial_z e^{-c^1(z;0)}]:
\prod_{i=1}^M e^{\pi \sqrt{-1} a_0^i},\\
X_{M+1+j}^{-}(z)&=&-:e^{-h^{M+1+j}(z;-1/2)}e^{c^j(z;0)}
[\partial_z e^{-c^{j+1}(z;0)}]:,
\end{eqnarray}
for $i=1,2,\cdots,M$ and $j=1,2,\cdots,N$.
\end{thm}

\subsection{Highest weight module}

In this section we study the space that the bosonizations act.
We introduce
the vacuum vector $|0 \rangle$ by
\begin{eqnarray}
a_n^i|0\rangle=b_n^j|0\rangle=c_n^j|0\rangle~~~(n \leq 0, 
i=1,2,\cdots, M+1, j=1,2,\cdots, N+1).
\end{eqnarray}
For $\lambda_a^i, \lambda_b^j, \lambda_c^j \in {\bf C}$ 
$(i=1,\cdots,M+1,j=1,\cdots,N+1)$,
we set the vector
\begin{eqnarray}
|\lambda_a^1, \cdots, \lambda_a^{M+1}, 
\lambda_b^1,\cdots,\lambda_b^{N+1},
\lambda_c^1,\cdots,\lambda_c^{N+1}\rangle=
e^{\sum_{i=1}^{M+1}\lambda_{a}^i Q_{a^i}
+\sum_{j=1}^{N+1}\lambda_b^j Q_{b^j}
+\sum_{j=1}^{N+1}\lambda_c^j Q_{c^j}}|0\rangle.
\end{eqnarray}
The Fock module ${\cal F}_{\lambda_a^1, \cdots, \lambda_a^{M+1}, 
\lambda_b^1,\cdots,\lambda_b^{N+1},
\lambda_c^1,\cdots,\lambda_c^{N+1}}$
is generated by acting creation operators
$a_{-n}^i, b_{-n}^j, c_{-n}^j$ $(n>0)$
over the vector $
|\lambda_a^1, \cdots, \lambda_a^{M+1}, 
\lambda_b^1,\cdots,\lambda_b^{N+1},
\lambda_c^1,\cdots,\lambda_c^{N+1}\rangle$.
In order to obtain the highest weight vectors 
of $U_q(\widehat{sl}(M+1|N+1))$,
we impose the conditions.
\begin{eqnarray}
&&h_i
|\lambda_a^1, \cdots, \lambda_a^{M+1}, 
\lambda_b^1,\cdots,\lambda_b^{N+1},
\lambda_c^1,\cdots,\lambda_c^{N+1}\rangle
=\lambda_i
|\lambda_a^1, \cdots, \lambda_a^{M+1}, 
\lambda_b^1,\cdots,\lambda_b^{N+1},
\lambda_c^1,\cdots,\lambda_c^{N+1}\rangle,\nonumber\\
&&e_i
|\lambda_a^1, \cdots, \lambda_a^{M+1}, 
\lambda_b^1,\cdots,\lambda_b^{N+1},
\lambda_c^1,\cdots,\lambda_c^{N+1}\rangle
=0~~~~~(i=1,2,\cdots,M+N+1).\nonumber
\end{eqnarray}
Solving these equations, we have two classes of solutions.\\
(1)~$|\Lambda_i \rangle~~~(i=1,2,\cdots,M+N+1)$.\\
For $i=1,2,\cdots,M+1$, we identify
\begin{eqnarray}
|\Lambda_i \rangle=|
\overbrace{\beta+1,\cdots,\beta+1}^{i},
\overbrace{\beta,\cdots,\beta}^{M+N+2-i},
\overbrace{0,\cdots,0}^{N+1}\rangle~~~~(\beta \in {\bf C}).
\end{eqnarray}
For $j=1,\cdots, N+1$, we identify
\begin{eqnarray}
|\Lambda_{M+1+j} \rangle=|
\overbrace{\beta+1,\cdots,\beta+1}^{M+1+j},
\overbrace{\beta,\cdots,\beta}^{N+1-j},
\overbrace{0,\cdots,0}^{j},
\overbrace{-1,\cdots,-1}^{N+1-j}\rangle~~~~(\beta \in {\bf C}).
\end{eqnarray}
(2)~$|(1-\alpha) \Lambda_0+\alpha \Lambda_{M+1}\rangle$ 
for $\alpha \in {\bf C}$.
We identify
\begin{eqnarray}
|(1-\alpha)\Lambda_0+\alpha \Lambda_{M+1}\rangle=|
\overbrace{\beta,\cdots,\beta}^{M+1},
\overbrace{\beta-\alpha,\cdots,\beta-\alpha}^{N+1},
\overbrace{-\alpha,\cdots,-\alpha}^{N+1}\rangle~~~~(\beta \in {\bf C}).
\end{eqnarray}
For $i=1,2,\cdots,M+1$, $j=1,2,\cdots,N+1$ and $\alpha,\beta \in {\bf C}$,
we set the spaces.
\begin{eqnarray}
&&{\cal F}_{(\Lambda_i,\beta)}=\bigoplus_{i_1,\cdots,i_{M+N+1}\in {\bf Z}}
{\cal F}_{
(\overbrace{\beta+1,\cdots,\beta+1}^{i},
\overbrace{\beta,\cdots,\beta}^{M+N+2-i},
\overbrace{0,\cdots,0}^{N+1})
\circ
(i_1,i_2,\cdots,i_{M+N+1})},
\\
&&{\cal F}_{(\Lambda_{M+1+j},
\beta)}=\bigoplus_{i_1,\cdots,i_{M+N+1}\in {\bf Z}}
{\cal F}_{
(\overbrace{\beta+1,\cdots,\beta+1}^{M+1+j},
\overbrace{\beta,\cdots,\beta}^{N+1-j},
\overbrace{0,\cdots,0}^{j},
\overbrace{-1,\cdots,-1}^{N+1-j})
\circ
(i_1,i_2,\cdots,i_{M+N+1})},\\
&&
{\cal F}_{(\alpha,\beta)}=
\bigoplus_{i_1,\cdots,i_{M+N+1}\in {\bf Z}}
{\cal F}_{
(
\overbrace{\beta,\cdots,\beta}^{M+1},
\overbrace{\beta-\alpha,\cdots,\beta-\alpha}^{N+1},
\overbrace{-\alpha,\cdots,-\alpha}^{N+1}
)
\circ
(i_1,i_2,\cdots,i_{M+N+1})}.
\end{eqnarray}
Here we have used the following abbreviation.
\begin{eqnarray}
&&(\lambda_a^1,\cdots,\lambda_a^{M+1},\lambda_b^1,\cdots,\lambda_b^{N+1},
\lambda_c^1,\cdots,\lambda_c^{N+1}) \circ
(i_1,i_2,\cdots,i_{M+N+1})
\nonumber\\
&=&
(\lambda_a^1,\cdots,\lambda_a^{M+1},
\lambda_b^1,\cdots,\lambda_b^{N+1},
\lambda_c^1,\cdots,\lambda_c^{N+1})\nonumber\\
&+&(i_1,i_2-i_1,\cdots,i_{M+1}-i_M,
i_{M+1}-i_{M+2},\cdots,i_{M+N}-i_{M+N+1},i_{M+N+1}\nonumber\\
&&,i_{M+1}-i_{M+2},\cdots,i_{M+N}-i_{M+N+1},i_{M+N+1}).
\end{eqnarray}
The actions of $U_q(\widehat{sl}(M+1|N+1))$ on
the spaces ${\cal F}_{(\Lambda_i,\beta)}, {\cal F}_{(\alpha,\beta)}$ are closed. 
However, these modules are not irreducible in general.
In order to obtain irreducible module,
we introduce $\xi$-$\eta$ system.
We introduce the operators $\xi_
m^j$ and $\eta_m^j$ $(j=1,2,\cdots, N+1;
m \in {\bf Z})$ by
\begin{eqnarray}
\xi^j(z)=\sum_{m \in {\bf Z}}\xi_m^j z^{-m}=:e^{-c^j(z)}:,~~~
\eta^j(z)=\sum_{m \in {\bf Z}}\eta_m^j z^{-m-1}=:e^{c^j(z)}:.
\end{eqnarray}
The Fourier components 
$\xi_m^j=\oint \frac{dz}{2\pi \sqrt{-1}}z^{m-1}\xi^j(z)$ and
$\eta_m^j=\oint \frac{dz}{2\pi \sqrt{-1}}z^{m}\eta^j(z)$ are well-defined
on the spaces ${\cal F}_{(\Lambda_i,\beta)},
{\cal F}_{(\alpha,\beta)}$ for $\alpha \in {\bf Z}$.
They satisfy the anti-commutation relations.
\begin{eqnarray}
\{\xi_m^j,\eta_n^j\}=\delta_{m+n,0},~~~
\{\xi_m^j,\xi_n^j\}=\{\eta_m^j,\eta_n^j\}=0~~~(j=1,2,\cdots,N+1).
\end{eqnarray}
Here we have used $\{a,b\}=ab+ba$.
They commute with each other.
\begin{eqnarray}
[\xi_m^j,\eta_n^{j'}]=
[\xi_m^j,\xi_n^{j'}]=[\eta_m^j,\eta_n^{j'}]=0~~~(1\leq j\neq j' \leq N+1).
\end{eqnarray}
We focus our attention on the operators $\eta_0^j, \xi_0^j$ satisfying
$(\eta_0^j)^2=0$, $(\xi_0^j)^2=0$.
They satisfy
\begin{eqnarray}
{\rm Im}(\eta_0^j)={\rm Ker}(\eta_0^j),~~~
{\rm Im}(\xi_0^j)={\rm Ker}(\xi_0^j).
\end{eqnarray}
The products $\eta_0^j \xi_0^j$ and $\xi_0^j \eta_0^j$ are 
projection operators, which satisfy
\begin{eqnarray}
\eta_0^j \xi_0^j+\xi_0^j \eta_0^j=1,
\end{eqnarray}
\begin{eqnarray}
(\eta_0^j \xi_0^j)^2=\eta_0^j \xi_0^j,~
(\xi_0^j \eta_0^j)^2=\xi_0^j \eta_0^j,~
(\eta_0^j \xi_0^j)(\xi_0^j \eta_0^j)=0,~
(\xi_0^j \eta_0^j)(\eta_0^j \xi_0^j)=0.
\end{eqnarray}
Hence we have a direct sum decomposition
for $i=1,2,\cdots,M+N+1, j=1,2,\cdots,N+1$.
\begin{eqnarray}
{\cal F}_{(\Lambda_i,\beta)}=
\eta_0^j \xi_0^j 
{\cal F}_{(\Lambda_i,\beta)}
\oplus
\xi_0^j \eta_0^j
{\cal F}_{(\Lambda_i,\beta)},~~~
{\cal F}_{(\alpha,\beta)}=
\eta_0^j \xi_0^j 
{\cal F}_{(\alpha,\beta)}
\oplus
\xi_0^j \eta_0^j
{\cal F}_{(\alpha,\beta)},
\end{eqnarray}
and
\begin{eqnarray}
{\rm Ker}(\eta_0^j)=\eta_0^j \xi_0^j {\cal F}_{(\Lambda_i,\beta)}
~~{\rm or}~~
\eta_0^j \xi_0^j {\cal F}_{(\alpha,\beta)},
~~~
{\rm Coker}(\eta_0^j)=\xi_0^j \eta_0^j {\cal F}_{(\Lambda_i,\beta)}
~~{\rm or}~~
\xi_0^j \eta_0^j {\cal F}_{(\alpha,\beta)}.
\end{eqnarray}
We set the operators $\eta_0$ and $\xi_0$ by
\begin{eqnarray}
\eta_0=\prod_{j=1}^{N+1}\eta_0^j,~~~
\xi_0=\prod_{j=1}^{N+1}\xi_0^j.
\end{eqnarray}
Following the conjectures in \cite{KSU},
we expect the following identifications.
\begin{eqnarray}
&&V(\Lambda_i)=
{\rm Coker}_{\eta_0}=
\xi_0 \eta_0 {\cal F}_{(\Lambda_i,\beta)}~~~(i=1,2,\cdots,M+N+1),\\
&&V((1-\alpha)\Lambda_0+\alpha \Lambda_{M+1})=\left\{
\begin{array}{cc}
{\rm Coker}(\eta_0)=\xi_0\eta_0 {\cal F}_{(\alpha,\beta)}
&(\alpha=0,1,2,\cdots),\\
{\rm Ker}(\eta_0)=\eta_0\xi_0 {\cal F}_{(\alpha,\beta)}
&(\alpha=-1,-2,\cdots).
\end{array}
\right.
\end{eqnarray}
Since the operators $\eta_0^j$ and $\xi_0^j$ commute with 
$U_q(\widehat{sl}(M+1|N+1))$ up to sign $\pm$,
we can regard 
${\rm Ker}(\eta_0)$ and ${\rm Coker}(\eta_0)$ as 
$U_q(\widehat{sl}(M+1|N+1))$-module.
In what follows
we will work on the space, that is expected to be the irreducible highest weight module $V(\Lambda_{M+1})$.
\begin{eqnarray}
"V(\Lambda_{M+1})=\xi_0 \eta_0
{\cal F}_{(1,\beta)}".\nonumber
\end{eqnarray}

\subsection{Vertex operator}

In this section we give the bosonization of the vertex operators
$\Phi_j^*(z)$, and give 
the integral representations of them.
We set the following combinations of the Drinfeld generators.
\begin{eqnarray}
&&
h_{i,m}^{*}=\sum_{j=1}^{M+N+1}
\frac{[\alpha_{i,j}m]_q[\beta_{i,j}m]_q}{[(M-N)m]_q[m]_q}h_{j,m},\\
&&
Q_{h_i^{*}}=
\sum_{j=1}^{M+N+1}\frac{\alpha_{i,j} \beta_{i,j}}{M-N}Q_{h_{j,0}},~~~
h_{i,0}^{*}=\sum_{j=1}^{M+N+1}
\frac{\alpha_{i,j}\beta_{i,j}}{M-N}h_{j,0}.
\end{eqnarray}
Here we have set
\begin{eqnarray}
\alpha_{i,j}&=&\left\{\begin{array}{cc}
{\rm Min}(i,j)&~({\rm Min}(i,j) \leq M+1),\\
2(M+1)-{\rm Min}(i,j)&~({\rm Min}(i,j)>M+1),
\end{array}
\right.\\
\beta_{i,j}&=&\left\{\begin{array}{cc}
M-N-{\rm Max}(i,j)&~({\rm Max}(i,j)\leq M+1),\\
-M-N-2+{\rm Max}(i,j)&~({\rm Max}(i,j)>M+1).
\end{array}\right.
\end{eqnarray}
We have the following commutation relations.
\begin{eqnarray}
&&[h_{i,m}^{*},h_{j,n}]=\delta_{i,j}\delta_{m+n,0}\frac{[m]_q^2}{m},
~~~[h_{i,m}^{*},h_{j,m}^{*}]=\delta_{m+n,0}\frac{
[\alpha_{i,j}m]_q[\beta_{i,j}m]_q
[m]_q}{m [(M-N)m]_q},\\
&&[h_{i,0}^{*},Q_{h_j}]=\delta_{i,j},
~~~[h_{i,0}^{*},Q_{h_j^{*}}]=\frac{\alpha_{i,j} \beta_{i,j}}{(M-N)}.
\end{eqnarray}

\begin{thm}~\cite{KSU}~~~The bosonic operator
$\phi^*(z)$ given below satisfies the same commutation relations as 
the vertex operator $\Phi^*(z)$.
In other words,
the bosonizations of the vertex operator $\Phi^*(z)$
on the space ${\cal F}_{(\alpha,\beta)}$, ${\cal F}_{(\Lambda_i,\beta)}$,
${\cal F}_{(\Lambda_{M+1+j},\beta)}$
are given by the followings.
\begin{eqnarray}
\phi^*(z)=\sum_{j=1}^{M+N+2}\phi^*_j(z)\otimes v_j^*.
\end{eqnarray}
Here the bosonic operators $\phi_j^*(z)$ $(j=1,2,\cdots,M+N+2)$
are defined iteratively by
\begin{eqnarray}
&&\phi^*_1(z)=:e^{h^{* 1}(qz;-1/2)}:
\prod_{k=1}^{M+1}e^{\pi \sqrt{-1} \frac{k-1}{M-N} a_0^k},
\label{boson:VO1}\\
&&\nu_j q^{\nu_j}\phi_{j+1}^*(z)=-[\phi_j^*(z),f_j]_{q^{\nu_j}}~~~
(j=1,2,\cdots,M+N+1).
\label{boson:VO2}
\end{eqnarray}
\end{thm}
The ${\bf Z}_2$-grading is given by $|\phi^*_j(z)|=\frac{\nu_j+1}{2}$
$(j=1,2,\cdots, M+N+2)$.

\begin{cor}~~~The bosonizations of the vertex operators
$\Phi^*_j(z)$ on the space
$\xi_0\eta_0 {\cal F}_{(\alpha,\beta)}$ $(\alpha=0,1,2,\cdots)$ 
are given by the following projection.
\begin{eqnarray}
\Phi_j^*(z)=\xi_0\eta_0\cdot \phi_j^*(z)\cdot \xi_0\eta_0~~~
(j=1,2,\cdots,M+N+2).
\end{eqnarray}
\end{cor}
In what follows
we call the bosonic operators
$\phi^*_j(z)$ the vertex operators.
We prepare the auxiliary operators
$X_{M+j,\epsilon}^{-}(w)$ $(\epsilon=\pm)$ by
\begin{eqnarray}
X_{M+j}^{-}(w)=\frac{1}{(q-q^{-1})w}(X_{M+j,+}^{-}(w)-X_{M+j,-}^{-}(w))~~~
(j=1,\cdots, N+1).
\end{eqnarray}
In other words, we set
\begin{eqnarray}
X_{M+1,\epsilon}^{-}(w)&=&:e^{-h^{M+1}(w)-c^1(q^\epsilon
w)}:\prod_{j=1}^M e^{\pi \sqrt{-1} a_0^j}~~~(\epsilon=\pm),
\\
X_{M+1+j,\epsilon}^{-}(w)&=&
-:e^{-h^{M+j+1}(w)+c^j(w)-c^{j+1}(q^\epsilon w)}:
~~~(\epsilon=\pm, j=2,3,\cdots,N+1).
\end{eqnarray}
Using the normal orderings in
appendix \ref{appendix2} we have the following normal orderings
for $j=1,2,\cdots,M$.
\begin{eqnarray}
&&:\phi_1^*(z)X_1^{-}(qw_1)\cdots X_j^{-}(qw_j):
X_{j+1}^{-}(qw_{j+1})=
e^{\frac{\pi \sqrt{-1}}{M}}\frac{1}{qw_j(1-qw_{j+1}/w_j)}::,\nonumber
\\
&&X_{j+1}^{-}(qw_{j+1})
:\phi_1^*(z)
X_1^{-}(qw_1)\cdots X_j^{-}(qw_j):=
-e^{\frac{\pi \sqrt{-1}}{M}}\frac{1}{qw_{j+1}(1-qw_{j}/w_{j+1})}::,
\nonumber
\\
&&:\phi_1^*(z)X_1^{-}(qw_1)\cdots X_M^{-}(qw_M):
X_{M+1}^{-}(qw_{M+1})=
-e^{\frac{\pi \sqrt{-1}}{M}}\frac{1}{qw_M(1-qw_{M+1}/w_M)}::,
\nonumber
\\
&&X_{M+1}^{-}(qw_{M+1})
:\phi_1^*(z)X_1^{-}(qw_1)\cdots X_M^{-}(qw_M):=
-e^{\frac{\pi \sqrt{-1}}{M}}\frac{1}{qw_{M+1}(1-qw_{M}/w_{M+1})}::.\nonumber
\end{eqnarray}
For $\epsilon=\pm$ and $j=1,2,\cdots,N+1$, we have
\begin{eqnarray}
&&qX_{M+j,+}^{-}(w_1)X_{M+j+1,\epsilon}^{-}(w_2)-
X_{M+j+1,\epsilon}^{-}(w_2)X_{M+j,+}^{-}(w_1)\nonumber\\
&=&\frac{(q^2-1)}{(1-qw_1/w_2)}
:X_{M+j,+}^{-}(w_1)X_{M+j+1,\epsilon}^{-}(w_2):,\nonumber
\\
&&
q X_{M+j,-}^{-}(w_1)X_{M+j+1,\epsilon}^{-}(w_2)-
X_{M+j+1,\epsilon}^{-}(w_2)X_{M+j,-}^{-}(w_1)\nonumber\\
&=&
\frac{(q^2-1)}{(1-w_1/qw_2)}
:X_{M+j,-}^{-}(w_1)X_{M+j+1,\epsilon}^{-}(w_2):.\nonumber
\end{eqnarray}
Using these normal orderings and (\ref{boson:VO2}),
we have the following integral representations.

\begin{prop}~~~The vertex operators 
$\phi_j^*(z)$ $(j=1,2, \cdots, M+N+2)$ have the following integral
representations.
\begin{eqnarray}
\phi_1^*(z)&=&:e^{h^{* 1}(qz)}:\prod_{k=1}^{M+1}e^{\pi \sqrt{-1} \frac{k-1}{M} a_0^k},
\label{boson:VO3}
\\
\phi_i^*(z)&=&e^{\frac{\pi \sqrt{-1}}{M}(i-1)}(q-q^{-1})^{i-1}
\prod_{k=1}^{i-1}
\int_{C_i} \frac{dw_k}{2\pi \sqrt{-1} w_k}
\frac{z^{-1}
w_{i-1}}{
\displaystyle
\prod_{k=0}^{i-2}(1-qw_k/w_{k+1})(1-qw_{k+1}/w_k)}\nonumber\\
&\times&
:\phi_1^*(z)X_1^{-}(qw_1)\cdots X_{i-1}^{-}(qw_{i-1}):~~~
(i=2,\cdots, M+1),
\label{boson:VO4}
\\
\phi_{M+2}^*(z)
&=&
e^{\frac{M\pi \sqrt{-1}}{M-N}}(q-q^{-1})^{M}
(qz)^{-1}
\sum_{\epsilon=\pm}
\epsilon \prod_{k=1}^{M+1}
\int_{C_{M+2}} \frac{dw_k}{2\pi \sqrt{-1} w_k}
\frac{1}{
\displaystyle
\prod_{k=0}^{M}
(1-qw_k/w_{k+1})(1-qw_{k+1}/w_k)}\nonumber\\
&\times&
:\phi_1^*(z)X_1^{-}(qw_1)\cdots X_M^{-}(qw_M)
X_{M+1,\epsilon}^-
(qw_{M+1}):,
\label{boson:VO5}
\\
\phi_{M+1+j}^*(z)
&=&e^{\frac{M \pi \sqrt{-1}}{M-N}}q^{j-1}(q-q^{-1})^M(qz)^{-1}
\sum_{\epsilon_1,\cdots,\epsilon_j=\pm}
\prod_{k=1}^j \epsilon_k \prod_{k=1}^{M+j}
\int_{C_{M+1+j}} \frac{dw_k}{2\pi \sqrt{-1}w_k}
\nonumber\\
&\times&
\frac{1}{
\displaystyle
\prod_{k=0}^M (1-qw_k/w_{k+1})(1-qw_{k+1}/w_k)
\prod_{k=1}^{j-1}(1-q^{\epsilon_k}w_{M+k}/w_{M+k+1})}\nonumber\\
&\times&
:\phi_1^*(z)X_1^{-}(qw_1)\cdots X_M^{-}(qw_M)
X_{M+1,\epsilon_1}^{-}(qw_{M+1})\cdots X_{M+j,\epsilon_j}^{-}(qw_{M+j}):,
\nonumber\\
&&~~~(j=1,2,\cdots, N+1).
\label{boson:VO6}
\end{eqnarray}
Here we have read $w_0=z$.
We take the integration contour $C_i$ $(i=1,2,\cdots,M+N+2)$
to be simple closed curve that encircles
$w_l=0, qw_{l-1}$ but not $q^{-1}w_{l-1}$ for $l=1,2,\cdots, i-1$.
\end{prop}

\section{Boundary state}
\label{section5}

In this section we give the bosonization of
the boundary state ${_{(i)}}\langle B|$.
The construction of the boundary state is the main result of this paper.
We give complete proof of this bosonization of the boundary state.

\subsection{Boundary state}

In this section we give the bosonization of
the boundary state ${_{(i)}}\langle B|$.
We use the highest weight vector $\langle \Lambda_{M+1}| \in V^*(\Lambda_{M+1})$ given by
\begin{eqnarray}
\langle \Lambda_{M+1}|=\langle 0
|e^{-\beta \sum_{i=1}^{M+1}Q_{a^i}
+(1-\beta)\sum_{j=1}^{N+1}Q_{b^j}
+\sum_{j=1}^{N+1}Q_{c^j}},
\end{eqnarray}
where $\langle 0|$ is the vacuum vector satisfying
\begin{eqnarray}
\langle 0|a_n^i=\langle 0|b_n^j=\langle 0|c_n^j=0~~~(n \geq 0, 
i=1,2,\cdots, M+1, j=1,2,\cdots, N+1).
\end{eqnarray}
We have
\begin{eqnarray}
\langle \Lambda_{M+1}|h_i=\delta_{i,M+1} \langle  \Lambda_{M+1}|,~~~
\langle \Lambda_{M+1}|c_0^j=-\langle \Lambda_{M+1}|,
\end{eqnarray}
for $i=1,2,\cdots,M+N+1$ and $j=1,2,\cdots,N+1$.
In what follows we use the auxiliary function
\begin{eqnarray}
\theta_m=\left\{\begin{array}{cc}
1& (m : {\rm even}),\\
0& (m : {\rm odd}).
\end{array}
\right.
\end{eqnarray}

~\\
\begin{dfn}~~~
We define the bosonic operators $G^{(i)}$ $(i=1,2,3)$ by
\begin{eqnarray}
G^{(i)}&=&-\frac{1}{2}\sum_{j=1}^{M+N+1}
\sum_{m=1}^\infty
\frac{m q^{-2m}}{[m]_q^2}h_m^j h_m^{* j}
-\sum_{j=1}^{N+1}
\sum_{m=1}^\infty
\frac{m q^{-2m}}{[m]_q^2}c_m^j c_m^j\nonumber\\
&+&\sum_{j=1}^{M+N+1}
\sum_{m=1}^\infty
\beta_{j,m}^{(i)}h_m^{* j}
+\sum_{j=1}^{N+1}
\sum_{m=1}^\infty
\gamma_{j,m} c_m^j.
\label{def:G}
\end{eqnarray}
Here we have set
\begin{eqnarray}
\gamma_{j,m}=-\frac{q^{-m}}{[m]_q}\theta_m~~~(j=1,2,\cdots,N+1).
\end{eqnarray}
Here 
we have set $\beta_{j,m}^{(i)}$~$(i=1,2,3)$ as followings.\\
{\rm CONDITION 1~:~}For $i=1$ we have set
\begin{eqnarray}
\beta_{j,m}^{(1)}=\left\{
\begin{array}{cc}
\frac{\displaystyle
q^{-3m/2}-q^{-m/2}}{
\displaystyle
[m]_q}\theta_m 
&(1\leq j \leq M),\\
\frac{\displaystyle
-2 q^{-3m/2}
}{\displaystyle
[m]_q}\theta_m &(j=M+1),\\
\frac{\displaystyle q^{-3m/2}+q^{-m/2}}{\displaystyle [m]_q}\theta_m
&
(M+2\leq j \leq M+N+1).
\end{array}
\right.
\end{eqnarray}
{\rm CONDITION 2~:~}For $i=2$ we have set
\begin{eqnarray}
\beta_{j,m}^{(2)}=\beta_{j,m}^{(1)}-\frac{\displaystyle
r^mq^{(\alpha_L-3/2) m}}{\displaystyle
[m]_q}\delta_{j,L}~~~~~(j=1,2,\cdots,M+N+1).
\end{eqnarray}
{\rm Condition 2.1~:~}For $L \leq M+1$ we have set
\begin{eqnarray}
\alpha_L=-L.
\end{eqnarray}
{\rm Condition 2.2~:~}For $M+2 \leq L \leq M+N+1$ we have set
\begin{eqnarray}
\alpha_L=L-2M-2.
\end{eqnarray}
{\rm CONDITION 3~:~}For $i=3$ we have set
\begin{eqnarray}
\beta_{j,m}^{(2)}=\beta_{j,m}^{(1)}-\frac{\displaystyle
r^mq^{(\alpha_L-3/2) m}}{\displaystyle
[m]_q}\delta_{j,L}-\frac{
\displaystyle q^{(\alpha_{L+K}-3/2) m}/r^m
}{
\displaystyle [m]_q
}\delta_{j,L+K}~~~~~(j=1,2,\cdots,M+N+1).
\end{eqnarray}
{\rm Condition 3.1~:~}For $L+K\leq M+1$ we have set
\begin{eqnarray}
(\alpha_L,\alpha_{L+K})=(-L,L-K).
\end{eqnarray}
{\rm Condition 3.2~:~}For $L\leq M+1 \leq L+K-1$ we have set
\begin{eqnarray}
(\alpha_L,\alpha_{L+K})=(-L,3L+K-2M-2).
\end{eqnarray}
{\rm Condition 3.3~:~}For $M+2 \leq L$ we have set
\begin{eqnarray}
(\alpha_L,\alpha_{L+K})=(L-2M-2,2M+K-L+2).
\end{eqnarray}
\end{dfn}
The following is {\bf main theorem} of this paper.

\begin{thm}~~~
The bosonization of the boundary state ${_{(i)}}\langle B|$ 
$(i=1,2,3)$ is given by
\begin{eqnarray}
{_{(i)}}\langle B|=\langle \Lambda_{M+1}|e^{G^{(i)}}.
\end{eqnarray}
Here the bosonic operator $G^{(i)}$ $(i=1,2,3)$ is given by (\ref{def:G}).
In other words the vector
${_{(i)}}\langle B|$ 
becomes the eigenvector of the transfer matrix $T_B^{(i)}(z)$
with the eigenvalue $1$.
\begin{eqnarray}
{_{(i)}}\langle B|T_B^{(i)}(z)={_{(i)}}\langle B|.
\end{eqnarray}
\end{thm}

\subsection{Excitation}

In this section we introduce other eigenvectors of $T_B^{(i)}(z)$,
that describes the excitations.

\begin{dfn}~~~
We define the type-II vertex operators 
$\Psi(z)$ and $\Psi^*(z)$
as the intertwiners of $U_q(\widehat{sl}(M+1|N+1))$-module if they exist.
\begin{eqnarray}
&&\Psi(z) : V(\lambda)\rightarrow V_z \otimes
V(\mu),~~~
\Psi^*(z) : V(\mu)\rightarrow 
V_z^{* S}\otimes V(\lambda),
\\
&&\Psi(z)\cdot x=\Delta(x)\cdot \Psi(z),~~~
\Psi^*(z)\cdot x=\Delta(x)\cdot \Psi^*(z),
\end{eqnarray}
for $x \in U_q(\widehat{sl}(M+1|N+1))$.
\end{dfn}
We expand the vertex operators
$\Psi(z)=\sum_{j=1}^{M+N+2}
v_j \otimes \Psi_j(z)$ and
$\Psi^*(z)=\sum_{j=1}^{M+N+2}
v_j^* \otimes \Psi_j^*(z)$.
The type-II vertex operator
$\Psi_\mu^*(\xi)$ and type-I vertex operators
$\Phi_j(z)$, $\Phi_j^*(z)$ satisfy the following commutation relations.
\begin{eqnarray}
\Psi_\mu^*(\xi)\Phi_j(z)&=&
\tau(\xi/z)\Phi_j(z)\Psi_\mu^*(\xi)(-1)^{[v_\mu][v_j]},
\label{com:type-IIVO}\\
\Psi_\mu^*(\xi)\Phi_j^*(z)&=&
\tau(\xi/z)\Phi_j^*(z)\Psi_\mu^*(\xi)(-1)^{[v_\mu][v_j]}.
\label{com:type-IIVO2}
\end{eqnarray}
Here we have set
\begin{eqnarray}
\tau(z)=-z^{\frac{1-M+N}{M-N}}
\exp\left(
-\sum_{m=1}^\infty
\frac{[(M-N-1)m]_q}{m[(M-N)m]_q}(z^m-z^{-m})
\right).
\end{eqnarray}

\begin{dfn}~~~We call the following vectors 
$~_{\mu_1,\mu_2,\cdots,\mu_n (i)}
\langle \xi_1,\xi_2,\cdots,\xi_n|$
the excitations.
We set
\begin{eqnarray}
&&~_{\mu_1,\mu_2,\cdots,\mu_n (i)}
\langle \xi_1,\xi_2,\cdots,\xi_n|={_{(i)}}\langle B|
\Psi_{\mu_1}^*(\xi_1)
\Psi_{\mu_2}^*(\xi_2)\cdots
\Psi_{\mu_n}^*(\xi_n),\nonumber
\label{def:excitation}
\end{eqnarray}
for $\mu_1,\mu_2,\cdots,\mu_n=1,2,\cdots,M+N+2$.
\end{dfn}

\begin{cor}~~~The excitations become 
the eigenvector of the transfer matrix $T_B^{(i)}(z)$.
\begin{eqnarray}
~_{\mu_1,\mu_2,\cdots,\mu_n (i)}
\langle \xi_1,\xi_2,\cdots,\xi_n|T_B^{(i)}(z)
=_{\mu_1,\mu_2,\cdots,\mu_n (i)}
\langle \xi_1,\xi_2,\cdots,\xi_n|\prod_{\mu=1}^n
\tau(\xi_\mu z)\tau(\xi_\mu/z).
\end{eqnarray}
\end{cor}
We expect that the excitations
(\ref{def:excitation}) are the basis of the space of the physical state of
the supersymmetry $U_q(\widehat{sl}(M+1|N+1))$ chain with a boundary.

\section{Proof of main theorem}
\label{section6}

In this section we give complete proof of the main theorem. 
We would like to show
\begin{eqnarray}
{_{(i)}}\langle B|\phi_j^*(z^{-1})K^{(i)}(z)_j^j ={_{(i)}}\langle B|
\phi_j^*(z^{-1})~~~(j=1,2,\cdots,M+N+2).
\label{boundary:case123}
\end{eqnarray}
It is convenient to use the following abbreviations.
\begin{eqnarray}
&&
h_+^i(z)=-\sum_{m=1}^\infty
\frac{h_{i,m}}{[m]_q}q^{\frac{m}{2}}z^{-m},~~~
h_-^i(z)=\sum_{m=1}^\infty
\frac{h_{i,-m}}{[m]_q}q^{\frac{m}{2}}z^m
~~~(i=1,2,\cdots,M+N+1),\\
&&
h_+^{* 1}(z)=-\sum_{m=1}^\infty
\frac{h_{1,m}^{*}}{[m]_q}q^{\frac{m}{2}}z^{-m},~~~
h_-^{* 1}(z)=\sum_{m=1}^\infty
\frac{h_{1,-m}^{*}}{[m]_q}q^{\frac{m}{2}}z^m,\\
&&c_+^j(z)=
-\sum_{m=1}^\infty
\frac{c_m^j}{[m]_q}z^{-m},~~~
c_-^j(z)=\sum_{m=1}^\infty
\frac{c_{-m}^j}{[m]_q}z^m~~~(j=1,2,\cdots,N+1).
\end{eqnarray}
We set the function $D(z,w)$ by
\begin{eqnarray}
D(z,w)=(1-qzw)(1-qz/w)(1-qw/z)(1-q/wz).
\end{eqnarray}
The function $D(z,w)$ is invariant under 
$(z,w)\to (1/z,w), (z,1/w), (1/z,1/w)$.

\begin{prop}~~~
The operator $G^{(i)}$ $(i=1,2,3)$ given in (\ref{def:G}) satisfies
\begin{eqnarray}
&&e^{G^{(i)}}h_{j,-m} e^{-G^{(i)}}
=h_{j,-m}-q^{-2m}h_{j,m}+\frac{[m]_q^2}{m}\beta_{j,m}^{(i)}
~~~(m>0, j=1,2,\cdots, M+N+1),\\
&&e^{G^{(i)}}c_{-m}^j e^{-G^{(i)}}
=c_{-m}^j-q^{-2m}c_m^j+\frac{[m]_q^2}{m}\gamma_{j,m}
~~~(m>0, j=1,2,\cdots, N+1).
\end{eqnarray}
\end{prop}

\begin{prop}~~~For the boundary conditions $i=1,2,3$,
we have
\begin{eqnarray}
&&{_{(i)}}\langle B|h_j=\delta_{j,M+1}{_{(i)}}\langle B|~~~(j=1,2,\cdots,M+N+1),\\
&&{_{(i)}}\langle B|h_{1,0}^*=-\frac{N+1}{M-N}{_{(i)}}\langle B|,\\
&&{_{(i)}}\langle B|c_0^j=-{_{(i)}}\langle B|~~~(j=1,2,\cdots,N+1).
\end{eqnarray}
\end{prop}
We show the relation (\ref{boundary:case123}) for each boundary condition
${_{(i)}}\langle B|$ $(i=1,2,3)$, case by case.

\subsection{Boundary condition 1}

In this section we show (\ref{boundary:case123}) 
for the boundary condition ${_{(1)}}\langle B|$.
Very explicitly we would like to show
\begin{eqnarray}
z^{\frac{M}{M-N}} \varphi^{(1)}(z^{-1}){_{(1)}}\langle B|\phi_j^*(z)=
z^{-\frac{M}{M-N}} \varphi^{(1)}(z){_{(1)}}\langle B|\phi_j^*(z^{-1})~~~~~(1\leq j \leq M+N+2).
\label{boundary:case1}
\end{eqnarray}
We would like to comment that
RHS (resp.LHS) is obtained from LHS (resp.RHS) 
under $z \to 1/z$.
The proof for $1\leq j \leq M+1$ is similar as those of non-super $\widehat{sl}(N)$ case.
The proof for $M+2 \leq j \leq M+N+2$ is different from those of non-super case.
We prepare propositions.
\begin{prop}~~~
The actions of 
$h_+^i(w)$, $h_-^{* 1}(w)$ and $c_-^j(w)$ 
on the boundary state ${_{(1)}}\langle B|$
are given as followings.
\begin{eqnarray}
&&{_{(1)}}\langle B|e^{-h_-^i(qw)}=g_i^{(1)}(w){_{(1)}}\langle B|e^{-h_+^i(q/w)}~~~
(i=1,2,\cdots,M+N+1),
\label{action:G1-1}\\
&&{_{(1)}}\langle B|e^{h^{* 1}_-(qw)}=\varphi^{(1)}(w){_{(1)}}\langle B|e^{h_+^{* 1}(q/w)},
\label{action:G1-2}\\
&&{_{(1)}}\langle B|e^{-c_-^j(qw)}=c_j^{(1)}(w){_{(1)}}\langle B|e^{-c_+^j(q/w)}
~~~(j=1,2,\cdots,N+1).
\label{action:G1-3}
\end{eqnarray}
Here we have set
\begin{eqnarray}
g_i^{(1)}(w)&=&\left\{\begin{array}{cc}
(1-w^2)& (1\leq i \leq M),\\
(1+w^2)& (i=M+1),\\
1& (M+2\leq i \leq M+N+1).
\end{array}
\right.
\label{def:g}
\end{eqnarray}
The function $\varphi^{(1)}(w)$ is given in (\ref{def:varphi1})
and $c_j^{(1)}(w)=1$ for $j=1,2,\cdots,N+1$.
\end{prop}

\begin{prop}~~~The following relation holds.
\begin{eqnarray}
\sum_{\epsilon=\pm} \int_C \frac{d w_1}{w_1}\frac{\epsilon q^{\epsilon} 
(1-q w_1 w_2)e^{-c_+^j(q^{1+\epsilon}w_1)-c_+^j(q^{1-\epsilon}/w_1)}}{
(1-q^{\epsilon} w_1 w_2)(1-q^{\epsilon}w_1/w_2)}
=\int_{C}\frac{d w_1}{w_1}
\frac{(-1+w_1^2)e^{-c_+^j(w_1)-c_+^j(q^2/w_1)}}{
(1-w_1 w_2/q)(1-w_1/qw_2)}.
\label{recursion:integral1}
\end{eqnarray}
Here the integration contour $C$ encircles
$w_1=0, q w_2^{\pm 1}$ not but $w_1=q^{-1}w_2^{\pm 1}$.
This integral is invariant
under $w_2 \to 1/w_2$.
\end{prop}

~\\
{\bf Proof for boundary condition 1.}
~~~We show main theorem for the boundary condition ${}_{(1)}\langle B|$.
We show the relation (\ref{boundary:case1}).
\\
$\bullet$~The case for $j=1$ : ${_{(1)}}\langle B|\phi_1^*(z)$.\\
Using the bosonization (\ref{boson:VO3}) and the relation 
(\ref{action:G1-2}),
we get LHS of (\ref{boundary:case1}) as following :
\begin{eqnarray}
z^{\frac{M}{M-N}}\varphi^{(1)}(1/z){_{(1)}}\langle B|\phi^*_1(z)
=
q^{-\frac{M}{M-N}}
\varphi^{(1)}(z)\varphi^{(1)}(1/z){_{(1)}}\langle B|e^{h_+^{*1}(qz)+h_+^{*1}(q/z)}
e^{Q_{h_1^*}}.
\end{eqnarray}
This is invariant under $z \to 1/z$. Hence LHS and RHS 
of (\ref{boundary:case1}) coincide.\\
$\bullet$~The case for $j=2$ : ${_{(1)}}\langle B|\phi_2^*(z)$.
\\
Using the bosonization (\ref{boson:VO4}),
the relations (\ref{action:G1-1}), (\ref{action:G1-2}),
and the normal orderings in appendix \ref{appendix2},
we get LHS of (\ref{boundary:case1}) as following :
\begin{eqnarray}
&&z^{\frac{M}{M-N}}\varphi^{(1)}(1/z){_{(1)}}\langle B|\phi^*_2(z)
\nonumber\\
&=&q^{-\frac{M}{M-N}-1}(q-q^{-1})
e^{\frac{\pi \sqrt{-1}}{M-N}}
\varphi^{(1)}(z)\varphi^{(1)}(1/z)\nonumber\\
&\times&
\int_{C_1}\frac{dw}{2\pi \sqrt{-1}w}
\frac{(1-w^2)(1-q/zw)}{D(z,w)}{_{(1)}}\langle B|
e^{Q_{h_1^*-h_1}}e^{h_+^{* 1}(q z)+h_+^{* 1}(q/z)-h_+^1(q w)+h_+^1(q/w)}.
\end{eqnarray}
We note that the integrand 
$\frac{\varphi^{(1)}(z)\varphi^{(1)}(1/z)}{D(z,w)}
e^{h_+^{* 1}(qz)+h_+^{* 1}(q/z)}$
is invariant $z \to 1/z$.
We have LHS$-$RHS of (\ref{boundary:case1}) as following :
\begin{eqnarray}
&&q^{-\frac{M}{M-N}}(q-q^{-1})
e^{\frac{\pi \sqrt{-1}}{M-N}}\varphi^{(1)}(z)\varphi^{(1)}(1/z)(z-z^{-1})\nonumber\\
&\times &
\int_{\widetilde{C}_1}\frac{dw}{2\pi \sqrt{-1}w}\frac{(w^{-1}-w)}{D(z,w)}
{_{(1)}}\langle B|e^{Q_{h_1^*-h_1}}e^{h_+^{*1}(q z)+h_+^{* 1}(q/z)
-h_+^1(qw)-h_+^1(q/w)}.
\end{eqnarray}
Here the integration contour 
$\widetilde{C}_1$ encircles $w=0,qz^{\pm 1}$ but not
$w=q^{-1}z^{\pm 1}$.
The integration contour $\widetilde{C}_1$ is invariant
under $w \to 1/w$.
The integrand 
$\frac{(w^{-1}-w)}{D(z,w)}e^{-h_+^1(qw)-h_+^1(q/w)}$
creates just signature $(-1)$ under $w \to 1/w$.
Hence we have LHS$-$RHS$=0$. \\
$\bullet$~The case for $3 \leq j \leq M+1$ : ${_{(1)}}\langle B|\phi_j^*(z)$.\\
Using the bosonization (\ref{boson:VO4}), the relations (\ref{action:G1-1}), (\ref{action:G1-2}) and
normal orderings in appendix \ref{appendix2},
we have LHS$-$RHS of (\ref{boundary:case1})
as following.
\begin{eqnarray}
&&z^{\frac{M}{M-N}}\varphi^{(1)}(1/z)
{_{(1)}}\langle B|
\phi^*_j(z)-z^{-\frac{M}{M-N}}\varphi^{(1)}(z)
{_{(1)}}\langle B|\phi_j^*(1/z)
\nonumber\\
&=&
q^{-\frac{M}{M-N}}(q-q^{-1})^{j-1}
\varphi^{(1)}(z)\varphi^{(1)}(1/z)(z-z^{-1})
e^{\frac{\pi \sqrt{-1} (j-1)}{M-N}}\nonumber\\
&\times&
\prod_{k=1}^{j-1}
\int_{\widetilde{C}_j}\frac{dw_k}{2\pi \sqrt{-1}w_k}
\frac{
\displaystyle \prod_{k=1}^{j-2}(1-q/w_k w_{k+1})}{
\displaystyle
\prod_{k=0}^{j-2}D(w_k,w_{k+1})}
(w_1^{-1}-w_1)
\prod_{k=2}^{j-1}(1-w_k^2)\nonumber\\
&\times&
{_{(1)}}\langle B|
e^{Q_{h_1^*-h_1 \cdots-h_{j-1}}}
e^{h_+^{*1}(qz)+h_+^{*1}(q/z)-\sum_{k=1}^{j-1}(
h_+^k(qw_k)+h_+^k(q/w_k))}.
\end{eqnarray}
Here the integration contour $\widetilde{C}_j$ encircles
$w_k=0, q w_{k-1}^{\pm 1}$ 
not but $w_k=q^{-1} w_{k-1}^{\pm 1}$ for $1\leq k \leq j-1$.
Let's study the changing of the variable $w_1 \to 1/w_1$.
We note that the integrand
$\frac{1}{D(z,w_1)}
e^{-h_+^1(qw_1)-h_+^1(q/w_1)}$ and 
the integration contour $\widetilde{C}_{j}$ 
are invariant under $w_1 \to 1/w_1$.
Taking into account of symmetrization
$\oint_{|w_1|=1}\frac{dw_1}{w_1}f(w_1)=\frac{1}{2}
\oint_{|w_1|=1}\frac{dw_1}{w_1}(f(w_1)+f(1/w_1))$
and relation
$(1-q/w_1 w_2)-(w_1 \leftrightarrow
 1/w_1)=(w_1^{-1}-w_1)(-q/w_2)$,
we symmetrize the variables $w_1,w_2,\cdots,w_{j-2}$, iteratively,
we have
\begin{eqnarray}
&&
(-q/2)^{j-2}q^{-\frac{M}{M-N}}(q-q^{-1})^{j-1}
\varphi^{(1)}(z)\varphi^{(1)}(1/z)(z-z^{-1})
e^{\frac{\pi \sqrt{-1} (j-1)}{M-N}}
\nonumber
\\
&\times&
\prod_{k=1}^{j-1}
\int_{\widetilde{C}_j}\frac{dw_k}{2\pi \sqrt{-1}w_k}
\frac{\displaystyle
\prod_{k=1}^{j-2}
(w_k^{-1}-w_k)^2 (w_{j-1}^{-1}-w_{j-1})}{
\displaystyle \prod_{k=0}^{j-2}D(w_k,w_{k+1})}\nonumber\\
&\times&
{_{(i)}}\langle B|
e^{Q_{h_1^*-h_1 \cdots-h_{j-1}}}
e^{h_+^{*1}(qz)+h_+^{*1}(q/z)-\sum_{k=1}^{j-1}(
h_+^k(qw_k)+h_+^k(q/w_k))}=0.
\end{eqnarray}
Here we have used 
$\oint_{|w_{j-1}|=1}
\frac{dw_{j-1}}{w_{j-1}}
(w_{j-1}^{-1}-w_{j-1})e^{-h_+^{j-1}(qw_{j-1})-h_+^{j-1}(q/w_{j-1})}=0.$
\\
$\bullet$~The case for $j=M+2$ : ${_{(1)}}\langle B|\phi_{M+2}^*(z)$.\\
Using the bosonization 
(\ref{boson:VO5}) , the relations (\ref{action:G1-1}), 
(\ref{action:G1-2}), (\ref{action:G1-3}), and
the normal orderings in appendix \ref{appendix2},
we have LHS$-$RHS of (\ref{boundary:case1})
as following.
\begin{eqnarray}
&&z^{\frac{M}{M-N}}\varphi^{(1)}(1/z)
{_{(1)}}\langle B|
\phi^*_{M+2}(z)-
z^{-\frac{M}{M-N}}\varphi^{(1)}(z)
{_{(1)}}\langle B|\phi_{M+2}^*(1/z)
\nonumber\\
&=&
q^{-\frac{M}{M-N}+1}(q-q^{-1})^M
\varphi^{(1)}(z)\varphi^{(1)}(1/z)(z-z^{-1})
e^{\frac{\pi \sqrt{-1}}{M-N}M}
\nonumber\\
&\times&
\sum_{\epsilon=\pm}
\epsilon 
\prod_{k=1}^{M+1} \int_{\widetilde{C}_{M+2}}
\frac{dw_k}{2\sqrt{-1} w_k}
\frac{(w_1^{-1}-w_1)(1+w_{M+1}^2)}
{\displaystyle \prod_{k=0}^M D(w_k,w_{k+1})}
\prod_{k=1}^M (1-q/w_k w_{k+1})
\prod_{k=2}^M (1-w_k^2)
\\
&\times&
{_{(1)}}\langle B|e^{Q_{h_1^*-h_1\cdots-h_{M+1}-c_1}}
e^{h_+^{* 1}(qz)+h_+^{* 1}(q/z)-\sum_{k=1}^{M+1}(h_+^k(qw_k)+h_+^k(q/w_k))
-c_+^1(q^{1+\epsilon}w_{M+1})-c_+^1(q^{1-\epsilon}/w_{M+1})}.\nonumber
\end{eqnarray}
Here the integration contour $\widetilde{C}_{M+2}$ encircles
$w_k=0, q w_{k-1}^{\pm 1}$ 
not but $w_k=q^{-1} w_{k-1}^{\pm 1}$ for $1 \leq k \leq M+1$.
Taking into account of
symmetrization $\oint_{|w|=1}\frac{dw}{w}f(w)=\frac{1}{2}\oint_{|w|=1}(f(w)+f(1/w))$
and relation
$(1-q/w_1w_2)-(w_1 \leftrightarrow w_1^{-1})=(-q/w_2)(w_1^{-1}-w_1)$,
we symmetrize the variables $w_1,w_2,\cdots,w_{M}$ iteratively.
Then we have
\begin{eqnarray}
&&(-q/2)^M
q^{-\frac{M}{M-N}+1}(q-q^{-1})^M
\varphi^{(1)}(z)\varphi^{(1)}(1/z)(z-z^{-1})
e^{\frac{\pi \sqrt{-1}}{M-N}M}
\nonumber\\
&\times&
\prod_{k=1}^{M+1} \int_{\widetilde{C}_{M+2}}
\frac{dw_k}{2\sqrt{-1} w_k}
\frac{\displaystyle
\prod_{k=1}^M(w_k^{-1}-w_k)^2
(w_{M+1}^{-1}+w_{M+1})}
{\displaystyle
\prod_{k=0}^M D(w_k,w_{k+1})}\nonumber\\
&\times&
{_{(1)}}\langle B|e^{Q_{h_1^*-h_1\cdots-h_{M+1}-c_1}}
e^{h_+^{* 1}(qz)+h_+^{* 1}(q/z)-\sum_{k=1}^{M+1}(h_+^k(qw_k)+h_+^k(q/w_k))}
\nonumber\\
&\times&\left(e^{-c_+^1(q^2w_{M+1})-c_+^1(1/w_{M+1})}-
e^{-c_+^1(w_{M+1})-c_+^1(q^2/w_{M+1})}\right)=0.
\end{eqnarray}
Here we have used
\begin{eqnarray}
\oint_{|w_{M+1}|=1} \frac{dw_{M+1}}{w_{M+1}}\frac{(w_{M+1}+w_{M+1}^{-1})}{
D(w_M,w_{M+1})}
\left(e^{-c_+^1(q^2w_{M+1})-c_+^1(1/w_{M+1})}-
e^{-c_+^1(w_{M+1})-c_+^1(q^2/w_{M+1})}\right)=0.\nonumber
\end{eqnarray}
$\bullet$~The case for $2 \leq j \leq N+1$ : ${_{(1)}}\langle B|\phi_{M+1+j}^*(z)$.\\
Using the bosonization 
(\ref{boson:VO6}), the relations (\ref{action:G1-1}), (\ref{action:G1-2}), (\ref{action:G1-3}),
and the normal orderings in appendix \ref{appendix2},
 we have LHS$-$RHS of 
(\ref{boundary:case1}) as following.
\begin{eqnarray}
&&z^{\frac{M}{M-N}}
\varphi^{(1)}(1/z){_{(1)}}\langle B|
\phi^*_{M+j+1}(z)-
z^{-\frac{M}{M-N}}\varphi^{(1)}(z)
{_{(1)}}\langle B|\phi_{M+1+j}^*(1/z)
\nonumber\\
&=&
q^{-\frac{M}{M-N}+1}(q-q^{-1})^M
\varphi^{(1)}(z)\varphi^{(1)}(1/z)(z-z^{-1})
e^{\frac{\pi \sqrt{-1} M}{M-N}}\nonumber\\
&\times&
\sum_{\epsilon_1,\cdots,\epsilon_j=\pm}
\epsilon_j
\prod_{k=1}^{M+j}\int_{\widetilde{C}_{M+1+j}}\frac{dw_k}{2\pi \sqrt{-1}w_k}
\frac{(w_1^{-1}-w_1)(1+w_{M+1}^2)}{
\displaystyle
\prod_{k=0}^M
D(w_k,w_{k+1})}\prod_{k=1}^M(1-q/w_kw_{k+1})\prod_{k=2}^M(1-w_k^2)\nonumber\\
&\times&
\prod_{k=1}^{j-1}
\frac{\epsilon_k q^{\epsilon_k}(1-qw_{M+k}w_{M+k+1})}{
(1-q^{\epsilon_k}w_{M+k}w_{M+k+1})(1-q^{\epsilon_k}w_{M+k}/w_{M+k+1})}
{_{(1)}}\langle B|
e^{Q_{h_1^*-h_1\cdots -h_{M+j}-c_j}}
e^{h_+^{*1}(qz)+h_+^{*1}(q/z)}
\nonumber
\\
&\times& e^{-\sum_{k=1}^{M+j}
(h_+^k(qw_k)+h_+^k(q/w_k))
+\sum_{k=1}^{j-1}
(c_+^k(q w_{M+k+1})+c_+^k(q/w_{M+k+1}))
-\sum_{k=1}^j
(c_+^k(q^{1+\epsilon_k}w_{M+k})+c_+^k(q^{1-\epsilon_k}/w_{M+k}))}.
\nonumber\\
\end{eqnarray}
Here the integration contour $\widetilde{C}_{M+1+j}$ encircles $w_k=0,qw_{k-1}^{\pm 1}$
not but $w_{k}=q^{-1}w_{k-1}^{\pm 1}$ for $1\leq k \leq M+j$.
Using relation
$(1-q/w_1w_2)-(w_1 \leftrightarrow w_1^{-1})=
(-q/w_2)(w_1^{-1}-w_1)$ 
we symmetrize the variables $w_1,w_2,\cdots,w_{M}$ iteratively,
we have
\begin{eqnarray}
&&
q^{-\frac{M}{M-N}}(-q/2)^M(q-q^{-1})^M
\varphi^{(1)}(z)\varphi^{(1)}(1/z)(z-z^{-1})
e^{\frac{\pi \sqrt{-1} M}{M-N}}\nonumber\\
&\times&
\sum_{\epsilon_1,\cdots,\epsilon_j=\pm}
\epsilon_j
\prod_{k=1}^{M+j}\int_{\widetilde{C}_{M+1+j}}\frac{dw_k}{2\pi \sqrt{-1}w_k}
\frac{\displaystyle
\prod_{k=1}^M
(w_k^{-1}-w_k)^2
(w_{M+1}^{-1}+w_{M+1})}
{\displaystyle
\prod_{k=0}^M
D(w_k,w_{k+1})}
\nonumber\\
&\times&
\prod_{k=1}^{j-1}
\frac{\epsilon_k q^{\epsilon_k}(1-qw_{M+k}w_{M+k+1})}{
(1-q^{\epsilon_k}w_{M+k}w_{M+k+1})(1-q^{\epsilon_k}w_{M+k}/w_{M+k+1})}
{_{(1)}}\langle B|
e^{Q_{h_1^*-h_1\cdots -h_{M+j}-c_j}}
e^{h_+^{*1}(qz)+h_+^{*1}(q/z)}
\\
&\times& e^{-\sum_{k=1}^{M+j}
(h_+^k(qw_k)+h_+^k(q/w_k))
+\sum_{k=1}^{j-1}
(c_+^k(q w_{M+k+1})+c_+^k(q/w_{M+k+1}))
-\sum_{k=1}^j
(c_+^k(q^{1+\epsilon_k}w_{M+k})+c_+^k(q^{1-\epsilon_k}/w_{M+k}))}.
\nonumber
\end{eqnarray}
Using the relation (\ref{recursion:integral1}) for the variables 
$w_{M+1},\cdots,w_{M+j-1}$ iteratively,
we have
\begin{eqnarray}
&&
q^{-\frac{M}{M-N}}(-q/2)^M(q-q^{-1})^M
\varphi^{(1)}(z)\varphi^{(1)}(1/z)(z-z^{-1})
(-1)^{j-1}e^{\frac{\pi \sqrt{-1} M}{M-N}}\nonumber\\
&\times&
\prod_{k=1}^{M+j}\int_{\widetilde{C}_{M+1+j}}\frac{dw_k}{2\pi \sqrt{-1}w_k}
\frac{\displaystyle
\prod_{k=1}^M
(w_k^{-1}-w_k)^2
(w_{M+1}^{-1}+w_{M+1})}
{\displaystyle
\prod_{k=0}^M D(w_k,w_{k+1})}\nonumber\\
&\times&
{_{(1)}}\langle B|
e^{Q_{h_1^*-h_1\cdots -h_{M+j}-c_j}}
\prod_{k=1}^{j-1}
\frac{(-1+w_{M+k}^2)e^{-c_+^k(w_{M+k})-c_+^k(q^2/w_{M+k})}}{
(1-w_{M+k}w_{M+k+1}/q)(1-w_{M+k}/qw_{M+k+1})}\nonumber\\
&\times&
e^{h_+^{*1}(qz)+h_+^{*1}(q/z)-\sum_{k=1}^{M+j}
(h_+^k(qw_k)+h_+^k(q/w_k))+
\sum_{k=1}^{j-1}(c_+^k(q w_{M+k+1})+c_+^k(q/w_{M+k+1}))}\nonumber\\
&\times&
\left(e^{-c_+^j(q^{2}w_{M+j})-c_+^j(1/w_{M+j})}-
e^{-c_+^j(q^{2}/w_{M+j})-c_+^j(w_{M+j})}\right)=0.
\end{eqnarray}
Here we have used
\begin{eqnarray}
\oint_{|w|=1} \frac{dw}{w}
e^{c_+^{j-1}(qw)+c_+^{j-1}(q/w)}
\left(e^{-c_+^j(q^2w)-c_+^j(1/w)}-
e^{-c_+^j(w)-c_+^j(q^2/w)}\right)f(w)=0,
\nonumber
\end{eqnarray}
where $f(w)=f(1/w)$.\\
Now we have shown the relation (\ref{boundary:case1})
for every $j=1,2,\cdots,M+N+2$.\\
{\bf Q.E.D.}

\subsection{Boundary condition 2}

In this section we study (\ref{boundary:case123}) for 
the boundary condition ${_{(2)}}\langle B|$.
Very explicitly we study
\begin{eqnarray}
z^{\frac{M}{M-N}}\varphi^{(2)}(z^{-1}){_{(2)}}\langle B|\phi_j^*(z)&=&
z^{-\frac{M}{M-N}} 
\varphi^{(2)}(z){_{(2)}}\langle B|\phi_j^*(z^{-1})~~~(1\leq j \leq L),
\label{boundary:case2-1}\\
z^{\frac{M}{M-N}}(1-rz)\varphi^{(2)}(z^{-1}){_{(2)}}\langle B|\phi_j^*(z)&=&
z^{-\frac{M}{M-N}} 
(1-r/z){_{(2)}}\langle B|\phi_j^*(z^{-1})~~(L+1 \leq j \leq M+N+2).\nonumber\\
\label{boundary:case2-2}
\end{eqnarray}
The structure of (\ref{boundary:case2-1}) is the same as those of
(\ref{boundary:case1}) for
the boundary condition ${_{(1)}}\langle B|$.
In this section we focus our attention on
the relation (\ref{boundary:case2-2}) that is
new for the boundary condition ${_{(2)}}\langle B|$.
We give proofs for following two conditions.
\begin{eqnarray}
{\rm Condition~2.1} &:& L \leq M+1,\nonumber\\
{\rm Condition~2.2} &:& M+2 \leq L \leq M+N+1.\nonumber
\end{eqnarray}

\begin{prop}~~~
The actions of $h_+^i(w)$, $h_-^{* 1}(w)$ and $c_-^j(w)$ on the boundary state
${_{(2)}}\langle B|$ are given as followings.
\begin{eqnarray}
&&{_{(2)}}\langle B|e^{-h_-^i(qw)}=g_i^{(2)}(w){_{(2)}}\langle B|e^{-h_+^i(q/w)}~~~
(i=1,2,\cdots,M+N+1),\label{action:G2-1}\\
&&{_{(2)}}\langle B|e^{h^{* 1}_-(qw)}=\varphi^{(2)}(w){_{(2)}}\langle B|e^{h_+^{* 1}(q/w)},
\label{action:G2-2}\\
&&{_{(2)}}\langle B|e^{-c_-^j(qw)}=c_j^{(2)}(w){_{(2)}}\langle B|e^{-c_+^j(q/w)}
~~~(j=1,2,\cdots,N+1).
\label{action:G2-3}
\end{eqnarray}
Here $\varphi^{(2)}(w)$ are given in
(\ref{def:varphi2-1}) and (\ref{def:varphi2-2}). 
We have set $c_j^{(2)}(w)=1$ $(j=1,2,\cdots,N+1)$.
We have set $g_i^{(2)}(w)$ $(i=1,2,\cdots,M+N+1)$ by
\begin{eqnarray}
g_i^{(2)}(w)&=&\left\{\begin{array}{cc}
g_i^{(1)}(w)& (1\leq i\neq L \leq M+N+1),\\
\frac{\displaystyle 1}{\displaystyle
(1-rq^{\alpha_L}w)}g_L^{(1)}(w)& (i=L),
\end{array}
\right.
\end{eqnarray}
where $g_i^{(1)}(w)$ is given by (\ref{def:g}).
The parameter $\alpha_L$ is given by followings.
\\
{\rm Condition 2.1~:~}For $1\leq L \leq M+1$ we have set
\begin{eqnarray}
\alpha_L=-L.
\end{eqnarray}
{\rm Condition 2.2~:~}For $M+2 \leq L \leq M+N+2$ we have set
\begin{eqnarray}
\alpha_L=-2M-2+L.
\end{eqnarray}
\end{prop}

\begin{prop}~~~The following relation holds.
\begin{eqnarray}
&&\sum_{\epsilon=\pm} \int_C 
\frac{d w_1}{w_1}\frac{\epsilon q^{\epsilon} 
(1-q^{\alpha} r w_1)(1-q w_1 w_2)e^{-c_+^j(q^{1+\epsilon}w_1)-c_+^j(q^{1-\epsilon}/w_1)}}{
(1-q^{\epsilon} w_1 w_2)(1-q^{\epsilon}w_1/w_2)}\nonumber\\
&=&
q(1-q^{\alpha+1}rw_2)\int_{C}\frac{d w_1}{w_1}
\frac{(-1+w_1^2)e^{-c_+^j(w_1)-c_+^j(q^2/w_1)}}{
(1-w_1 w_2/q)(1-w_1/qw_2)},
\label{recursion:integral2}
\end{eqnarray}
where the integration contour $C$ encircles
$w_1=0, q w_2^{\pm 1}$ not but $w_1=q^{-1}w_2^{\pm 1}$.
\\
The integral
$\int_{C}\frac{d w_1}{w_1}
\frac{(-1+w_1^2)e^{-c_+^j(w_1)-c_+^j(q^2/w_1)}}{
(1-w_1 w_2/q)(1-w_1/qw_2)}$ in RHS is invariant under
$w_2 \to w_2^{-1}$.
\end{prop}

~\\
{\bf Proof for boundary condition 2.1}~~~
We show the main theorem for 
the boundary condition ${_{(2)}}\langle B|$ 
and $1\leq L\leq M+1$.
Here we show the relation (\ref{boundary:case2-2}).
\\
$\bullet$~The case for $L+1\leq j \leq M+1$ : 
${_{(2)}}\langle B|\phi_{j}^*(z)$.\\
Using the bosonization (\ref{boson:VO4}), the relations
(\ref{action:G2-1}), (\ref{action:G2-2}), and the normal ordering
in appendix \ref{appendix2}, we have
LHS$-$RHS of (\ref{boundary:case2-2}) as following :
\begin{eqnarray}
&&z^{\frac{M}{M-N}}(1-rz)\varphi^{(2)}(1/z)
{_{(2)}}\langle B|
\phi^*_j(z)-z^{-\frac{M}{M-N}}(1-r/z)\varphi^{(2)}(z)
{_{(2)}}\langle B|\phi_j^*(1/z)
\nonumber\\
&=&
q^{-\frac{M}{M-N}}(q-q^{-1})^{j-1}
\varphi^{(2)}(z)\varphi^{(2)}(1/z)(z-z^{-1})
e^{\frac{\pi \sqrt{-1} (j-1)}{M-N}}\nonumber\\
&\times&
\prod_{k=1}^{j-1}
\int_{\widetilde{C}_j}\frac{dw_k}{2\pi \sqrt{-1}w_k}
\frac{\displaystyle
\prod_{k=1}^{j-2}(1-q/w_kw_{k+1})}{
\displaystyle
\prod_{k=0}^{j-2}D(w_k,w_{k+1})}
\frac{(w_1^{-1}-w_1)(1-rq^{-1}w_1)}{(1-rq^{-L}w_L)}
\prod_{k=2}^{j-1}(1-w_k^2)\nonumber\\
&\times&
{_{(2)}}\langle B|
e^{Q_{h_1^*-h_1 \cdots-h_{j-1}}}
e^{h_+^{*1}(qz)+h_+^{*1}(q/z)-\sum_{k=1}^{j-1}(
h_+^k(qw_k)+h_+^k(q/w_k))}.
\end{eqnarray}
Here the integration contour $\widetilde{C}_j$ encircles
$w_k=0, q w_{k-1}^{\pm 1}$ 
not but $w_k=q^{-1} w_{k-1}^{\pm 1}$ for 
$1\leq k \leq j-1$.
Taking into account of symmetrization
$\oint_{|w|=1}\frac{dw}{w}f(w)=\frac{1}{2}\oint_{|w|=1}\frac{dw}{w}
(f(w)+f(1/w))$ and relation
$(1-q/w_1w_2)(1-rq^{-\alpha}w_1)+(w_1 \leftrightarrow w_1^{-1})
=(w_1^{-1}-w_1)(-q/w_2)(1-rq^{-\alpha-1}w_2)$, 
we symmetrize the variables $w_1,w_2,\cdots,w_{L-1}$, iteratively.
Using relation
$(1-q/w_Lw_{L+1})-(w_L\leftrightarrow w_L^{-1})=(-q/w_{L+1})(w_L^{-1}-w_L)$,
we symmetrize the variables $w_L,w_{L+1},\cdots,w_{j-1}$, iteratively,
we have
\begin{eqnarray}
&&
q^{-\frac{M}{M-N}}(q-q^{-1})^{j-1}(-q/2)^{j-1}
\varphi^{(2)}(z)\varphi^{(2)}(1/z)(z-z^{-1})
e^{\frac{\pi \sqrt{-1} (j-1)}{M-N}}\nonumber\\
&\times&
\prod_{k=1}^{j-1}
\int_{\widetilde{C}_j}\frac{dw_k}{2\pi \sqrt{-1}w_k}
\frac{\displaystyle \prod_{k=1}^{j-2}(w_k^{-1}-w_k)^2}{
\displaystyle
\prod_{k=0}^{j-2}D(w_k,w_{k+1})}
(w_{j-1}^{-1}-w_{j-1})
\nonumber\\
&\times&
{_{(2)}}\langle B|
e^{Q_{h_1^*-h_1 \cdots-h_{j-1}}}
e^{h_+^{*1}(qz)+h_+^{*1}(q/z)-\sum_{k=1}^{j-1}(
h_+^k(qw_k)+h_+^k(q/w_k))}=0.
\end{eqnarray}
Here we have used $\oint_{|w|=1}\frac{dw}{w}(w^{-1}-w)f(w)=0$,
where $f(w)=f(1/w)$.
\\
$\bullet$~The case for $j=L+1=M+2$ : ${_{(2)}}\langle B|\phi_{M+2}^*(z)$.\\ 
Using the bosonization (\ref{boson:VO5}), relations (\ref{action:G2-1}), (\ref{action:G2-2}), 
(\ref{action:G2-3}), and
the normal orderings in appendix \ref{appendix2}, we have
LHS$-$RHS of (\ref{boundary:case2-2}) as following :
\begin{eqnarray}
&&z^{\frac{M}{M-N}}\varphi^{(2)}(1/z)(1-r z){_{(2)}}\langle B|
\phi_{M+2}^*(z)-
z^{-\frac{M}{M-N}}
\varphi^{(2)}(z)(1-r/z){_{(2)}}\langle B|\phi_{M+2}^*(1/z)
\nonumber\\
&=&
q^{-\frac{M}{M-N}+1}(q-q^{-1})^M \varphi^{(2)}(z)\varphi^{(2)}(1/z)(z-z^{-1})
e^{\frac{\pi \sqrt{-1}}{M-N}M}
\nonumber\\
&\times&
\sum_{\epsilon=\pm}
\epsilon \prod_{k=1}^{M+1}
\int_{\widetilde{C}_{M+2}}\frac{dw_k}{2\pi\sqrt{-1}w_k}
\frac{\displaystyle
\prod_{k=1}^M(1-q/w_kw_{k+1})
}{\displaystyle \prod_{k=0}^M D(w_k,w_{k+1})}
\frac{(w_1^{-1}-w_1)(1-rq^{-1}w_1)(1+w_{M+1}^2)}{(1-rq^{-L}w_{L})}\nonumber
\\
&\times&
\prod_{k=2}^M(1-w_k^2)
\times
{_{(2)}}\langle B|e^{Q_{h_1^*-h_1\cdots -h_L-c_1}}\nonumber\\
&\times&
e^{h_+^{*1}(qz)+h_+^{*1}(q/z)-\sum_{k=1}^L(h_+^k(qw_k)+h_+^k(q/w_k))
-c_+^1(q^{1+\epsilon}w_{M+1})-c_+^1(q^{1-\epsilon}/w_{M+1})}.
\end{eqnarray}
Here the integration contour 
$\widetilde{C}_{M+2}$ encircles $w_k=0,qw_{k-1}^{\pm 1}$ not but
$w_k=q^{-1}w_{k-1}^{\pm 1}$ for $1\leq k \leq M+1$.
Using relation $(1-rq^\alpha w_1)(1-q/w_1w_2)-
(w_1\leftrightarrow w_1^{-1})=
(w_1-w_1^{-1})(1-rq^{\alpha-1}w_2)$,
we symmetrize the variables $w_1,w_2,\cdots,w_L=w_{M+1}$, iteratively, 
we have
\begin{eqnarray}
&&
q^{-\frac{M}{M-N}+1}(q-q^{-1})^M(-q/2)^M \varphi^{(2)}(z)\varphi^{(2)}(1/z)
(z-z^{-1})
e^{\frac{\pi \sqrt{-1}}{M-N}M}
\nonumber\\
&\times&
\int_{\widetilde{C}_{M+2}}\frac{dw_k}{2\pi\sqrt{-1}w_k}
\frac{\displaystyle \prod_{k=1}^M(w_k^{-1}-w_k)^2}
{\displaystyle \prod_{k=0}^M D(w_k, w_{k+1})}
(w_{M+1}^{-1}+w_{M+1})
\times
{_{(2)}}\langle B|e^{Q_{h_1^*-h_1\cdots -h_L-c_1}}\\
&\times&
e^{h_+^{*1}(qz)+h_+^{*1}(q/z)-\sum_{k=1}^L(h_+^k(qw_k)+h_+^k(q/w_k))}
\left(
e^{-c_+^1(q^{2}w_{M+1})-c_+^1(1/w_{M+1})}-
e^{-c_+^1(w_{M+1})-c_+^1(q^{2}/w_{M+1})}
\right)=0.\nonumber
\end{eqnarray}
Here we have used $\oint_{|w|=1}\frac{dw}{w}
(e^{-c_+^1(q^2w)-c_+^1(1/w)}-e^{-c_+^1(w)-c_+^1(q^2/w)})f(w)=0$,
where $f(w)=f(1/w)$.
\\
$\bullet$~The case for $M+3 \leq j \leq M+N+2$ :
${_{(2)}}\langle B|\phi_j^*(z)$.\\
Using the bosonization 
(\ref{boson:VO6}), the relations 
(\ref{action:G2-1}), (\ref{action:G2-2}), (\ref{action:G2-3}), and the normal orderings in appendix \ref{appendix2},
we have LHS$-$RHS of 
(\ref{boundary:case1}) for $1\leq i \leq N+1$.
\begin{eqnarray}
&&z^{\frac{M}{M-N}}(1-rz)
\varphi^{(2)}(1/z){_{(2)}}\langle B|
\phi^*_{M+1+i}(z)-
z^{-\frac{M}{M-N}}(1-r/z)\varphi^{(2)}(z)
{_{(2)}}\langle B|\phi_{M+1+i}^*(1/z)
\nonumber\\
&=&
q^{-\frac{M}{M-N}+1}(q-q^{-1})^M
\varphi^{(2)}(z)\varphi^{(2)}(1/z)(z-z^{-1})
e^{\frac{\pi \sqrt{-1} M}{M-N}}\nonumber\\
&\times&
\sum_{\epsilon_1,\cdots,\epsilon_i=\pm}
\epsilon_i
\prod_{k=1}^{M+i}
\int_{\widetilde{C}_{M+1+i}}\frac{dw_k}{2\pi \sqrt{-1}w_k}
\frac{\displaystyle 
\prod_{k=1}^M(1-q/w_kw_{k+1})}{
\displaystyle
\prod_{k=0}^M
D(w_k,w_{k+1})}
\frac{(w_1^{-1}-w_1)(1-rq^{-1}w_1)(1+w_{M+1}^2)}
{(1-rq^{-L}w_L)}
\nonumber\\
&\times&
\prod_{k=2}^M(1-w_k^2)
\prod_{k=1}^{i-1}
\frac{\epsilon_k q^{\epsilon_k}(1-qw_{M+k}w_{M+k+1})}{
(1-q^{\epsilon_k}w_{M+k}w_{M+k+1})(1-q^{\epsilon_k}w_{M+k}/w_{M+k+1})}
\nonumber\\
&\times&
{_{(2)}}\langle B|
e^{Q_{h_1^*-h_1\cdots -h_{M+i}-c_i}}
e^{h_+^{*1}(qz)+h_+^{*1}(q/z)
-\sum_{k=1}^{M+i}
(h_+^k(qw_k)+h_+^k(q/w_k))}\nonumber\\
&\times&
e^{\sum_{k=1}^{i-1}
(c_+^k(q w_{M+k+1})+c_+^k(q/w_{M+k+1}))
-\sum_{k=1}^i
(c_+^k(q^{1+\epsilon_k}w_{M+k})+c_+^k(q^{1-\epsilon_k}/w_{M+k}))}.
\end{eqnarray}
Here the integration contour $\widetilde{C}_{M+1+i}$
encircles $w_k=0,qw_{k-1}^{\pm 1}$
not but $w_{k}=q^{-1}w_{k-1}^{\pm 1}$ for $1\leq k \leq M+i$.
Using relation
$(1-rq^{\alpha}w_1)(1-q/w_1w_2)-(w_1 \leftrightarrow w_1^{-1})
=(1-rq^{\alpha-1}w_2)(w_1-w_1^{-1})$,
we symmetrize the variables 
$w_1,w_2,\cdots,w_{L-1}$, iteratively.
Using relation
$(1-q/w_{L}w_{L+1})-(w_L \leftrightarrow w_L^{-1})=(-q/w_{L+1})(w_L^{-1}-w_L)$
we symmetrize the variables $w_L,\cdots, w_M$, iteratively.
Then we have
\begin{eqnarray}
&&
q^{-\frac{M}{M-N}+1}(q-q^{-1})^M(-q/2)^M
\varphi^{(2)}(z)\varphi^{(2)}(1/z)(z-z^{-1})
e^{\frac{\pi \sqrt{-1} M}{M-N}}\nonumber\\
&\times&
\sum_{\epsilon_1,\cdots,\epsilon_i=\pm}
\epsilon_i
\prod_{k=1}^{M+i}
\int_{\widetilde{C}_{M+1+i}}\frac{dw_k}{2\pi \sqrt{-1}w_k}
\frac{\displaystyle \prod_{k=1}^M(w_k^{-1}-w_k)^2}{
\displaystyle
\prod_{k=0}^MD(w_k,w_{k+1})}
(w_{M+1}^{-1}+w_{M+1})\\
&\times&
\prod_{k=1}^{i-1}
\frac{\epsilon_k q^{\epsilon_k}(1-qw_{M+k}w_{M+k+1})}{
(1-q^{\epsilon_k}w_{M+k}w_{M+k+1})(1-q^{\epsilon_k}w_{M+k}/w_{M+k+1})}
{_{(2)}}\langle B|
e^{Q_{h_1^*-h_1\cdots -h_{M+i}-c_i}}
e^{h_+^{*1}(qz)+h_+^{*1}(q/z)}
\nonumber
\\
&\times& e^{-\sum_{k=1}^{M+i}
(h_+^k(qw_k)+h_+^k(q/w_k))
+\sum_{k=1}^{i-1}
(c_+^k(q w_{M+k+1})+c_+^k(q/w_{M+k+1}))
-\sum_{k=1}^i
(c_+^k(q^{1+\epsilon_k}w_{M+k})+c_+^k(q^{1-\epsilon_k}/w_{M+k}))}.
\nonumber
\end{eqnarray}
Using relation (\ref{recursion:integral1}) for variables 
$w_{M+1},\cdots,w_{i-1}$, iteratively,
we have
\begin{eqnarray}
&&
q^{-\frac{M}{M-N}+1}(q-q^{-1})^M(-q/2)^M
\varphi^{(2)}(z)\varphi^{(2)}(1/z)(z-z^{-1})
e^{\frac{\pi \sqrt{-1} M}{M-N}}\nonumber\\
&\times&
\prod_{k=1}^{M+i}
\int_{\widetilde{C}_{M+1+i}}\frac{dw_k}{2\pi \sqrt{-1}w_k}
\frac{\displaystyle \prod_{k=1}^M(w_k^{-1}-w_k)^2}{
\displaystyle
\prod_{k=0}^M D(w_k,w_{k+1})}
(w_{M+1}^{-1}+w_{M+1})
\times
{_{(2)}}\langle B|
e^{Q_{h_1^*-h_1\cdots -h_{M+i}-c_i}}
\nonumber
\\
&\times&
\prod_{k=1}^{i-1}
\frac{(-1+w_{M+k}^2)e^{-c_+^k(w_{M+k})-c_+^k(q^2/w_{M+k})}}{
(1-w_{M+k}w_{M+k+1}/q)(1-w_{M+k}/qw_{M+k+1})}\nonumber\\
&\times&
e^{h_+^{*1}(qz)+h_+^{*1}(q/z)
-\sum_{k=1}^{M+i}
(h_+^k(qw_k)+h_+^k(q/w_k))
+\sum_{k=1}^{i-1}
(c_+^k(q w_{M+k+1})+c_+^k(q/w_{M+k+1}))}\nonumber\\
&\times&\left(
e^{-c_+^i(q^{2}w_{M+i})-c_+^i(1/w_{M+i})}
-e^{-c_+^i(w_{M+i})-c_+^i(q^2/w_{M+i})}\right)=0.
\end{eqnarray}
Here we have used
\begin{eqnarray}
\oint_{|w|=1} \frac{dw}{w}
e^{c_+^{i-1}(qw)+c_+^{i-1}(q/w)}
\left(e^{-c_+^i(q^2w)-c_+^i(1/w)}-
e^{-c_+^i(w)-c_+^i(q^2/w)}\right)f(w)=0,
\nonumber
\end{eqnarray}
where $f(w)=f(1/w)$.\\
Now we have shown (\ref{boundary:case2-2})
for every $j=L+1,\cdots,M+N+2$.
\\
{\bf Q.E.D.}

~\\
{\bf Proof for boundary condition 2.2}~~~
We show the main theorem 
for the boundary condition ${_{(2)}}\langle B|$ 
and $M+2 \leq L \leq M+N+1$.
Here we show (\ref{boundary:case2-2}) that is
new for the boundary condition ${_{(2)}}\langle B|$.
We use the integral relations (\ref{recursion:integral1}) 
and (\ref{recursion:integral2}).
\\
$\bullet$~The case for $L+1\leq j \leq M+N+2$ : 
${_{(2)}}\langle B|\phi_{j}^*(z)$.\\
Using the bosonization 
(\ref{boson:VO6}), the relations (\ref{action:G2-1}), (\ref{action:G2-2}),
(\ref{action:G2-3}), and the normal orderings in appendix
{\ref{appendix2}}, we have LHS$-$RHS of 
(\ref{boundary:case1}) for $1\leq i \leq N+1$.
\begin{eqnarray}
&&z^{\frac{M}{M-N}}(1-rz)
\varphi^{(2)}(1/z){_{(2)}}\langle B|
\phi^*_{M+1+i}(z)-
z^{-\frac{M}{M-N}}(1-r/z)\varphi^{(2)}(z)
{_{(2)}}\langle B|\phi_{M+1+i}^*(1/z)
\nonumber\\
&=&
q^{-\frac{M}{M-N}+1}(q-q^{-1})^M
\varphi^{(2)}(z)\varphi^{(2)}(1/z)(z-z^{-1})
e^{\frac{\pi \sqrt{-1} M}{M-N}}\nonumber\\
&\times&
\sum_{\epsilon_1,\cdots,\epsilon_i=\pm}
\epsilon_i
\prod_{k=1}^{M+i}
\int_{\widetilde{C}_{M+1+i}}\frac{dw_k}{2\pi \sqrt{-1}w_k}
\frac{\displaystyle
\prod_{k=1}^M(1-q/w_kw_{k+1})
}{
\displaystyle
\prod_{k=0}^M
D(w_k,w_{k+1})}
\frac{(w_1^{-1}-w_1)(1-rq^{-1}w_1)(1+w_{M+1}^2)}{(1-rq^{-2M-2+L}w_L)}
\nonumber\\
&\times&
\prod_{k=2}^M(1-w_k^2)
\prod_{k=1}^{i-1}
\frac{\epsilon_k q^{\epsilon_k}(1-qw_{M+k}w_{M+k+1})}{
(1-q^{\epsilon_k}w_{M+k}w_{M+k+1})(1-q^{\epsilon_k}w_{M+k}/w_{M+k+1})}
\nonumber\\
&\times&
{_{(2)}}\langle B|
e^{Q_{h_1^*-h_1\cdots -h_{M+i}-c_i}}
e^{h_+^{*1}(qz)+h_+^{*1}(q/z)-\sum_{k=1}^{M+i}
(h_+^k(qw_k)+h_+^k(q/w_k))}\nonumber\\
&\times&
e^{\sum_{k=1}^{i-1}
(c_+^k(q w_{M+k+1})+c_+^k(q/w_{M+k+1}))
-\sum_{k=1}^i
(c_+^k(q^{1+\epsilon_k}w_{M+k})+c_+^k(q^{1-\epsilon_k}/w_{M+k}))}.
\end{eqnarray}
Here the integration contour $\widetilde{C}_{M+1+i}$
encircles $w_k=0,qw_{k-1}^{\pm 1}$
not but $w_{k}=q^{-1}w_{k-1}^{\pm 1}$ for $1\leq k \leq M+i$.
Using relation $(1-q/w_1w_2)(1-rq^{-\alpha}w_1)-
(w_1 \leftrightarrow 
w_1^{-1})
=(w_1-w_1^{-1})(1-rq^{-\alpha-1}w_2)$,
we symmetrize the variables $w_1,w_2,\cdots,w_{M}$ iteratively.
We have
\begin{eqnarray}
&&
q^{-\frac{M}{M-N}+1}(q-q^{-1})^M(-q/2)^M
\varphi^{(2)}(z)\varphi^{(2)}(1/z)(z-z^{-1})
e^{\frac{\pi \sqrt{-1} M}{M-N}}\nonumber\\
&\times&
\sum_{\epsilon_1,\cdots,\epsilon_i=\pm}
\epsilon_i
\prod_{k=1}^{M+i}
\int_{\widetilde{C}_{M+1+i}}\frac{dw_k}{2\pi \sqrt{-1}w_k}
\frac{\displaystyle
\prod_{k=1}^M(w_k^{-1}-w_k)^2}{
\displaystyle
\prod_{k=0}^M
D(w_k,w_{k+1})}
\frac{(w_{M+1}^{-1}+w_{M+1})(1-rq^{-M-1}w_{M+1})}{(1-rq^{-2M-2+L}w_L)}
\nonumber\\
&\times&
\prod_{k=1}^{i-1}
\frac{\epsilon_k q^{\epsilon_k}(1-qw_{M+k}w_{M+k+1})}{
(1-q^{\epsilon_k}w_{M+k}w_{M+k+1})(1-q^{\epsilon_k}w_{M+k}/w_{M+k+1})}
\nonumber\\
&\times&
{_{(2)}}\langle B|
e^{Q_{h_1^*-h_1\cdots -h_{M+i}-c_i}}
e^{h_+^{*1}(qz)+h_+^{*1}(q/z)-\sum_{k=1}^{M+i}
(h_+^k(qw_k)+h_+^k(q/w_k))}\nonumber\\
&\times&
e^{\sum_{k=1}^{i-1}
(c_+^k(q w_{M+k+1})+c_+^k(q/w_{M+k+1}))
-\sum_{k=1}^i
(c_+^k(q^{1+\epsilon_k}w_{M+k})+c_+^k(q^{1-\epsilon_k}/w_{M+k}))}.
\end{eqnarray}
We use the relation (\ref{recursion:integral2}) for the variables
$w_{M+1},\cdots,w_{L-1}$ iteratively.
We use the relation (\ref{recursion:integral1}) 
for the variables $w_{L},\cdots,w_{M+i-1}$ iteratively.
Then we have
\begin{eqnarray}
&&
q^{-\frac{M}{M-N}+1}(q-q^{-1})^M(-q/2)^M
q^{L-M-1}
\varphi^{(2)}(z)\varphi^{(2)}(1/z)(z-z^{-1})
e^{\frac{\pi \sqrt{-1} M}{M-N}}\nonumber\\
&\times&
\prod_{k=1}^{M+i}
\int_{\widetilde{C}_{M+1+i}}\frac{dw_k}{2\pi \sqrt{-1}w_k}
\frac{\displaystyle
\prod_{k=1}^M(w_k^{-1}-w_k)^2}{
\displaystyle
\prod_{k=0}^M
D(w_k,w_{k+1})}
(w_{M+1}^{-1}+w_{M+1})
\nonumber\\
&\times&
{_{(2)}}\langle B|
e^{Q_{h_1^*-h_1\cdots -h_{M+i}-c_i}}
\prod_{k=1}^{i-1}
\frac{(-1+w_{M+k}^2)e^{-c_+^k(w_{M+k})-c_+^k(q^2/w_{M+k})}}{
(2- w_{M+k}/qw_{M+k+1})(1-w_{M+k}w_{M+k+1}/q)}\nonumber\\
&\times&
e^{h_+^{*1}(qz)+h_+^{*1}(q/z)-\sum_{k=1}^{M+i}
(h_+^k(qw_k)+h_+^k(q/w_k))+
\sum_{k=1}^{i-1}
(c_+^k(q w_{M+k+1})+c_+^k(q/w_{M+k+1}))}\nonumber\\
&\times&\left(
e^{-c_+^i(q^{2}w_{M+i})-c_+^i(1/w_{M+i})}
-e^{-c_+^i(w_{M+i})-c_+^i(q^2/w_{M+i})}\right)=0.
\end{eqnarray}
Here we have used
\begin{eqnarray}
\oint_{|w|=1} \frac{dw}{w}
e^{c_+^{i-1}(qw)+c_+^{i-1}(q/w)}
\left(e^{-c_+^i(q^2w)-c_+^i(1/w)}-
e^{-c_+^i(w)-c_+^i(q^2/w)}\right)f(w)=0,
\nonumber
\end{eqnarray}
where $f(w)=f(1/w)$.\\
{\bf Q.E.D.}\\
Now we have shown (\ref{boundary:case2-2}) for every $j=L+1,\cdots,M+N+2$.

\subsection{Boundary condition 3}

In this section we study (\ref{boundary:case123}) for 
the boundary condition ${_{(3)}}\langle B|$.
Very explicitly we study
\begin{eqnarray}
z^{\frac{M}{M-N}}\varphi^{(3)}(z^{-1}){_{(3)}}\langle B|\phi_j^*(z)&=&
z^{-\frac{M}{M-N}} 
\varphi^{(3)}(z){_{(3)}}\langle B|\phi_j^*(z^{-1})~~~
(1\leq j \leq L),
\label{boundary:case3-1}\\
z^{\frac{M}{M-N}}
(1-rz)\varphi^{(3)}(z^{-1}){_{(3)}}\langle B|\phi_j^*(z)&=&
z^{-\frac{M}{M-N}} 
(1-r/z)\varphi^{(3)}(z){_{(3)}}\langle B|
\phi_j^*(z^{-1})~~(L+1 \leq j \leq L+K),
\nonumber\\
\label{boundary:case3-2}\\
z^{\frac{M}{M-N}+1}\varphi^{(3)}(z^{-1}){_{(3)}}\langle B|\phi_j^*(z)&=&
z^{-\frac{M}{M-N}-1} \varphi^{(3)}(z){_{(3)}}\langle B|\phi_j^*(z^{-1})~~~
(L+K+1 \leq j \leq M+N+2).\nonumber\\
\label{boundary:case3-3}
\end{eqnarray}
The structures of (\ref{boundary:case3-1}) and
(\ref{boundary:case3-2}) are the same as those of
(\ref{boundary:case2-1}) and (\ref{boundary:case2-2})
for the boundary condition ${_{(2)}}\langle B|$.
In this section we focus our attention on
the relation (\ref{boundary:case3-3}) that is
new for the boundary condition ${_{(3)}}\langle B|$.
We give proofs for following three conditions.
\begin{eqnarray}
{\rm Condition~3.1} &:& L+K \leq M+1,\nonumber\\
{\rm Condition~3.2} &:& L \leq M+1 \leq L+K-1,\nonumber\\
{\rm Condition~3.3} &:& M+2 \leq L.\nonumber
\end{eqnarray}

\begin{prop}~~~
The actions of $h_+^i(w)$, $h_-^{* 1}(w)$ and $c_-^j(w)$ on the boundary state
${_{(3)}}\langle B|$ are given as followings.
\begin{eqnarray}
&&{_{(3)}}\langle B|e^{-h_-^i(qw)}
=g_i^{(3)}(w){_{(3)}}\langle B|e^{-h_+^i(q/w)}~~~
(i=1,2,\cdots,M+N+1),\label{action:G3-1}\\
&&{_{(3)}}\langle B|e^{h^{* 1}_-(qw)}=\varphi^{(3)}(w)
{_{(3)}}\langle B|e^{h_+^{* 1}(q/w)},\label{action:G3-2}\\
&&{_{(3)}}\langle B|e^{-c_-^j(qw)}=c_j^{(3)}(w)
{_{(3)}}\langle B|e^{-c_+^j(q/w)}
~~~(j=1,2,\cdots,N+1).\label{action:G3-3}
\end{eqnarray}
Here $\varphi^{(3)}(w)$ are given in 
(\ref{def:varphi3-1}), (\ref{def:varphi3-2}) and (\ref{def:varphi3-3}). 
We have set $c_j^{(3)}(w)=1$ $(j=1,2,\cdots,N+1)$.
We have set $g_i^{(3)}(w)$ $(i=1,2,\cdots,M+N+1)$ by
\begin{eqnarray}
g_i^{(3)}(w)&=&\left\{\begin{array}{cc}
g_i^{(1)}(w)& (1\leq i\neq L, L+K \leq M+N+1),\\
\frac{\displaystyle 1}{\displaystyle
(1-rq^{\alpha_L}w)}g_L^{(1)}(w)& (i=L),\\
\frac{\displaystyle
1}{\displaystyle
(1-q^{\alpha_{L+K}}w/r)}g_{L+K}^{(1)}(w)& (i=L+K),
\end{array}
\right.
\end{eqnarray}
where $g_i^{(1)}(w)$ is given by (\ref{def:g}).
The parameters $(\alpha_L,\alpha_{L+K})$ are given by followings.
\\
{\rm Condition 3.1~:~}For $L+K\leq M+1$ we have set
\begin{eqnarray}
(\alpha_L,\alpha_{L+K})=(-L, L-K).
\end{eqnarray}
{\rm Condition 3.2~:~}For $L\leq M+1 \leq L+K-1$ we have set
\begin{eqnarray}
(\alpha_L,\alpha_{L+K})=
(-L, 3L+K-2M-2),
\end{eqnarray}
{\rm Condition 3.3~:~}For $M+2 \leq L$ we have set
\begin{eqnarray}
(\alpha_L, \alpha_{L+K})=
(L-2M-2, 2M+K-L+2).
\end{eqnarray}
\end{prop}

\begin{prop}~~~We have the following two relations.
\begin{eqnarray}
&&\sum_{\epsilon=\pm} \int_C 
\frac{d w_1}{w_1}
\frac{\epsilon q^{\epsilon} 
w_1(1-q w_1 w_2)e^{-c_+^j(q^{1+\epsilon}w_1)-c_+^j(q^{1-\epsilon}/w_1)}}{
(1-rq^{\alpha}w_1)(1-q^{\epsilon} w_1 w_2)(1-q^{\epsilon}w_1/w_2)}\\
&=&
-r q^{\alpha-1}(1-q^{-\alpha+1}w_2/r)
\int_C \frac{dw_1}{w_1}\frac{(-1+w_1^2)e^{-c_+^j(w_1)-c_+^j(q^2/w_1)}}{
(1-w_1w_2/q)(1-w_1/qw_2)(1-rq^\alpha w_1)(1-rq^\alpha/w_1)},\nonumber
\label{recursion:integral3}
\\
&&\sum_{\epsilon=\pm} \int_C 
\frac{d w_1}{w_1}\frac{\epsilon q^{\epsilon} 
w_1(1-q w_1 w_2)e^{-c_+^j(q^{1+\epsilon}w_1)-c_+^j(q^{1-\epsilon}/w_1)}}{
(1-q^{\epsilon} w_1 w_2)(1-q^{\epsilon}w_1/w_2)}\nonumber\\
&=&
w_2
\int_C \frac{dw_1}{w_1}\frac{(-1+w_1^2)e^{-c_+^j(w_1)-c_+^j(q^2/w_1)}}{
(1-w_1w_2/q)(1-w_1/qw_2)}.
\label{recursion:integral4}
\end{eqnarray}
Here the integration contour $C$ encircles
$w_1=0, q w_2^{\pm 1}$ not but $w_1=q^{-1}w_2^{\pm 1}$.
Here the integrals
$
\int_C \frac{dw_1}{w_1}\frac{(-1+w_1^2)e^{-c_+^j(w_1)-c_+^j(q^2/w_1)}}{
(1-w_1w_2/q)(1-w_1/qw_2)(1-rq^\alpha w_1)(1-rq^\alpha/w_1)}$
and
$
\int_C \frac{dw_1}{w_1}\frac{(-1+w_1^2)e^{-c_+^j(w_1)-c_+^j(q^2/w_1)}}{
(1-w_1w_2/q)(1-w_1/qw_2)}$ in RHS
are invariant under $w_2 \to w_2^{-1}$.
\end{prop}

~\\
{\bf Proof for boundary condition 3.1}~~~
We show the main theorem for 
the boundary condition ${_{(3)}}\langle B|$ and $L+K \leq M+1$.
Here we show the relation (\ref{boundary:case3-3}).
\\
$\bullet$~The case for $1\leq L+K+1 \leq j \leq M+1$ :
${_{(3)}}\langle B|\phi_j^*(z)$.
\\
Using the bosonization (\ref{boson:VO4}), 
the relations (\ref{action:G3-1}), (\ref{action:G3-2}), (\ref{action:G3-3}),
and the normal orderings in appendix \ref{appendix2},
we have LHS$-$RHS of (\ref{boundary:case3-3}) as following :
\begin{eqnarray} 
&&z^{\frac{M}{M-N}+1}\varphi^{(3)}(1/z){_{(3)}}
\langle B|\phi^*_{M+1+i}(z)-z^{-\frac{M}{M-N}-1}\varphi^{(3)}(z)
{_{(3)}}\langle B|\phi_{M+1+i}(1/z)\nonumber\\
&=&q^{-\frac{M}{M-N}-1}\varphi^{(3)}(z)\varphi^{(3)}(1/z)(z-z^{-1})
e^{\frac{\pi \sqrt{-1}}{M-N}(j-1)}\nonumber\\
&\times&
\prod_{k=1}^{j-1}
\int_{\widetilde{C}_j}\frac{dw_k}{2\pi \sqrt{-1}w_k}
\frac{\displaystyle
\prod_{k=1}^{j-2}(1-q/w_kw_{k+1})}{
\displaystyle
\prod_{k=0}^{j-2}
D(w_k,w_{k+1})}
\frac{
\displaystyle
\prod_{k=1}^{j-1}(1-w_k^2)}{(1-q^{-L}rw_L)(q^{L-K}w_{L+K}/r)}\nonumber\\
&\times&
{_{(3)}}\langle B|e^{Q_{h_1^*-h_1\cdots-h_{j-1}}}
e^{h_+^*(qz)+h_+^*(q/z)-\sum_{k=1}^{j-1}(h_+^k(qw_k)+h_+^k(q/w_k))}.
\end{eqnarray}
Here the integration contour $\widetilde{C}_{j}$
encircles $w_k=0,qw_{k-1}^{\pm 1}$
not but $w_{k}=q^{-1}w_{k-1}^{\pm 1}$ for $1\leq k \leq j$.
Taking into account of symmetrization
$\oint \frac{dw}{w}f(w)=\frac{1}{2}\oint(f(w)+f(1/w))$
and relation 
$w_1(1-q/w_1w_2)-(w_1 \leftrightarrow w_1^{-1})=(w_1-w_1^{-1})$,
we symmetrize the variables
$w_1,w_2,\cdots,w_{L-1}$ iteratively.
Using relation
$\frac{w_L(1-q/w_Lw_{L+1})}{(1-rq^{-L}w_L)}-
(w_L \leftrightarrow w_L^{-1})=\frac{(w_L-w_L^{-1})(1-rq^{-L+1}/w_{L+1})}{
(1-rq^{-L}w_L)(1-rq^{-L}/w_L)}$, we symmetrize the variable $w_L$.
Using relation
$(1-rq^\alpha/w_{L+1})w_{L+1}(1-q/w_{L+1}w_{L+2})-(w_{L+1}\leftrightarrow 
w_{L+1}^{-1})=
(w_{L+1}-w_{L+1}^{-1})(1-rq^{\alpha+1}/w_{L+2})
$, we symmetrize the variable 
$w_{L+1},\cdots,w_{L+K-1}$ iteratively.
Using relation $(1-q/w_{L+K+1}w_{L+K+2})-
(w_{L+K+1} \leftrightarrow w_{L+K+1}^{-1})=(-q/w_{L+K+2})(w_{L+K+1}^{-1}-
w_{L+K+1})$, we symmetrize the variables $w_{L+K+1},\cdots, w_{j-2}$
iteratively.
Then we have
\begin{eqnarray} 
&&q^{-\frac{M}{M-N}+j-2L-3}(-1/2)^{j-2}
\varphi^{(3)}(z)\varphi^{(3)}(1/z)(z-z^{-1})
e^{\frac{\pi \sqrt{-1}}{M-N}(j-1)}\nonumber\\
&\times&
\prod_{k=1}^{j-1}
\int_{\widetilde{C}_j}\frac{dw_k}{2\pi \sqrt{-1}w_k}
\frac{\displaystyle
\prod_{k=1}^{j-2}(w_k^{-1}-w_k)^2}{
\displaystyle
\prod_{k=0}^{j-2}
D(w_k,w_{k+1})}
\frac{
(w_{j-1}^{-1}-w_{j-1})}
{(1-q^{-L}rw_L)(q^{-L}r/w_L)}\nonumber\\
&\times&
{_{(3)}}\langle B|e^{Q_{h_1^*-h_1\cdots-h_{j-1}}}
e^{h_+^*(qz)+h_+^*(q/z)-\sum_{k=1}^{j-1}(h_+^k(qw_k)+h_+^k(q/w_k))}=0.
\end{eqnarray}
Here we have used $\oint_{|w|=1} \frac{dw}{w}(w-w^{-1})f(w)=0$,
where $f(w)=f(1/w)$.
~\\
$\bullet$~The case for $1\leq i \leq N+1$ : 
${_{(3)}}\langle B|\phi_{M+1+i}^*(z)$.
\\
Using the bosonization (\ref{boson:VO6}), 
the relations (\ref{action:G3-1}), (\ref{action:G3-2}), (\ref{action:G3-3}),
and normal orderings in appendix \ref{appendix2},
we have LHS$-$RHS of (\ref{boundary:case3-3}) as following :
\begin{eqnarray} 
&&z^{\frac{M}{M-N}+1}\varphi^{(3)}(1/z){_{(3)}}\langle B|\phi^*_{M+1+i}(z)
-z^{-\frac{M}{M-N}-1}\varphi^{(3)}(z)
{_{(3)}}\langle B|\phi_{M+1+i}(1/z)\nonumber\\
&=&
q^{-\frac{M}{M-N}-1}(q-q^{-1})^M
\varphi^{(3)}(z)\varphi^{(3)}(1/z)(z-z^{-1})e^{\frac{\pi \sqrt{-1}}{M-N}M}\nonumber\\
&\times&
\sum_{\epsilon_1,\cdots,\epsilon_i=\pm}
\epsilon_i \prod_{k=1}^{M+i}
\int_{\widetilde{C}_{M+1+i}}\frac{dw_k}{2\pi \sqrt{-1}w_k}
\frac{\displaystyle 
\prod_{k=1}^M(1-q/w_kw_{k+1})}{\displaystyle \prod_{k=0}^M D(w_k,w_{k+1})}
\prod_{k=1}^M(1-w_k^2)
\nonumber\\ 
&\times&
\frac{(1+w_{M+1}^2)}
{(1-rq^{-L}w_L)
(1-q^{L-K}w_{L+K}/r)}
\prod_{k=1}^{i-1}\frac{\epsilon_k q^{\epsilon_k}(1-q w_{M+k}w_{M+k+1})}{
(1-q^{\epsilon_k}w_{M+k}w_{M+k+1})(1-q^{\epsilon_k}w_{M+k}/w_{M+k+1})}\nonumber\\
&\times&{_{(3)}}\langle B|e^{Q_{h_1^*-h_1\cdots -h_{M+i}-c_i}}
e^{h_+^{*1}(qz)+h_+^{*1}(q/z)-\sum_{k=1}^{M+i}(h_+^k(qw_k)+h_+^k(q/w_k))}\nonumber\\
&\times&
e^{
\sum_{k=1}^{i-1}(c_+^k(q w_{M+k})+c_+^k(q/w_{M+k}))-
\sum_{k=1}^i
(c_+^k(q^{1+\epsilon_k}w_{M+k})+c_+^k(q^{1-\epsilon_k}/w_{M+k}))}.
\end{eqnarray}
Here the integration contour $\widetilde{C}_{M+1+i}$
encircles $w_k=0,qw_{k-1}^{\pm 1}$
not but $w_{k}=q^{-1}w_{k-1}^{\pm 1}$ for $1\leq k \leq M+i$.
Using relation 
$w_1(1-q/w_1w_2)-(w_1 \leftrightarrow w_1^{-1})=(w_1-w_1^{-1})$,
we symmetrize the variables
$w_1,w_2,\cdots,w_{L-1}$ iteratively.
Using relation
$\frac{w_L(1-q/w_Lw_{L+1})}{(1-rq^{-L}w_L)}-
(w_L \leftrightarrow w_L^{-1})=\frac{(w_L-w_L^{-1})(1-rq^{-L+1}/w_{L+1})}{
(1-rq^{-L}w_L)(1-rq^{-L}/w_L)}$, we symmetrize the variable $w_L$.
Using relation
$(1-rq^s/w_{L+1})w_{L+1}(1-q/w_{L+1}w_{L+2})-(w_{L+1}\leftrightarrow 
w_{L+1}^{-1})=
(w_{L+1}-w_{L+1}^{-1})(1-rq^{s+1}/w_{L+2})
$, we symmetrize the variable 
$w_{L+1},\cdots,w_{L+K-1}$ iteratively.
Using relation $(1-q/w_{L+K+1}w_{L+K+2})-
(w_{L+K+1} \leftrightarrow w_{L+K+1}^{-1})=(-q/w_{L+K+2})(w_{L+K+1}^{-1}-
w_{L+K+1})$, we symmetrize the variables $w_{L+K+1},\cdots, w_{M-1}$
iteratively.
Then we have
\begin{eqnarray}
&&
q^{-\frac{M}{M-N}-M-2L-2}(q-q^{-1})^M(-1/2)^M
\varphi^{(3)}(z)\varphi^{(3)}(1/z)(z-z^{-1})e^{\frac{\pi \sqrt{-1}}{M-N}M}\nonumber\\
&\times&
\sum_{\epsilon_1,\cdots,\epsilon_i=\pm}
\epsilon_i \prod_{k=1}^{M+i}
\int_{\widetilde{C}_{M+1+i}}\frac{dw_k}{2\pi \sqrt{-1}w_k}
\frac{\displaystyle 
\prod_{k=1}^M(w_k^{-1}-w_k)^2}
{\displaystyle \prod_{k=0}^M D(w_k,w_{k+1})}
\frac{(w_{M+1}^{-1}+w_{M+1})}
{(1-r q^{-L}w_L)
(1-r q^{-L}/w_L)}
\nonumber\\
&\times&
\prod_{k=1}^{i-1}
\frac{\epsilon_k q^{\epsilon_k}(1-q w_{M+k}w_{M+k+1})}{
(1-q^{\epsilon_k}w_{M+k}w_{M+k+1})(1-q^{\epsilon_k}w_{M+k}/w_{M+k+1})}\nonumber\\
&\times&{_{(3)}}\langle B|e^{Q_{h_1^*-h_1\cdots -h_{M+i}-c_i}}
e^{h_+^{*1}(qz)+h_+^{*1}(q/z)-\sum_{k=1}^{M+i}(h_+^k(qw_k)+h_+^k(q/w_k))}\nonumber\\
&\times&
e^{
\sum_{k=1}^{i-1}(c_+^k(q w_{M+k})+c_+^k(q/w_{M+k}))-
\sum_{k=1}^i
(c_+^k(q^{1+\epsilon_k}w_{M+k})+c_+^k(q^{1-\epsilon_k}/w_{M+k}))}.
\end{eqnarray}
Using the relation (\ref{recursion:integral1}) for the variables 
$w_{M+1},\cdots,w_{M+i-1}$, iteratively, we have
\begin{eqnarray} 
&&
q^{-\frac{M}{M-N}+M-2L-2}(q-q^{-1})^M(-1/2)^{M-1}
\varphi^{(3)}(z)\varphi^{(3)}(1/z)(z-z^{-1})e^{\frac{\pi \sqrt{-1}}{M-N}M}
\nonumber\\
&\times&
\prod_{k=1}^{M+i}
\int_{\widetilde{C}_{M+1+i}}\frac{dw_k}{2\pi \sqrt{-1}w_k}
\frac{\displaystyle 
\prod_{k=1}^M(w_k^{-1}-w_k)^2}{
\displaystyle \prod_{k=0}^M D(w_k,w_{k+1})}
\frac{(w_{M+1}^{-1}+w_{M+1})}
{(1-rq^{-L}w_L)
(1- q^{L-K}w_{L+K}/r)}\nonumber\\
&\times&
{_{(3)}}\langle B|
e^{Q_{h_1^*-h_1\cdots -h_{M+i}-c_i}}
e^{h_+^{*1}(qz)+h_+^{*1}(q/z)-\sum_{k=1}^{M+i}(h_+^k(qw_k)+h_+^k(q/w_k))
+\sum_{k=1}^{i-1}(c_+^k(q w_{M+k})+c_+^k(q/w_{M+k}))}\nonumber\\
&\times&
\frac{\displaystyle
\prod_{k=1}^{i-1}(-1+w_{M+k}^2)
e^{-c_+^k(w_{M+k})-c_+^k(q^2/w_{M+k})}
}{
(1-q^{\epsilon_k}w_{M+k}w_{M+k+1})(1-q^{\epsilon_k}w_{M+k}/w_{M+k+1})}\nonumber
\\
&\times&\left(
e^{-c_+^i(q^2w_{M+i})-c_+^i(q/w_{M+i})-e^{-c_+^i(w_{M+i})-c_+^i(q^2/w_{M+i})}}
\right)=0.
\end{eqnarray}
Here we have used $\oint_{|w|=1}\frac{dw}{w}e^{c_+^{i-1}(qw)+c_+^{i-1}(1/w)}
(e^{-c_+^i(q^2w)-c_+^i(1/w)}-e^{-c_+^i(w)-c_+^i(q^2/w)})f(w)=0$,
where $f(w)=f(1/w)$.
\\
{\bf Q.E.D.}

~\\
{\bf Proof for boundary condition 3.2}~~~
We show the main theorem for 
the boundary condition ${_{(3)}}\langle B|$ and $L \leq M+1 \leq L+K-1$.
Here we show the relation (\ref{boundary:case3-3}).
\\
$\bullet$~
The case for $L+K-M\leq i \leq N+1$ : ${_{(3)}}\langle B|\phi_{M+1+i}^*(z)$.
\\
Using the bosonization (\ref{boson:VO6}), 
the relations (\ref{action:G3-1}), (\ref{action:G3-2}), (\ref{action:G3-3}),
and normal orderings in appendix \ref{appendix2},
we have LHS$-$RHS of (\ref{boundary:case3-3}) as following :
\begin{eqnarray} 
&&z^{\frac{M}{M-N}+1}\varphi^{(3)}(1/z){_{(3)}}
\langle B|\phi^*_{M+1+i}(z)-
z^{\frac{M}{M-N}-1}\varphi^{(3)}(z)
{_{(3)}}\langle B|\phi_{M+1+i}^*(1/z)\nonumber\\
&=&
q^{-\frac{M}{M-N}-1}(q-q^{-1})^M
\varphi^{(3)}(z)\varphi^{(3)}(1/z)(z-z^{-1})
e^{\frac{\pi \sqrt{-1}}{M-N}M}\nonumber\\
&\times&
\sum_{\epsilon_1,\cdots,\epsilon_i=\pm}
\epsilon_i \prod_{k=1}^{M+i}
\int_{\widetilde{C}_{M+1+i}}\frac{dw_k}{2\pi \sqrt{-1}w_k}
\frac{\displaystyle 
\prod_{k=1}^M(1-q/w_kw_{k+1})}{\displaystyle \prod_{k=0}^M D(w_k,w_{k+1})}
\prod_{k=1}^M(1-w_k^2)
\nonumber\\ 
&\times&
\frac{(1+w_{M+1}^2)}{(1-rq^{-L}w_L)
(1-q^{3L+K-2M+K-2}w_{L+K}/r)}
\prod_{k=1}^{i-1}\frac{\epsilon_k q^{\epsilon_k}(1-q w_{M+k}w_{M+k+1})}{
(1-q^{\epsilon_k}w_{M+k}w_{M+k+1})(1-q^{\epsilon_k}w_{M+k}/w_{M+k+1})}\nonumber\\
&\times&{_{(3)}}\langle B|e^{Q_{h_1^*-h_1\cdots -h_{M+i}-c_i}}
e^{h_+^{*1}(qz)+h_+^{*1}(q/z)-\sum_{k=1}^{M+i}(h_+^k(qw_k)+h_+^k(q/w_k))}\nonumber\\
&\times&
e^{
\sum_{k=1}^{i-1}(c_+^k(q w_{M+k})+c_+^k(q/w_{M+k}))-
\sum_{k=1}^i
(c_+^k(q^{1+\epsilon_k}w_{M+k})+c_+^k(q^{1-\epsilon_k}/w_{M+k}))}.
\end{eqnarray}
Here the integration contour $\widetilde{C}_{M+1+i}$
encircles $w_k=0,qw_{k-1}^{\pm 1}$
not but $w_{k}=q^{-1}w_{k-1}^{\pm 1}$ for $1\leq k \leq M+i$.
Using relation $w_1(1-q/w_1w_2)-(w_1 \leftrightarrow w_1^{-1})=
(w_1-w_1^{-1})$, we symmetrize the variables $w_1,w_2,\cdots,w_{L-1}$,
iteratively.
Using relation 
$\frac{w_L(1-q/w_Lw_{L+1})}{(1-rq^{-L}w_L)}-
(w_L \leftrightarrow w_L^{-1})
=\frac{(w_L^{-1}-w_L)(1-rq^{-L+1}/w_{L+1})}
{(1-rq^{-L}w_L)(1-rq^{-L}/w_L)}$, 
we symmetrize the variables $w_L$.
Using relation 
$w_{L+1}(1-rq^{-\alpha}/w_{L+1})(1-q/w_{L+1}w_{L+2})-
(w_{L+1}\leftrightarrow w_{L+1}^{-1})=(w_{L+1}-w_{L+1}^{-1})
(1-rq^{-\alpha+1}/w_{L+2})$, we symmetrize the variables $w_{L+1},\cdots,w_M$, iteratively.
Then we have
\begin{eqnarray} 
&&
q^{-\frac{M}{M-N}-1}(-1/2)^M(q-q^{-1})^M
\varphi^{(3)}(z)\varphi^{(3)}(1/z)(z-z^{-1})
e^{\frac{\pi \sqrt{-1}}{M-N}M}\nonumber\\
&\times&
\sum_{\epsilon_1,\cdots,\epsilon_i=\pm}
\epsilon_i \prod_{k=1}^{M+i}
\int_{\widetilde{C}_{M+1+i}}\frac{dw_k}{2\pi \sqrt{-1}w_k}
\frac{\displaystyle 
\prod_{k=1}^M(w_k^{-1}-w_k)^2}{\displaystyle \prod_{k=0}^M D(w_k,w_{k+1})}
\frac{(1-rq^{M-2L+1}/w_{M+1})}{(1-rq^{-L}w_L)(1-rq^{-L}/w_L)}
\nonumber\\ 
&\times&
\frac{(1+w_{M+1}^2)}{(1-q^{3L+K-2M-2}w_{L+K}/r)}
\prod_{k=1}^{i-1}\frac{\epsilon_k q^{\epsilon_k}(1-q w_{M+k}w_{M+k+1})}{
(1-q^{\epsilon_k}w_{M+k}w_{M+k+1})(1-q^{\epsilon_k}w_{M+k}/w_{M+k+1})}\nonumber\\
&\times&{_{(3)}}\langle B|e^{Q_{h_1^*-h_1\cdots -h_{M+i}-c_i}}
e^{h_+^{*1}(qz)+h_+^{*1}(q/z)-\sum_{k=1}^{M+i}(h_+^k(qw_k)+h_+^k(q/w_k))}\nonumber\\
&\times&
e^{
\sum_{k=1}^{i-1}(c_+^k(q w_{M+k})+c_+^k(q/w_{M+k}))-
\sum_{k=1}^i
(c_+^k(q^{1+\epsilon_k}w_{M+k})+c_+^k(q^{1-\epsilon_k}/w_{M+k}))}.
\end{eqnarray}
Using the relation (\ref{recursion:integral2}), we symmetrize the variables $w_{M+1},\cdots,w_{L+K-1}$ iteratively.
Using the relation (\ref{recursion:integral1}), we symmetrize the variables
$w_{L+K},\cdots,w_{M+i}$ iteratively.
Then we have
\begin{eqnarray} 
&&
(-r)q^{-\frac{M}{M-N}+K-L-1}(-1/2)^M(q-q^{-1})^M
\varphi^{(3)}(z)\varphi^{(3)}(1/z)(z-z^{-1})
e^{\frac{\pi \sqrt{-1}}{M-N}M}\nonumber\\
&\times&
\sum_{\epsilon_1,\cdots,\epsilon_i=\pm}
\epsilon_i \prod_{k=1}^{M+i}
\int_{\widetilde{C}_{M+1+i}}\frac{dw_k}{2\pi \sqrt{-1}w_k}
\frac{\displaystyle 
\prod_{k=1}^M(w_k^{-1}-w_k)^2}{\displaystyle \prod_{k=0}^M D(w_k,w_{k+1})}
\frac{(w_{M+1}^{-1}+w_{M+1})}{(1-rq^{-L}w_L)(1-rq^{-L}/w_L)}
\nonumber\\ 
&\times&{_{(3)}}\langle B|e^{Q_{h_1^*-h_1\cdots -h_{M+i}-c_i}}
e^{h_+^{*1}(qz)+h_+^{*1}(q/z)-\sum_{k=1}^{M+i}(h_+^k(qw_k)+h_+^k(q/w_k))+
\sum_{k=1}^{i-1}(c_+^k(q w_{M+k})+c_+^k(q/w_{M+k}))}\nonumber\\
&\times&
\prod_{k=1}^{i-1}\frac{(-1-q w_{M+k}^2)e^{-c_+^k(w_{M+k})-c_+^k(q^2/w_{M+k})}
}{(1-w_{M+k}w_{M+k+1}/q)(1-w_{M+k}/qw_{M+k+1})}\nonumber\\
&\times&
\left(
e^{
-c_+^i(q^{2}w_{M+i})-c_+^i(1/w_{M+i})}
-
e^{
-c_+^i(w_{M+i})-c_+^i(q^{2}/w_{M+i})}
\right)=0.
\end{eqnarray}
Here we have used
\begin{eqnarray}
\oint_{|w|=1} \frac{dw}{w}
e^{c_+^{i-1}(qw)+c_+^{i-1}(q/w)}
\left(e^{-c_+^i(q^2w)-c_+^i(1/w)}-
e^{-c_+^i(w)-c_+^i(q^2/w)}\right)f(w)=0,
\nonumber
\end{eqnarray}
where $f(w)=f(1/w)$.\\
{\bf Q.E.D.}

~\\
{\bf Proof for boundary condition 3.3}~~~
We show the main theorem for 
the boundary condition 
${_{(3)}}\langle B|$ and $1\leq M+1 \leq L-1$.
Here we show the relation (\ref{boundary:case3-3}).
\\
$\bullet$~The case for $L+K-M\leq i \leq N+1$ : ${_{(3)}}\langle B|
\phi_{M+1+i}^*(z)$.\\
Using the bosonization (\ref{boson:VO6}) and the relations
(\ref{action:G3-1}), (\ref{action:G3-2}), (\ref{action:G3-3}),
and normal orderings in appendix \ref{appendix2}, 
we have LHS$-$RHS of (\ref{boundary:case3-3}) as following :
\begin{eqnarray} 
&&z^{\frac{M}{M-N}+1}\varphi^{(3)}(1/z){_{(3)}}\langle B|\phi^*_{M+1+i}(z)
-z^{-\frac{M}{M-N}-1}\varphi^{(3)}(z)
{_{(3)}}\langle B|\phi_{M+1+i}(1/z)\nonumber\\
&=&
q^{-\frac{M}{M-N}-1}(q-q^{-1})^M
\varphi^{(3)}(z)\varphi^{(3)}(1/z)(z-z^{-1})e^{\frac{\pi \sqrt{-1}}{M-N}M}\nonumber\\
&\times&
\sum_{\epsilon_1,\cdots,\epsilon_i=\pm}
\epsilon_i \prod_{k=1}^{M+i}
\int_{\widetilde{C}_{M+1+i}}\frac{dw_k}{2\pi \sqrt{-1}w_k}
\frac{\displaystyle 
\prod_{k=1}^M(1-q/w_kw_{k+1})}{\displaystyle \prod_{k=0}^M D(w_k,w_{k+1})}
\prod_{k=1}^M(1-w_k^2)
\nonumber\\ 
&\times&
\frac{(1+w_{M+1}^2)}{(1-rq^{L-2M-2}w_L)
(1-q^{2M+K-L+2}w_{L+K}/r)}
\prod_{k=1}^{i-1}\frac{\epsilon_k q^{\epsilon_k}(1-q w_{M+k}w_{M+k+1})}{
(1-q^{\epsilon_k}w_{M+k}w_{M+k+1})(1-q^{\epsilon_k}w_{M+k}/w_{M+k+1})}\nonumber\\
&\times&{_{(3)}}\langle B|e^{Q_{h_1^*-h_1\cdots -h_{M+i}-c_i}}
e^{h_+^{*1}(qz)+h_+^{*1}(q/z)-\sum_{k=1}^{M+i}(h_+^k(qw_k)+h_+^k(q/w_k))}\nonumber\\
&\times&
e^{
\sum_{k=1}^{i-1}(c_+^k(q w_{M+k})+c_+^k(q/w_{M+k}))-
\sum_{k=1}^i
(c_+^k(q^{1+\epsilon_k}w_{M+k})+c_+^k(q^{1-\epsilon_k}/w_{M+k}))}.
\end{eqnarray}
Here the integration contour $\widetilde{C}_{M+1+i}$
encircles $w_k=0,qw_{k-1}^{\pm 1}$
not but $w_{k}=q^{-1}w_{k-1}^{\pm 1}$ for $1\leq k \leq M+i$.
Using relation
$w_1(1-q/w_1w_2)-(w_1 \leftrightarrow w_1^{-1})=
(w_1-w_1^{-1})$, 
we symmetrize the variables $w_1,w_2,\cdots,w_M$, iteratively.
We have
\begin{eqnarray} 
&&
q^{-\frac{M}{M-N}-1}(q-q^{-1})^M(-1/2)^M
\varphi^{(3)}(z)\varphi^{(3)}(1/z)(z-z^{-1})
e^{\frac{\pi \sqrt{-1}}{M-N}M}\nonumber\\
&\times&
\sum_{\epsilon_1,\cdots,\epsilon_i=\pm}
\epsilon_i \prod_{k=1}^{M+i}
\int_{\widetilde{C}_{M+1+i}}\frac{dw_k}{2\pi \sqrt{-1}w_k}
\frac{\displaystyle \prod_{k=1}^M
(w_k^{-1}-w_k)^2}{\displaystyle \prod_{k=0}^M D(w_k,w_{k+1})}
\frac{
(w_{M+1}^{-1}+w_{M+1})
}{(1-rq^{L-2M-2}w_L)(1-q^{2M+K-L+2}w_{L+K}/r)}\nonumber\\
&\times&
{_{(3)}}\langle B|e^{Q_{h_1^*-h_1\cdots -h_{M+i}-c_i}}
e^{h_+^{*1}(qz)+h_+^{*1}(q/z)
-\sum_{k=1}^{M+i}(h_+^k(qw_k)+h_+^k(q/w_k))
+\sum_{k=1}^{i-1}(c_+^k(q w_{M+k})+c_+^k(q/w_{M+k}))}
\nonumber\\
&\times&
w_{M+1}\prod_{k=1}^{i-1}\frac{\epsilon_k q^{\epsilon_k}(1-q w_{M+k}w_{M+k+1})
e^{-c_+^k(q^{1+\epsilon_k}w_{M+k})-c_+^k(q^{1-\epsilon_k}/w_{M+k})}
}{
(1-q^{\epsilon_k}w_{M+k}w_{M+k+1})
(1-q^{\epsilon_k}w_{M+k}/w_{M+k+1})}
e^{
-c_+^i(q^{1+\epsilon_i}w_{M+i})-c_+^i(q^{1-\epsilon_i}/w_{M+i})}.
\nonumber\\
\end{eqnarray}
We use the relation (\ref{recursion:integral4}) for the variables
$w_{M+1},\cdots, w_{L-1}$, iteratively.
We use the relation ({\ref{recursion:integral3}}) for the variable $w_L$.
Using the relation (\ref{recursion:integral2}),
we symmetrize
the variables $w_{L+1},\cdots,w_{L+K-1}$, iteratively.
Using the relation (\ref{recursion:integral1}),
we symmetrize the variables $w_{L+K},\cdots, w_{M+i-1}$, iteratively.
Then we have
\begin{eqnarray} 
&&
(-r)q^{-\frac{M}{M-N}+K+L-2M-5}(q-q^{-1})^M(-1/2)^M
\varphi^{(3)}(z)\varphi^{(3)}(1/z)(z-z^{-1})
e^{\frac{\pi \sqrt{-1}}{M-N}M}\nonumber\\
&\times&
\prod_{k=1}^{M+i}
\int_{\widetilde{C}_{M+1+i}}\frac{dw_k}{2\pi \sqrt{-1}w_k}
\frac{\displaystyle \prod_{k=1}^M
(w_k^{-1}-w_k)^2}{\displaystyle \prod_{k=0}^M D(w_k,w_{k+1})}
\frac{(w_{M+1}^{-1}+w_{M+1})}
{(1-rq^{L-2M-2}w_L)(1-rq^{L-2M-2}/w_L)}
\nonumber\\
&\times&
{_{(3)}}\langle B|e^{Q_{h_1^*-h_1\cdots -h_{M+i}-c_i}}
e^{h_+^{*1}(qz)+h_+^{*1}(q/z)
-\sum_{k=1}^{M+i}(h_+^k(qw_k)+h_+^k(q/w_k))
+\sum_{k=1}^{i-1}(c_+^k(q w_{M+k})+c_+^k(q/w_{M+k}))}
\nonumber\\
&\times&
\prod_{k=1}^{i-1}\frac{(-1+w_{M+k}^2)
e^{-c_+^k(w_{M+k})-c_+^k(q^{2}/w_{M+k})}
}{
(1-w_{M+k}w_{M+k+1}/q)
(1-q w_{M+k}/w_{M+k+1})}\nonumber\\
&\times&
\left(
e^{
-c_+^i(q^{2}w_{M+i})-c_+^i(1/w_{M+i})}
-
e^{
-c_+^i(w_{M+i})-c_+^i(q^{2}/w_{M+i})}
\right)=0.
\end{eqnarray}
Here we have used
\begin{eqnarray}
\oint_{|w|=1} \frac{dw}{w}
e^{c_+^{i-1}(qw)+c_+^{i-1}(q/w)}
\left(e^{-c_+^i(q^2w)-c_+^i(1/w)}-
e^{-c_+^i(w)-c_+^i(q^2/w)}\right)f(w)=0,
\nonumber
\end{eqnarray}
where $f(w)=f(1/w)$.\\
{\bf Q.E.D.}\\
Now we have shown (\ref{boundary:case3-3}) 
for every $j=L+K+1,\cdots,M+N+2$.

\subsection*{Acknowledgement}

This work is supported by the Grant-in-Aid for Scientific Research {\bf C} (21540228)
from Japan Society for Promotion of Science.

\begin{appendix}

\section{Figure}
\label{appendix0}

In this appendix we summarize the figures that we use in the section 
\ref{section2}.

~\\

~~~~~~~~~~~~~~~~~~~~~~~~~~~~
\unitlength 0.1in
\begin{picture}( 20.8500, 12.8000)( 18.8000,-15.2500)
\put(31.0000,-8.0000){\makebox(0,0){$k_2$}}%
\put(41.0000,-8.0000){\makebox(0,0){$j_2$}}%
\put(36.0000,-13.1000){\makebox(0,0){$k_1$}}%
\put(36.0000,-3.1000){\makebox(0,0){$j_1$}}%
\put(26.1000,-8.0000){\makebox(0,0){$R(z)_{k_1,k_2}^{j_1,j_2}=$}}%
\put(31.0000,-8.0000){\makebox(0,0){$k_2$}}%
\put(41.0000,-8.0000){\makebox(0,0){$j_2$}}%
\put(36.0000,-13.1000){\makebox(0,0){$k_1$}}%
\put(36.0000,-3.1000){\makebox(0,0){$j_1$}}%
\put(26.1000,-8.0000){\makebox(0,0){$R(z)_{k_1,k_2}^{j_1,j_2}=$}}%
%
\special{pn 8}%
\special{pa 3610 400}%
\special{pa 3610 1200}%
\special{fp}%
\special{sh 1}%
\special{pa 3610 1200}%
\special{pa 3630 1134}%
\special{pa 3610 1148}%
\special{pa 3590 1134}%
\special{pa 3610 1200}%
\special{fp}%
\special{pa 4010 800}%
\special{pa 3210 800}%
\special{fp}%
\special{sh 1}%
\special{pa 3210 800}%
\special{pa 3278 820}%
\special{pa 3264 800}%
\special{pa 3278 780}%
\special{pa 3210 800}%
\special{fp}%
\put(37.4000,-10.0000){\makebox(0,0){$z$}}%
\put(36.1000,-15.9000){\makebox(0,0){Fig.1. $R$-matrix}}%
\put(31.0000,-8.0000){\makebox(0,0){$k_2$}}%
\put(41.0000,-8.0000){\makebox(0,0){$j_2$}}%
\put(36.0000,-13.1000){\makebox(0,0){$k_1$}}%
\put(36.0000,-3.1000){\makebox(0,0){$j_1$}}%
\put(26.1000,-8.0000){\makebox(0,0){$R(z)_{k_1,k_2}^{j_1,j_2}=$}}%
\put(31.0000,-8.0000){\makebox(0,0){$k_2$}}%
\put(41.0000,-8.0000){\makebox(0,0){$j_2$}}%
\put(36.0000,-13.1000){\makebox(0,0){$k_1$}}%
\put(36.0000,-3.1000){\makebox(0,0){$j_1$}}%
\put(26.1000,-8.0000){\makebox(0,0){$R(z)_{k_1,k_2}^{j_1,j_2}=$}}%
%
\special{pn 8}%
\special{pa 3610 400}%
\special{pa 3610 1200}%
\special{fp}%
\special{sh 1}%
\special{pa 3610 1200}%
\special{pa 3630 1134}%
\special{pa 3610 1148}%
\special{pa 3590 1134}%
\special{pa 3610 1200}%
\special{fp}%
\special{pa 4010 800}%
\special{pa 3210 800}%
\special{fp}%
\special{sh 1}%
\special{pa 3210 800}%
\special{pa 3278 820}%
\special{pa 3264 800}%
\special{pa 3278 780}%
\special{pa 3210 800}%
\special{fp}%
\end{picture}%

~\\

~\\
~~~~~~~~~~~~~~~~
\unitlength 0.1in
\begin{picture}( 42.8500, 14.3500)(  1.0500,-18.3500)
%
\special{pn 8}%
\special{pa 1010 400}%
\special{pa 1010 1600}%
\special{fp}%
\special{pa 1010 400}%
\special{pa 810 600}%
\special{fp}%
\special{pa 1010 600}%
\special{pa 810 800}%
\special{fp}%
\special{pa 1010 800}%
\special{pa 810 1000}%
\special{fp}%
\special{pa 1010 1000}%
\special{pa 810 1200}%
\special{fp}%
\special{pa 1010 1200}%
\special{pa 810 1400}%
\special{fp}%
\special{pa 1010 1400}%
\special{pa 810 1600}%
\special{fp}%
\put(4.9000,-10.0000){\makebox(0,0){$K^+(z)_k^j=$}}%
\put(12.0000,-6.3000){\makebox(0,0){$k$}}%
\put(11.9000,-13.7000){\makebox(0,0){$j$}}%
\put(15.7000,-13.9000){\makebox(0,0){$z^{-1}$}}%
\put(15.7000,-6.1000){\makebox(0,0){$z$}}%
%
\special{pn 8}%
\special{pa 4190 410}%
\special{pa 4190 410}%
\special{fp}%
\special{pa 4190 1610}%
\special{pa 4190 410}%
\special{fp}%
\special{pa 4390 410}%
\special{pa 4190 610}%
\special{fp}%
\special{pa 4390 610}%
\special{pa 4190 810}%
\special{fp}%
\special{pa 4390 810}%
\special{pa 4190 1010}%
\special{fp}%
\special{pa 4390 1010}%
\special{pa 4190 1210}%
\special{fp}%
\special{pa 4390 1210}%
\special{pa 4190 1410}%
\special{fp}%
\special{pa 4390 1410}%
\special{pa 4190 1610}%
\special{fp}%
%
\special{pn 8}%
\special{pa 3790 610}%
\special{pa 4190 1010}%
\special{fp}%
\special{sh 1}%
\special{pa 4190 1010}%
\special{pa 4158 950}%
\special{pa 4152 972}%
\special{pa 4130 978}%
\special{pa 4190 1010}%
\special{fp}%
\special{pa 4190 1010}%
\special{pa 3790 1410}%
\special{fp}%
\special{sh 1}%
\special{pa 3790 1410}%
\special{pa 3852 1378}%
\special{pa 3828 1372}%
\special{pa 3824 1350}%
\special{pa 3790 1410}%
\special{fp}%
\put(35.1000,-10.1000){\makebox(0,0){$K(z)_k^j=$}}%
\put(36.8000,-6.1000){\makebox(0,0){$z$}}%
\put(36.4000,-14.1000){\makebox(0,0){$z^{-1}$}}%
\put(39.9000,-6.1000){\makebox(0,0){$j$}}%
\put(40.1000,-14.1000){\makebox(0,0){$k$}}%
\put(25.2000,-19.0000){\makebox(0,0){Fig.2. $K$-matrix}}%
%
\special{pn 8}%
\special{pa 1400 1400}%
\special{pa 1000 1000}%
\special{fp}%
\special{sh 1}%
\special{pa 1000 1000}%
\special{pa 1034 1062}%
\special{pa 1038 1038}%
\special{pa 1062 1034}%
\special{pa 1000 1000}%
\special{fp}%
\special{pa 1000 1000}%
\special{pa 1400 600}%
\special{fp}%
\special{sh 1}%
\special{pa 1400 600}%
\special{pa 1340 634}%
\special{pa 1362 638}%
\special{pa 1368 662}%
\special{pa 1400 600}%
\special{fp}%
\end{picture}%

~\\

~\\
~~~~~~~~~~~~~~~~
\unitlength 0.1in
\begin{picture}( 44.1000, 18.9000)(  4.0000,-21.0500)
%
\special{pn 8}%
\special{pa 600 410}%
\special{pa 600 1610}%
\special{fp}%
\special{pa 600 410}%
\special{pa 400 610}%
\special{fp}%
\special{pa 600 610}%
\special{pa 400 810}%
\special{fp}%
\special{pa 600 810}%
\special{pa 400 1010}%
\special{fp}%
\special{pa 600 1010}%
\special{pa 400 1210}%
\special{fp}%
\special{pa 600 1210}%
\special{pa 400 1410}%
\special{fp}%
\special{pa 600 1410}%
\special{pa 400 1610}%
\special{fp}%
%
\special{pn 8}%
\special{pa 4610 420}%
\special{pa 4610 1620}%
\special{fp}%
\special{pa 4810 420}%
\special{pa 4610 620}%
\special{fp}%
\special{pa 4810 620}%
\special{pa 4610 820}%
\special{fp}%
\special{pa 4810 820}%
\special{pa 4610 1020}%
\special{fp}%
\special{pa 4810 1020}%
\special{pa 4610 1220}%
\special{fp}%
\special{pa 4810 1220}%
\special{pa 4610 1420}%
\special{fp}%
\special{pa 4810 1420}%
\special{pa 4610 1620}%
\special{fp}%
%
\special{pn 8}%
\special{pa 1010 400}%
\special{pa 1010 1600}%
\special{fp}%
\special{sh 1}%
\special{pa 1010 1600}%
\special{pa 1030 1534}%
\special{pa 1010 1548}%
\special{pa 990 1534}%
\special{pa 1010 1600}%
\special{fp}%
\special{pa 1410 400}%
\special{pa 1410 1600}%
\special{fp}%
\special{sh 1}%
\special{pa 1410 1600}%
\special{pa 1430 1534}%
\special{pa 1410 1548}%
\special{pa 1390 1534}%
\special{pa 1410 1600}%
\special{fp}%
\special{pa 4210 400}%
\special{pa 4210 1600}%
\special{fp}%
\special{sh 1}%
\special{pa 4210 1600}%
\special{pa 4230 1534}%
\special{pa 4210 1548}%
\special{pa 4190 1534}%
\special{pa 4210 1600}%
\special{fp}%
\special{pa 3810 400}%
\special{pa 3810 1600}%
\special{fp}%
\special{sh 1}%
\special{pa 3810 1600}%
\special{pa 3830 1534}%
\special{pa 3810 1548}%
\special{pa 3790 1534}%
\special{pa 3810 1600}%
\special{fp}%
%
\special{pn 8}%
\special{pa 610 1000}%
\special{pa 642 992}%
\special{pa 674 982}%
\special{pa 704 972}%
\special{pa 736 962}%
\special{pa 768 954}%
\special{pa 798 944}%
\special{pa 830 934}%
\special{pa 862 926}%
\special{pa 892 916}%
\special{pa 924 908}%
\special{pa 988 888}%
\special{pa 1018 880}%
\special{pa 1050 870}%
\special{pa 1082 862}%
\special{pa 1112 854}%
\special{pa 1144 844}%
\special{pa 1176 836}%
\special{pa 1206 826}%
\special{pa 1270 810}%
\special{pa 1300 802}%
\special{pa 1364 786}%
\special{pa 1394 778}%
\special{pa 1490 754}%
\special{pa 1520 746}%
\special{pa 1552 738}%
\special{pa 1584 732}%
\special{pa 1614 724}%
\special{pa 1646 718}%
\special{pa 1678 710}%
\special{pa 1708 704}%
\special{pa 1772 692}%
\special{pa 1802 686}%
\special{pa 1898 668}%
\special{pa 1928 662}%
\special{pa 1960 658}%
\special{pa 1992 652}%
\special{pa 2022 648}%
\special{pa 2054 642}%
\special{pa 2086 638}%
\special{pa 2116 634}%
\special{pa 2180 626}%
\special{pa 2210 622}%
\special{pa 2242 620}%
\special{pa 2274 616}%
\special{pa 2304 614}%
\special{pa 2336 612}%
\special{pa 2368 608}%
\special{pa 2400 606}%
\special{pa 2430 606}%
\special{pa 2494 602}%
\special{pa 2524 602}%
\special{pa 2556 600}%
\special{pa 2650 600}%
\special{pa 2682 602}%
\special{pa 2712 602}%
\special{pa 2744 604}%
\special{pa 2776 604}%
\special{pa 2808 606}%
\special{pa 2838 608}%
\special{pa 2902 612}%
\special{pa 2932 616}%
\special{pa 2964 618}%
\special{pa 2996 622}%
\special{pa 3026 624}%
\special{pa 3090 632}%
\special{pa 3120 636}%
\special{pa 3152 640}%
\special{pa 3184 646}%
\special{pa 3214 650}%
\special{pa 3246 654}%
\special{pa 3310 666}%
\special{pa 3340 670}%
\special{pa 3404 682}%
\special{pa 3434 688}%
\special{pa 3466 694}%
\special{pa 3498 702}%
\special{pa 3528 708}%
\special{pa 3560 714}%
\special{pa 3592 722}%
\special{pa 3622 728}%
\special{pa 3654 736}%
\special{pa 3686 742}%
\special{pa 3716 750}%
\special{pa 3812 774}%
\special{pa 3842 782}%
\special{pa 3906 798}%
\special{pa 3936 806}%
\special{pa 4000 822}%
\special{pa 4030 832}%
\special{pa 4094 848}%
\special{pa 4124 858}%
\special{pa 4156 866}%
\special{pa 4188 876}%
\special{pa 4220 884}%
\special{pa 4250 894}%
\special{pa 4282 902}%
\special{pa 4314 912}%
\special{pa 4344 922}%
\special{pa 4376 930}%
\special{pa 4408 940}%
\special{pa 4438 950}%
\special{pa 4470 958}%
\special{pa 4502 968}%
\special{pa 4532 978}%
\special{pa 4564 986}%
\special{pa 4596 996}%
\special{pa 4610 1000}%
\special{sp}%
%
\special{pn 8}%
\special{pa 4610 1000}%
\special{pa 4580 1010}%
\special{pa 4548 1020}%
\special{pa 4516 1028}%
\special{pa 4484 1038}%
\special{pa 4454 1048}%
\special{pa 4422 1056}%
\special{pa 4390 1066}%
\special{pa 4360 1076}%
\special{pa 4328 1084}%
\special{pa 4296 1094}%
\special{pa 4266 1104}%
\special{pa 4234 1112}%
\special{pa 4202 1122}%
\special{pa 4172 1130}%
\special{pa 4140 1140}%
\special{pa 4108 1148}%
\special{pa 4078 1156}%
\special{pa 4046 1166}%
\special{pa 3982 1182}%
\special{pa 3952 1192}%
\special{pa 3888 1208}%
\special{pa 3858 1216}%
\special{pa 3794 1232}%
\special{pa 3764 1240}%
\special{pa 3732 1248}%
\special{pa 3700 1254}%
\special{pa 3670 1262}%
\special{pa 3638 1270}%
\special{pa 3606 1276}%
\special{pa 3576 1284}%
\special{pa 3512 1296}%
\special{pa 3480 1304}%
\special{pa 3450 1310}%
\special{pa 3386 1322}%
\special{pa 3356 1328}%
\special{pa 3324 1334}%
\special{pa 3292 1338}%
\special{pa 3262 1344}%
\special{pa 3230 1348}%
\special{pa 3198 1354}%
\special{pa 3168 1358}%
\special{pa 3072 1370}%
\special{pa 3042 1374}%
\special{pa 2978 1382}%
\special{pa 2948 1384}%
\special{pa 2916 1388}%
\special{pa 2884 1390}%
\special{pa 2854 1392}%
\special{pa 2790 1396}%
\special{pa 2760 1398}%
\special{pa 2728 1398}%
\special{pa 2696 1400}%
\special{pa 2540 1400}%
\special{pa 2508 1398}%
\special{pa 2476 1398}%
\special{pa 2446 1396}%
\special{pa 2414 1394}%
\special{pa 2382 1394}%
\special{pa 2352 1390}%
\special{pa 2288 1386}%
\special{pa 2258 1382}%
\special{pa 2226 1380}%
\special{pa 2162 1372}%
\special{pa 2132 1368}%
\special{pa 2068 1360}%
\special{pa 2038 1356}%
\special{pa 2006 1352}%
\special{pa 1974 1346}%
\special{pa 1944 1342}%
\special{pa 1880 1330}%
\special{pa 1850 1324}%
\special{pa 1786 1312}%
\special{pa 1756 1306}%
\special{pa 1692 1294}%
\special{pa 1660 1286}%
\special{pa 1630 1280}%
\special{pa 1598 1272}%
\special{pa 1566 1266}%
\special{pa 1536 1258}%
\special{pa 1504 1250}%
\special{pa 1472 1244}%
\special{pa 1442 1236}%
\special{pa 1378 1220}%
\special{pa 1348 1212}%
\special{pa 1284 1196}%
\special{pa 1252 1186}%
\special{pa 1222 1178}%
\special{pa 1190 1170}%
\special{pa 1158 1160}%
\special{pa 1128 1152}%
\special{pa 1096 1144}%
\special{pa 1064 1134}%
\special{pa 1034 1126}%
\special{pa 1002 1116}%
\special{pa 970 1108}%
\special{pa 940 1098}%
\special{pa 908 1090}%
\special{pa 876 1080}%
\special{pa 846 1070}%
\special{pa 814 1062}%
\special{pa 750 1042}%
\special{pa 720 1034}%
\special{pa 656 1014}%
\special{pa 626 1004}%
\special{pa 610 1000}%
\special{sp}%
%
\special{pn 8}%
\special{pa 2400 400}%
\special{pa 2810 400}%
\special{fp}%
\special{sh 1}%
\special{pa 2810 400}%
\special{pa 2744 380}%
\special{pa 2758 400}%
\special{pa 2744 420}%
\special{pa 2810 400}%
\special{fp}%
\special{pa 2810 1600}%
\special{pa 2410 1600}%
\special{fp}%
\special{sh 1}%
\special{pa 2410 1600}%
\special{pa 2478 1620}%
\special{pa 2464 1600}%
\special{pa 2478 1580}%
\special{pa 2410 1600}%
\special{fp}%
\put(25.4000,-10.3000){\makebox(0,0){$\cdots$}}%
\put(10.1000,-2.8000){\makebox(0,0){$P$}}%
\put(38.1000,-3.0000){\makebox(0,0){$2$}}%
\put(42.1000,-3.1000){\makebox(0,0){$1$}}%
\put(25.9000,-2.8000){\makebox(0,0){$z$}}%
\put(25.8000,-17.7000){\makebox(0,0){$z^{-1}$}}%
\put(25.7000,-21.7000){\makebox(0,0){Fig.3. Transfer matrix $T_B^{fin}(z)$}}%
\end{picture}%

~\\

~\\
~~~~~~~~~~~
\unitlength 0.1in
\begin{picture}( 44.3500, 16.1000)(  3.7500,-18.4500)
%
\special{pn 8}%
\special{pa 4610 420}%
\special{pa 4610 1620}%
\special{fp}%
\special{pa 4810 420}%
\special{pa 4610 620}%
\special{fp}%
\special{pa 4810 620}%
\special{pa 4610 820}%
\special{fp}%
\special{pa 4810 820}%
\special{pa 4610 1020}%
\special{fp}%
\special{pa 4810 1020}%
\special{pa 4610 1220}%
\special{fp}%
\special{pa 4810 1220}%
\special{pa 4610 1420}%
\special{fp}%
\special{pa 4810 1420}%
\special{pa 4610 1620}%
\special{fp}%
\put(25.4000,-10.3000){\makebox(0,0){$\cdots$}}%
\put(38.1000,-3.0000){\makebox(0,0){$2$}}%
\put(42.1000,-3.1000){\makebox(0,0){$1$}}%
%
\special{pn 8}%
\special{pa 800 600}%
\special{pa 864 604}%
\special{pa 896 604}%
\special{pa 928 606}%
\special{pa 960 606}%
\special{pa 1024 610}%
\special{pa 1056 610}%
\special{pa 1120 614}%
\special{pa 1152 614}%
\special{pa 1248 620}%
\special{pa 1280 620}%
\special{pa 1344 624}%
\special{pa 1376 624}%
\special{pa 1504 632}%
\special{pa 1536 632}%
\special{pa 1728 644}%
\special{pa 1760 644}%
\special{pa 1952 656}%
\special{pa 1984 660}%
\special{pa 2176 672}%
\special{pa 2208 676}%
\special{pa 2238 678}%
\special{pa 2270 680}%
\special{pa 2302 684}%
\special{pa 2366 688}%
\special{pa 2398 692}%
\special{pa 2430 694}%
\special{pa 2462 698}%
\special{pa 2494 700}%
\special{pa 2558 708}%
\special{pa 2590 710}%
\special{pa 2620 714}%
\special{pa 2652 718}%
\special{pa 2684 720}%
\special{pa 2844 740}%
\special{pa 2874 742}%
\special{pa 2970 754}%
\special{pa 3002 760}%
\special{pa 3066 768}%
\special{pa 3096 772}%
\special{pa 3160 780}%
\special{pa 3192 786}%
\special{pa 3256 794}%
\special{pa 3286 798}%
\special{pa 3318 804}%
\special{pa 3382 812}%
\special{pa 3414 818}%
\special{pa 3446 822}%
\special{pa 3476 828}%
\special{pa 3508 832}%
\special{pa 3540 838}%
\special{pa 3572 842}%
\special{pa 3604 848}%
\special{pa 3634 852}%
\special{pa 3666 858}%
\special{pa 3698 862}%
\special{pa 3730 868}%
\special{pa 3762 872}%
\special{pa 3792 878}%
\special{pa 3824 884}%
\special{pa 3856 888}%
\special{pa 3920 900}%
\special{pa 3950 904}%
\special{pa 4014 916}%
\special{pa 4046 920}%
\special{pa 4078 926}%
\special{pa 4108 932}%
\special{pa 4140 938}%
\special{pa 4172 942}%
\special{pa 4204 948}%
\special{pa 4234 954}%
\special{pa 4266 960}%
\special{pa 4298 964}%
\special{pa 4362 976}%
\special{pa 4392 982}%
\special{pa 4424 988}%
\special{pa 4456 992}%
\special{pa 4488 998}%
\special{pa 4518 1004}%
\special{pa 4582 1016}%
\special{pa 4610 1020}%
\special{sp}%
%
\special{pn 8}%
\special{pa 810 1410}%
\special{pa 842 1410}%
\special{pa 938 1404}%
\special{pa 970 1404}%
\special{pa 1034 1400}%
\special{pa 1066 1400}%
\special{pa 1162 1394}%
\special{pa 1194 1394}%
\special{pa 1258 1390}%
\special{pa 1290 1390}%
\special{pa 1418 1382}%
\special{pa 1450 1382}%
\special{pa 1610 1372}%
\special{pa 1642 1372}%
\special{pa 2090 1344}%
\special{pa 2122 1340}%
\special{pa 2154 1338}%
\special{pa 2184 1336}%
\special{pa 2248 1332}%
\special{pa 2280 1328}%
\special{pa 2344 1324}%
\special{pa 2376 1320}%
\special{pa 2440 1316}%
\special{pa 2472 1312}%
\special{pa 2504 1310}%
\special{pa 2536 1306}%
\special{pa 2568 1304}%
\special{pa 2600 1300}%
\special{pa 2630 1298}%
\special{pa 2662 1294}%
\special{pa 2694 1292}%
\special{pa 2790 1280}%
\special{pa 2822 1278}%
\special{pa 2918 1266}%
\special{pa 2948 1262}%
\special{pa 2980 1260}%
\special{pa 3172 1236}%
\special{pa 3202 1232}%
\special{pa 3298 1220}%
\special{pa 3330 1214}%
\special{pa 3394 1206}%
\special{pa 3424 1202}%
\special{pa 3456 1198}%
\special{pa 3488 1192}%
\special{pa 3584 1180}%
\special{pa 3614 1174}%
\special{pa 3678 1166}%
\special{pa 3710 1160}%
\special{pa 3774 1152}%
\special{pa 3804 1146}%
\special{pa 3868 1138}%
\special{pa 3900 1132}%
\special{pa 3932 1128}%
\special{pa 3964 1122}%
\special{pa 3994 1118}%
\special{pa 4026 1114}%
\special{pa 4058 1108}%
\special{pa 4090 1104}%
\special{pa 4122 1098}%
\special{pa 4154 1094}%
\special{pa 4184 1088}%
\special{pa 4216 1084}%
\special{pa 4248 1078}%
\special{pa 4280 1074}%
\special{pa 4312 1068}%
\special{pa 4342 1064}%
\special{pa 4374 1058}%
\special{pa 4406 1054}%
\special{pa 4438 1048}%
\special{pa 4470 1044}%
\special{pa 4500 1038}%
\special{pa 4532 1032}%
\special{pa 4564 1028}%
\special{pa 4596 1022}%
\special{pa 4610 1020}%
\special{sp}%
%
\special{pn 8}%
\special{pa 4210 410}%
\special{pa 4210 1610}%
\special{fp}%
\special{sh 1}%
\special{pa 4210 1610}%
\special{pa 4230 1544}%
\special{pa 4210 1558}%
\special{pa 4190 1544}%
\special{pa 4210 1610}%
\special{fp}%
%
\special{pn 8}%
\special{pa 3800 410}%
\special{pa 3800 1620}%
\special{fp}%
\special{sh 1}%
\special{pa 3800 1620}%
\special{pa 3820 1554}%
\special{pa 3800 1568}%
\special{pa 3780 1554}%
\special{pa 3800 1620}%
\special{fp}%
%
\special{pn 8}%
\special{pa 820 430}%
\special{pa 1420 430}%
\special{fp}%
\special{sh 1}%
\special{pa 1420 430}%
\special{pa 1354 410}%
\special{pa 1368 430}%
\special{pa 1354 450}%
\special{pa 1420 430}%
\special{fp}%
\special{pa 1420 1630}%
\special{pa 820 1630}%
\special{fp}%
\special{sh 1}%
\special{pa 820 1630}%
\special{pa 888 1650}%
\special{pa 874 1630}%
\special{pa 888 1610}%
\special{pa 820 1630}%
\special{fp}%
\put(6.0000,-4.2000){\makebox(0,0){$z$}}%
\put(6.0000,-16.4000){\makebox(0,0){$z^{-1}$}}%
\put(25.2000,-19.1000){\makebox(0,0){Fig.4. Transfer matrix $\widetilde{T}_B^{(i)}(z)$}}%
\end{picture}%

~\\

~\\
~~~~~~~~~~~~~~~
\unitlength 0.1in
\begin{picture}( 42.3500, 15.1000)(  4.8000,-17.4500)
\put(38.1000,-3.0000){\makebox(0,0){$2$}}%
\put(42.1000,-3.1000){\makebox(0,0){$1$}}%
%
\special{pn 8}%
\special{pa 4210 410}%
\special{pa 4210 1610}%
\special{fp}%
\special{sh 1}%
\special{pa 4210 1610}%
\special{pa 4230 1544}%
\special{pa 4210 1558}%
\special{pa 4190 1544}%
\special{pa 4210 1610}%
\special{fp}%
%
\special{pn 8}%
\special{pa 3800 410}%
\special{pa 3800 1620}%
\special{fp}%
\special{sh 1}%
\special{pa 3800 1620}%
\special{pa 3820 1554}%
\special{pa 3800 1568}%
\special{pa 3780 1554}%
\special{pa 3800 1620}%
\special{fp}%
%
\special{pn 8}%
\special{pa 1200 1000}%
\special{pa 4600 1000}%
\special{fp}%
\special{sh 1}%
\special{pa 4600 1000}%
\special{pa 4534 980}%
\special{pa 4548 1000}%
\special{pa 4534 1020}%
\special{pa 4600 1000}%
\special{fp}%
\special{pa 4600 1000}%
\special{pa 4600 1000}%
\special{fp}%
\put(47.9000,-10.0000){\makebox(0,0){$j$}}%
\put(27.8000,-8.2000){\makebox(0,0){$\cdots$}}%
\put(13.9000,-8.2000){\makebox(0,0){$z$}}%
\put(7.9000,-10.1000){\makebox(0,0){$\widetilde{\Phi}_j(z)=$}}%
\put(25.9000,-18.1000){\makebox(0,0){Fig.5. Vertex operator $\widetilde{\Phi}_j(z)$}}%
\end{picture}%

~\\
~~~~~~~~~~~~~~~
\unitlength 0.1in
\begin{picture}( 42.9500, 15.1000)(  4.2000,-17.4500)
\put(38.1000,-3.0000){\makebox(0,0){$2$}}%
\put(42.1000,-3.1000){\makebox(0,0){$1$}}%
%
\special{pn 8}%
\special{pa 4210 410}%
\special{pa 4210 1610}%
\special{fp}%
\special{sh 1}%
\special{pa 4210 1610}%
\special{pa 4230 1544}%
\special{pa 4210 1558}%
\special{pa 4190 1544}%
\special{pa 4210 1610}%
\special{fp}%
%
\special{pn 8}%
\special{pa 3800 410}%
\special{pa 3800 1620}%
\special{fp}%
\special{sh 1}%
\special{pa 3800 1620}%
\special{pa 3820 1554}%
\special{pa 3800 1568}%
\special{pa 3780 1554}%
\special{pa 3800 1620}%
\special{fp}%
\put(47.9000,-10.0000){\makebox(0,0){$j$}}%
\put(27.8000,-8.2000){\makebox(0,0){$\cdots$}}%
\put(13.9000,-8.2000){\makebox(0,0){$z$}}%
\put(8.0000,-10.0000){\makebox(0,0){$\widetilde{\Phi}_j^*(z)=$}}%
%
\special{pn 8}%
\special{pa 4600 1010}%
\special{pa 1200 1010}%
\special{fp}%
\special{sh 1}%
\special{pa 1200 1010}%
\special{pa 1268 1030}%
\special{pa 1254 1010}%
\special{pa 1268 990}%
\special{pa 1200 1010}%
\special{fp}%
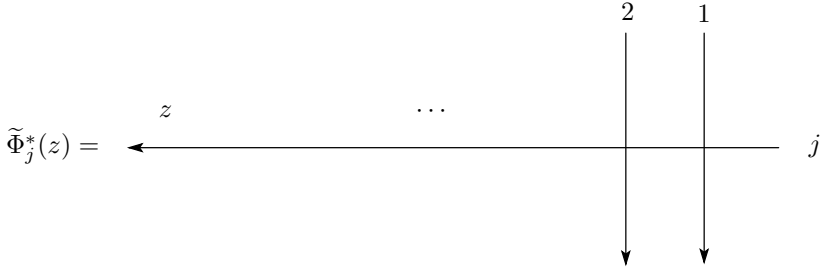
\put(24.9000,-18.1000){\makebox(0,0){Fig.6. dual Vertex operator $\widetilde{\Phi}_j^*(z)$}}%
\end{picture}%

\section{$K$-matrix}
\label{appendix1}

In this appendix, we classify diagonal solutions of the
boundary Yang-Baxter equation associated with $U_q(\widehat{sl}(M+1|N+1))$.
Let us set the vector space 
$V=\oplus_{j=1}^{M+N+2}{\bf C}v_j$.
Let us consider
the $R$-matrix $\bar{R}(z) \in {\rm End}(V \otimes V)$
introduced in (\ref{def:R-matrix1}), 
(\ref{def:R-matrix2}), (\ref{def:R-matrix3}) and (\ref{def:R-matrix4}). 
Non-zero elements of the $R$-matrix are restricted to the followings.
\begin{eqnarray}
\bar{R}(z)_{i,j}^{i,j}\neq 0,~~\bar{R}(z)_{i,j}^{j,i}\neq 0,~~~
(1\leq i,j \leq M+N+2).
\end{eqnarray}
Let us study the $K$-matrix $\bar{K}(z) \in {\rm End}(V)$ 
defined as followings.
\begin{eqnarray}
\bar{K}(z) \in {\rm End}(V),
~~~\bar{K}(z)v_j=\sum_{k=1}^{M+N+2}v_k \bar{K}(z)_k^j,
\end{eqnarray}
where we assume the diagonal matrix.
\begin{eqnarray}
\bar{K}(z)_k^j=\delta_{j,k}\bar{K}(z)_j^j.
\end{eqnarray}
The graded boundary Yang-Baxter equation
\begin{eqnarray}
\bar{K}_2(z_2)
\bar{R}_{21}(z_1z_2)
\bar{K}_1(z_1)
\bar{R}_{12}(z_1/z_2)=
\bar{R}_{21}(z_1/z_2)
\bar{K}_1(z_1)
\bar{R}_{12}(z_1z_2)
\bar{K}_2(z_2),
\end{eqnarray}
is equivalent to the following two relations
for $1\leq j < k \leq M+N+2$.
\begin{eqnarray}
\bar{R}(z)_{j,k}^{j,k}=\bar{R}(z)_{k,j}^{k,j},
\label{app:K-matrix1}
\end{eqnarray}
\begin{eqnarray}
&&\bar{R}(z_1/z_2)_{j,k}^{j,k}
\left(\bar{R}(z_1 z_2)_{k,j}^{j,k}
\bar{K}(z_1)_k^k \bar{K}(z_2)_k^k 
-
\bar{R}(z_1 z_2)_{j,k}^{k,j}
\bar{K}(z_1)_j^j \bar{K}(z_2)_j^j
\right)\nonumber\\
&+&
\bar{R}(z_1 z_2)_{j,k}^{j,k}\left(
\bar{R}(z_1/z_2)_{j,k}^{k,j}
\bar{K}(z_2)_k^k 
\bar{K}(z_1)_j^j
-\bar{R}(z_1/z_2)_{k,j}^{j,k}
\bar{K}(z_1)_k^k \bar{K}(z_2)_j^j
\right)=0.\label{app:K-matrix2}
\end{eqnarray}
The first condition (\ref{app:K-matrix1}) holds
for (\ref{def:R-matrix2}), (\ref{def:R-matrix3}), (\ref{def:R-matrix4}).
The second condition (\ref{app:K-matrix2}) is written as following.
\begin{eqnarray}
\left(1-\frac{z_1}{z_2}\right)
\left(z_1 z_2-\frac{\bar{K}(z_1)_j^j}{
\bar{K}(z_1)_k^k}\frac{\bar{K}(z_2)_j^j}{
\bar{K}(z_2)_k^k}\right)
+\left(1-z_1 z_2\right)
\left(\frac{
\bar{K}(z_1)_j^j}{\bar{K}(z_1)_k^k}
-\frac{z_1}{z_2}\frac{\bar{K}(z_2)_j^j}{\bar{K}(z_2)_k^k}\right)=0.
\label{app:K-matrix3}
\end{eqnarray}
Differentiating partially (\ref{app:K-matrix3}), at $(z_1,z_2)=(z,1)$,
with respect to $z_2$,
we have the following necessary condition.
\begin{eqnarray}
\frac{\bar{K}(z)_j^j}{
\bar{K}(z)_k^k}=\frac{\displaystyle 1-\beta z}{\displaystyle 1-\beta/z}
~~~~~(\beta \in {\bf C}).
\end{eqnarray}
This satisfies (\ref{app:K-matrix3}) for all $\beta \in {\bf C}$.
Taking into account of
simultaneous compatibility for $1\leq j<k \leq M+N+2$,
we have the following three kinds of general diagonal solutions of the boundary Yang-Baxter equation
associated with $U_q(\widehat{sl}(M+1|N+1))$.
\\~\\
{CASE 1}~:~One diagonal element. Dirichlet boundary condition.
\begin{eqnarray}
\bar{K}(z)_k^j=\delta_{j,k}.
\end{eqnarray}
{CASE 2}~:~Two different diagonal elements. We assume $1\leq L \leq M+N+1$ and $r \in {\bf C}$.
\begin{eqnarray}
\bar{K}(z)_k^j=\left\{\begin{array}{cc}
1& (1\leq j=k \leq L),\\
\frac{\displaystyle 1-r/z}{\displaystyle 1-r z}& (L+1\leq j=k \leq M+N+2),\\
0& (1\leq j \neq k \leq M+N+2).
\end{array}
\right.
\end{eqnarray}
{CASE 3}~:~Three different diagonal elements. We assume $1\leq L$,
$1\leq K$, $L+K \leq M+N+1$ and $r \in {\bf C}$.
\begin{eqnarray}
\bar{K}(z)_k^j=
\left\{\begin{array}{cc}
1& (1\leq j=k \leq L),\\
\frac{\displaystyle 1-r/z}{\displaystyle 1-r z}&
(L+1 \leq j=k \leq L+K),\\
z^{-2}& (L+K+1 \leq j=k \leq M+N+2),\\
0& (1\leq j\neq k \leq M+N+2).
\end{array}
\right.\label{def:tildeK3}
\end{eqnarray}
{\it Note. 
In the earlier studies \cite{LYH, BR},
Case I and Case II have been studied.
However
Case III is missing in the earlier studies \cite{LYH, BR}.
For instance, we have new solution for $U_q(\widehat{sl}(2|1))$,
which has three different diagonal elements.}
\begin{eqnarray}
\bar{K}(z)=\left(
\begin{array}{ccc}
1&0&0\\
0&\frac{\displaystyle 1-r/z}{\displaystyle 1-r z}&0\\
0&0&z^{-2}
\end{array}\right).
\end{eqnarray}

\section{Normal Ordering}
\label{appendix2}

In this appendix we summarize the normal orderings.
The following normal orderings are convenient
to calculations in a proof of main theorem.
\begin{eqnarray}
e^{h_+^{*1}(z)}e^{-h_-^1(w)}
&=&\frac{1}{(1-qw/z)}::,\\
e^{-h_+^1(w)}e^{h_-^{*1}(z)}
&=&\frac{1}{(1-qz/w)}::,\\
e^{h_+^{*1}(z)}e^{-h_-^j(w)}
&=&1::~~~(2\leq j \leq M+N+1),\\
e^{-h_+^j(w)}e^{h_-^{*1}(z)}~&=&
1::~~~(2\leq j \leq M+N+1),\\
e^{-h_+^j(w_1)}e^{-h_-^{j+1}(w_2)}
&=&
\frac{1}{(1-qw_2/w_1)}::~~~(1\leq j \leq M),\\
e^{-h_+^{j+1}(w_1)}e^{-h_-^j(w_2)}&=&
\frac{1}{(1-qw_2/w_1)}::
~~~(1\leq j \leq M),\\
e^{-h_+^{M+j}(w_1)}e^{-h_-^{M+j+1}(w_2)}&=&
(1-qw_2/w_1)::~~~(1\leq j \leq M),\\
e^{-h_+^{M+j+1}(w_1)}e^{-h_-^{M+j}(w_2)}&=&
(1-qw_2/w_1)::~~~(1\leq j \leq N).
\end{eqnarray}
The following normal orderings are convenient to
get the integral representations of the vertex operators.
\begin{eqnarray}
\phi_1^*(z)X_1^{-}(qw)
&=&
e^{\frac{\pi \sqrt{-1}}{M}}\frac{1}{qz(1-qw/z)}::,\\
X_1^{-}(qw)\phi_1^*(z)&=&
-e^{\frac{\pi \sqrt{-1}}{M}}\frac{1}{qw(1-qz/w)}
::,\\
\phi_1^*(z)X_j^{-}(w)
&=&1::~~~(2\leq j \leq M),\\
X_j^{-}(w)\phi_1^*(z)&=&
1 ::~~~(2\leq j \leq M),\\
\phi_1^*(z)X_{M+1,\epsilon}^{-}(w)
&=&-1::~~~(\epsilon=\pm),\\
X_{M+1,\epsilon}^{-}(w)\phi_1^*(z)&=&1 ::~~~(\epsilon=\pm),
\\
X_j^{-}(qw_1)X_{j+1}^{-}(qw_2)&=&\frac{1}{qw_1(1-qw_2/w_1)}::~~~~
(1\leq j \leq M),
\\
X_{j+1}^{-}(qw_1)X_j^{-}(qw_2)&=&\frac{-1}{qw_1(1-qw_2/w_1)}::~~~~
(1\leq j \leq M),
\\
X_j^{-}(qw_1)X_k^{-}(qw_2)&=&1
::~~~(|j-k|\geq 2),
\\
X_{M+j,+}^{-}(w_1)X_{M+j+1,\epsilon}^{-}(w_2)
&=&
\frac{(1-qw_2/w_1)}{q(1-w_2/qw_1)}::~~~(\epsilon=\pm),
\\
X_{M+1+j,+}^{-}(w_1)X_{M+j,\epsilon}^{-}(w_2)
&=&1::~~~(\epsilon=\pm),
\\
X_{M+j,-}^{-}(w_1)X_{M+j+1,\epsilon}^{-}(w_2)&=&
q::~~~(\epsilon=\pm),
\\
X_{M+j+1,-}^{-}(w_1)X_{M+j,\epsilon}^{-}(w_2)&=&
\frac{(1-qw_1/w_2)}{(1-w_1/qw_2)}::~~~(\epsilon=\pm).
\end{eqnarray}

\end{appendix}

\end{document}